\documentclass[journal,draftclsnofoot,onecolumn]{IEEEtran}
\usepackage{graphicx}
\usepackage{amssymb}
\usepackage{subfigure}
\usepackage{caption}
\renewcommand{\theequation}{\arabic{section}.\arabic{equation}}
\newtheorem{theorem}{Theorem}

\newtheorem{lemma}{Lemma}

\newtheorem{corollary}{Corollary}

\begin{document}

\title{Secure Communication over Finite State Multiple-Access Wiretap Channel with Delayed Feedback}

\author{Bin~Dai,~\IEEEmembership{Member,~IEEE,}
Zheng~Ma,~\IEEEmembership{Member,~IEEE,}
Ming~Xiao,~\IEEEmembership{Senior Member,~IEEE,}
Xiaohu~Tang,~\IEEEmembership{Senior Member,~IEEE}
and~Pingzhi~Fan,~\IEEEmembership{Fellow,~IEEE}
\thanks{B. Dai is with the
School of Information Science and Technology,
Southwest JiaoTong University, Chengdu, China,
and with
the National Mobile Communications Research Laboratory, Southeast University, Nanjing, China,
e-mail: daibin@home.swjtu.edu.cn.}
\thanks{Z. Ma, X. Tang and P. Fan are with the
School of Information Science and Technology,
Southwest JiaoTong University, Chengdu, China,
e-mail: zma@home.swjtu.edu.cn, xhutang@swjtu.edu.cn, pzfan@swjtu.edu.cn.}
\thanks{M. Xiao is with the
School of Electrical
Engineering and the ACCESS Linnaeus Center, Royal Institute of Technology, Sweden,
e-mail: mingx@kth.se.}
}

\maketitle

\begin{abstract}

Recently, it has been shown that the time-varying multiple-access channel (MAC) with perfect channel state information (CSI)
at the receiver and delayed feedback CSI at the transmitters can be modeled as
the finite state MAC (FS-MAC) with delayed state feedback, where the time variation of the channel is characterized by the
statistics of the underlying state process. To study the fundamental limit of the secure transmission over multi-user wireless communication systems,
we re-visit the FS-MAC with delayed state feedback by considering an external eavesdropper, which we call
the finite state multiple-access wiretap channel (FS-MAC-WT) with delayed feedback.
The main contribution of this paper is to show that taking full advantage of the delayed channel output feedback helps to increase the secrecy rate region
of the FS-MAC-WT with delayed state feedback, and the results of this paper are further illustrated by a degraded Gaussian fading example.

\end{abstract}


\section{Introduction \label{secI}}

In the future 5G network, a huge amount of private information, e.g. personal financial data and medical records, will be transmitted through
wireless channels. Due to the broadcast nature of the wireless communication,
information transmitted in the wireless channels is more vulnerable to eavesdropping, and thus the secure transmission
over the wireless channels is one of the most pressing problems in the design of 5G network. The physical layer security (PLS)
is a useful tool to solve the secure transmission problem in the 5G network, and it was founded by Wyner \cite{Wy} in his milestone paper on the wiretap channel.
In \cite{Wy}, Wyner introduced secrecy criteria into a physically degraded broadcast channel,
and proposed the notion of secrecy capacity to characterize the maximum achievable secrecy rate.
Secrecy capacities of the discrete memoryless and Gaussian cases of the physically degraded
wiretap channel are respectively determined in \cite{Wy} and \cite{CH}.
Later, Csisz$\acute{a}$r et al. \cite{CK} extended Wyner's physically degraded model \cite{Wy} to the
general broadcast channel with confidential messages (BC-CM), where an additional common message was transmitted together with the confidential message,
and this common message was allowed to be decoded by the eavesdropper. Secrecy capacity regions of the discrete memoryless and Gaussian cases of the BC-CM
are respectively determined in \cite{CK} and \cite{LPS}. The coding schemes proposed in \cite{Wy} and \cite{CK} have become standard techniques
for the theory of PLS.

Based on the work of \cite{Wy} and \cite{CK}, recently,
the wireless fading channel is modeled as the parallel wiretap channel \cite{LPS, LPS1}, where the transition probability of the channel
depends on the channel state information (CSI), the CSI is assumed to be i.i.d. generated, the channel is discrete memoryless for a given CSI,
the overall channel can be decomposed into several sub-channels, and the transition probability of each sub-channel is with respect to a certain value of the CSI.
Liang et al. \cite{LPS, LPS1} established the secrecy capacity of this
parallel wiretap channel model, and further derived the secrecy capacity of the corresponding fading wiretap channel. Here note that the fading wiretap channel
in \cite{LPS, LPS1} is also assumed to be equipped with i.i.d. generated CSI, and the CSI is known by the legitimate receiver and the transmitter.
Besides the work of \cite{LPS, LPS1}, other related works in the wiretap channel with i.i.d. CSI are in \cite{MVL}-\cite{CG}, and the recent results on
the PLS of multi-user channel models in the presence of i.i.d. CSI are in \cite{1}-\cite{dainew2}.

In practical wireless fading channels, the CSI
at each time instant is not independent of each other.
A practical model for the wireless fading channel with CSI was provided in \cite{wang} and \cite{zhang}, which was called the finite state Markov channel (FSMC).
The CSI in the FSMC is not i.i.d., and in fact it goes through a Markov process.
The capacity of the FSMC was first studied by Goldsmith et al.
\cite{god}, where the channel capacity was characterized in a multi-letter form.
A single-letter form of the capacity of the FSMC was investigated by Viswanathan \cite{vis}.
In \cite{vis}, Viswanathan investigated
the scenario that the CSI of the FSMC is entirely known by the receiver, and the receiver sends the CSI together with the received channel output back
to the transmitter through a noiseless feedback channel. Since this feedback is often not instantaneous, Viswanathan assumed that the transmitter
gets the feedback CSI and channel output after some time delay. The communication scenario described in \cite{vis}
is called the FSMC with delayed feedback, and the capacity of this model was determined in a single-letter form. Moreover,
Viswanathan further found out that the feedback channel output does not help to increase the channel capacity, which is similar to
Shannon's classical fact that the channel output feedback makes no contribution to the capacity of a discrete memoryless channel (DMC) \cite{coverx}.
Later, Basher et al. \cite{bash} extended Viswanathan's work \cite{vis} to a multiple-access situation, which was called
the finite state multiple-access channel (FS-MAC) with delayed state feedback. In this extended model,
the receiver sends the state back to the two
transmitters via two noiseless feedback channels, respectively, and the transmitters receive the state after some time delay.
The capacity region of this extended model is also determined in a single-letter form.

For the upcoming 5G wireless networks,
establishing more practical PLS models for the mobile
communication systems attracts researchers' interest. In \cite{san1,san},
a multi-letter form of the secrecy capacity of the wiretap channel with memory CSI is given,
which is not computable. Single-letter form of the secrecy capacity of the wiretap channel with dependent CSI remains open.
Recently, Dai et al. \cite{dainew}
re-visited the wiretap channel with dependent CSI
by considering the situation that
the CSI goes through a Markov process, it is entirely obtained by the legal receiver and the eavesdropper, and the transmitter
obtains the CSI via a feedback channel after some time delay. Dai et al. \cite{dainew} determined the secrecy capacity (in a single-letter form) of this model
for a degraded case.

In this paper, establishing a more practical PLS model
for the up-link of the wireless communication systems motivates us to study
the finite state multiple-access wiretap channel (FS-MAC-WT)
with delayed feedback,
see the following Figure \ref{f2}. The transition probability of the channel is governed by a state $S$ which
goes through a Markov process. At time $i$, the legal receiver obtains $Y_{i}$ and $S_{i}$, and delivers them to the
transmitters via two noiseless feedback channels with delay times $d_{1}$ and $d_{2}$, respectively. The $i$-th time channel encoders
produce the channel inputs on the basis of the transmitted messages and the delayed feedback.
In addition, at time $i$, an eavesdropper receives $Z_{i}$ and also perfectly obtains $S_{i}$.
The delay times $d_{1}$ and $d_{2}$ are supposed to be entirely known by
the legal receiver, the eavesdropper and the transmitters.
Here note that the FS-MAC-WT with delayed feedback in Figure \ref{f2} combines Wyner's wiretap channel \cite{Wy} with
Basher et al.'s FS-MAC with delayed state feedback \cite{bash}.
Unlike Viswanathan's fact that the feedback channel output does not help to increase the channel capacity \cite{vis},
we find out that
the full use of the delayed feedback channel output may increase the achievable secrecy rate region of the FS-MAC-WT with delayed state feedback,
where the ``full use'' indicates that the feedback channel output can not only be used to produce secret keys known by the legal receiver
and the transmitters \footnote{The idea of using noiseless feedback to produce secret key encrypting the transmitted message is from Ahlswede and Cai's
work on the wiretap channel with noiseless feedback \cite{AC}}, but also be used to allow the transmitters to cooperate with each other.
The main contribution of this paper is to provide inner and outer bounds on the secrecy capacity region of
the FS-MAC-WT with delayed feedback.
From a degraded Gaussian fading example, we show the
effects of feedback delay and channel memory on the secrecy sum rate of the FS-MAC-WT with delayed feedback, and show that the channel output feedback
enhances the capacity bounds for the FS-MAC-WT with only delayed state feedback.

\begin{figure}[htb]
\centering
\includegraphics[scale=0.5]{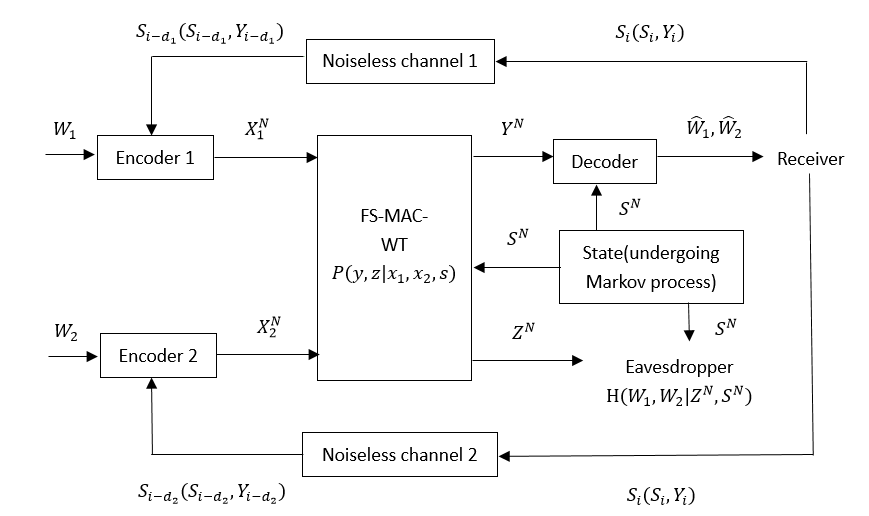}
\caption{The FS-MAC-WT with delayed feedback}
\label{f2}
\end{figure}

Throughout this paper, random variables are written in uppercase letter (e.g. $V$), real values are written in lowercase letter (e.g. $v$),
and alphabets are written in calligraphic letter (e.g. $\mathcal{V}$). The random vector and its value are written in a similar way.
The probability $Pr\{V=v\}$ is shortened by $P_{V}(v)$. In addition, for the remainder of this paper, the base of the logarithm is $2$.
The outline of this paper is organized as follows.
The definitions and the main results are given in Section \ref{secII};
a degraded Gaussian fading example is shown in Section \ref{secIII}; and
a summary of this work is given in Section \ref{secIV}.

\section{Definitions and the Main Results}\label{secII}

Now we consider the FS-MAC-WT with state at the legal receiver and delayed feedback at the transmitters with
delays $d_{1}$ and $d_{2}$, respectively, as shown in Figure \ref{f2}. The remainder of this section is organized as follows.
Subsection \ref{secII.1} is about the definitions of the channel, Subsection \ref{secII.2} is about the code description, and
Subsection \ref{secII.3} is about the main results.

\subsection{Channel Model}\label{secII.1}

The channel consists of two transmitters, one legal receiver and an eavesdropper.
Each transmitter $j\in\{1,2\}$ chooses a message $W_{j}$ with equal probability from the set $\{1,2,...,2^{NR_{j}}\}$ and
independent of the other transmitter. At each time instant, the channel is in one of a finite number of states $\mathcal{S}=\{s_{1},s_{2},...,s_{k}\}$.
In each state, the channel is discrete memoryless with inputs alphabet $\mathcal{X}_{1}$, $\mathcal{X}_{2}$ and
outputs alphabet $\mathcal{Y}$, $\mathcal{Z}$. Let $S_{i}$, $X_{1,i}$, $X_{2,i}$, $Y_{i}$ and $Z_{i}$ be the channel inputs and outputs at time
$i$ ($i\in\{1,2,...,N\}$). The channel transition probability at time $i$ is governed by the state $S_{i}$
and is given by $P_{Y,Z|X_{1},X_{2},S}(y_{i},z_{i}|x_{1,i},x_{2,i},s_{i})$. Since the channel is discrete memoryless, we have
\begin{equation}\label{e202}
P_{Y^{N},Z^{N}|X_{1}^{N},X_{2}^{N},S^{N}}(y^{N},z^{N}|x_{1}^{N},x_{2}^{N},s^{N})=\prod_{i=1}^{N}P_{Y,Z|X_{1},X_{2},S}(y_{i},z_{i}|x_{1,i},x_{2,i},s_{i}).
\end{equation}
The channel state process $\{S_{i}\}$ ($i\in\{1,2,...,N\}$) is a stationary irreducible aperiodic ergodic Markov chain,
and given the previous states, the current state is independent of the channel inputs and outputs, i.e.,
\begin{equation}\label{e203}
Pr\{S_{i}=s_{i}|X_{1}^{i}=x_{1}^{i},X_{2}^{i}=x_{2}^{i},Y^{i}=y^{i},Z^{i}=z^{i},S^{i-1}=s^{i-1}\}=Pr\{S_{i}=s_{i}|S_{i-1}=s_{i-1}\}.
\end{equation}
The state process $\{S_{i}\}$ is also independent of the messages $W_{1}$ and $W_{2}$, and hence
\begin{equation}\label{e203.xxs}
Pr\{S^{N}=s^{N},W_{1}=w_{1},W_{2}=w_{2}\}=\prod_{i=1}^{N}Pr\{S_{i}=s_{i}|S_{i-1}=s_{i-1}\}Pr\{W_{1}=w_{1}\}Pr\{W_{2}=w_{2}\}.
\end{equation}
Denote the
one-step transition probability matrix of the process $\{S_{i}\}$ by $K$, and the steady probability
of the state process $\{S_{i}\}$ by $\pi$. Then the
joint probability mass function $Pr\{S_{i}=s_{l}, S_{i-d}=s_{j}\}$ can be expressed by
\begin{equation}\label{e204}
\pi_{d}(S_{i}=s_{l}, S_{i-d}=s_{j})=\pi(s_{j})K^{d}(s_{l},s_{j}),
\end{equation}
where $s_{l}$ and $s_{j}$ are the $l$-th and $j$-th elements of the state alphabet $\mathcal{S}$, respectively,
and $K^{d}(s_{l},s_{j})$ is the $(l,j)$-th element of the $d$-step transition
probability matrix $K^{d}$ of the channel state process $\{S_{i}\}$.
\textbf{Without loss of generality, suppose that $d_{1}\geq d_{2}$, which indicates that at time $i$, the delayed
feedback state $S_{i-d_{1}}$ obtained by the transmitter $1$ is also known by the transmitter $2$ because $i-d_{1}\leq i-d_{2}$}, hence we have
\begin{equation}\label{e204.xxs}
Pr\{S_{i}=s_{l}, S_{i-d_{1}}=s_{j}, S_{i-d_{2}}=s_{v}\}=\pi(s_{j})K^{d_{1}-d_{2}}(s_{v},s_{j})K^{d_{2}}(s_{l},s_{v}),
\end{equation}
where $s_{l}, s_{j}, s_{v}\in \mathcal{S}$.

\subsection{Code Description}\label{secII.2}

For the FS-MAC-WT with delayed feedback, an ($N,2^{NR_{1}},2^{NR_{2}},d_{1},d_{2},\Delta,P_{e}$) code is composed of
\begin{itemize}
\item Two message sets $\mathcal{W}_{1}=\{1,2,...,2^{NR_{1}}\}$ and $\mathcal{W}_{2}=\{1,2,...,2^{NR_{2}}\}$.

\item At time instant $i$ ($i\in\{1,2,...,N\}$), the channel encoder $f_{ji}$ ($j\in\{1,2\}$) depends only on the
message $W_{j}$ and the delayed feedback $S^{i-d_{j}}$ (or $S^{i-d_{j}}$ and $Y^{i-d_{j}}$).
For the FS-MAC-WT with delayed state feedback,
the channel input $X_{j,i}$ ($j\in\{1,2\}$) at time instant $i$ is defined by
\begin{equation}\label{e206.xx1}
X_{j,i}=
\left\{
\begin{array}{ll}
f_{j,i}(W_{j}), & 1\leq i\leq d_{j}\\
f_{j,i}(W_{j},S^{i-d_{j}}), & d_{j}+1\leq i\leq N,
\end{array}
\right.
\end{equation}
and for the FS-MAC-WT with delayed state and channel output feedback,
the channel input $X_{j,i}$ ($j\in\{1,2\}$) is defined by
\begin{equation}\label{e206}
X_{j,i}=
\left\{
\begin{array}{ll}
f_{j,i}(W_{j}), & 1\leq i\leq d_{j}\\
f_{j,i}(W_{j},S^{i-d_{j}},Y^{i-d_{j}}), & d_{j}+1\leq i\leq N,
\end{array}
\right.
\end{equation}
where the channel encoder $f_{j,i}$ ($j\in\{1,2\}$) at time instant $i$
is stochastic, i.e., the encoding function is a random mapping
(not deterministic).

\item The channel decoder is a mapping $\psi$
\begin{equation}\label{e207}
\psi: \,\, \mathcal{Y}^{N}\times \mathcal{S}^{N}\rightarrow \mathcal{W}_{1}\times \mathcal{W}_{2},
\end{equation}
that maps the legal receiver's channel output $Y^{N}$ and the state $S^{N}$ into the message sets.
The average decoding error probability $P_{e}$ is denoted by
\begin{equation}\label{e208}
P_{e}=\frac{1}{2^{N(R_{1}+R_{2})}}\sum_{w_{1}=1}^{2^{NR_{1}}}\sum_{w_{2}=1}^{2^{NR_{2}}}\sum_{s^{N}}
P_{S^{N}}(s^{n})Pr\{\psi(y^{N},s^{N})\neq (w_{1},w_{2})|(w_{1},w_{2}) \,\,\mbox{was sent}\}.
\end{equation}

\item
Since state $S^{N}$ is also perfectly known by the eavesdropper, his equivocation about the messages
is denoted by
\begin{equation}\label{e210}
\Delta=\frac{1}{N}H(W_{1},W_{2}|Z^{N},S^{N}).
\end{equation}
Applying similar criteria in \cite{Wy} and \cite{CK}, we define an achievable secrecy rate pair $(R_{1}, R_{2})$ as follows.
Given a pair $(R_{1}, R_{2})$, if for arbitrarily small $\epsilon$,
there exists a
sequence of ($N,2^{NR_{1}},2^{NR_{2}},d_{1},d_{2},\Delta,P_{e}$) codes satisfying
\begin{eqnarray}\label{e202}
&&\frac{\log\parallel \mathcal{W}_{1}\parallel}{N}\geq R_{1}-\epsilon,
\frac{\log\parallel \mathcal{W}_{2}\parallel}{N}\geq R_{2}-\epsilon,
\Delta\geq R_{1}+R_{2}-\epsilon, \,\,P_{e}\leq \epsilon,
\end{eqnarray}
the pair $(R_{1}, R_{2})$ is an achievable secrecy rate pair.
Here we note that the joint secrecy ensures the individual secrecy, i.e., $\frac{1}{N}H(W_{1},W_{2}|Z^{N},S^{N})\geq R_{1}+R_{2}-\epsilon$ implies that
$\frac{1}{N}H(W_{j}|Z^{N},S^{N})\geq R_{j}-\epsilon$ for $j=1, 2$. The proof of this property is in \cite[p. 5691, Lemma 15]{keg}, and thus we omit it here.
\end{itemize}

\subsection{Main Results}\label{secII.3}

The secrecy capacity region consists of all achievable secrecy rate pairs.
Denote the secrecy capacity region of the FS-MAC-WT with delayed state and channel output feedback by $\mathcal{C}_{sf}$, and
the secrecy capacity region of the FS-MAC-WT with only delayed state feedback by $\mathcal{C}_{s}$.
In the remainder of this subsection, the following Theorems \ref{T3} and \ref{T4} provide bounds on $\mathcal{C}_{sf}$,
and Theorems \ref{T1} and \ref{T2} give bounds on $\mathcal{C}_{s}$.

\begin{theorem}\label{T3}
An inner bound $\mathcal{C}^{in}_{sf}$ on $\mathcal{C}_{sf}$ is given by
\begin{eqnarray*}
&&\mathcal{C}^{in}_{sf}=\{(R_{1}, R_{2}): 0\leq R_{1}\leq I(X_{1};Y|X_{2},S,\tilde{S}_{1},\tilde{S}_{2},Q),\\
&&0\leq R_{2}\leq I(X_{2};Y|X_{1},S,\tilde{S}_{1},\tilde{S}_{2},Q),\\
&&0\leq R_{1}+R_{2}\leq \min\{I(X_{1};Y|X_{2},S,\tilde{S}_{1},\tilde{S}_{2},Q)+I(X_{2};Y|X_{1},S,\tilde{S}_{1},\tilde{S}_{2},Q),
I(X_{1},X_{2};Y|S,\tilde{S}_{1},\tilde{S}_{2})\}\\
&&-I(X_{1},X_{2};Z|S,\tilde{S}_{1},\tilde{S}_{2})
+\min\{I(X_{1},X_{2};Z|S,\tilde{S}_{1},\tilde{S}_{2}),H(Y|Z,X_{1},X_{2},S,\tilde{S}_{1},\tilde{S}_{2})\},
\end{eqnarray*}
where the joint probability
\begin{eqnarray}\label{dota1}
&&P_{QS\tilde{S}_{1}\tilde{S}_{2}X_{1}X_{2}YZ}(q,s,\tilde{s}_{1},\tilde{s}_{2},x_{1},x_{2},y,z)\nonumber\\
&&=P_{YZ|X_{1}X_{2}S}(y,z|x_{1},x_{2},s)P_{X_{1}|\tilde{S}_{1},Q}(x_{1}|\tilde{s}_{1},q)
P_{X_{2}|\tilde{S}_{1},\tilde{S}_{2},Q}(x_{2}|\tilde{s}_{1},\tilde{s}_{2},q)\cdot \nonumber\\
&&P_{Q|\tilde{S}_{1}}(q|\tilde{s}_{1})K^{d_{2}}(s,\tilde{s}_{2})K^{d_{1}-d_{2}}(\tilde{s}_{2},\tilde{s}_{1})\pi(s_{1}),
\end{eqnarray}
and the cardinality of $Q$ is bounded by $|\mathcal{Q}|\leq 2$.
\end{theorem}

\begin{IEEEproof}
The proof of $|\mathcal{Q}|\leq 2$ is directly from the support lemma \cite[pp. 631-633]{network}, and thus we omit it here.
The inner bound $\mathcal{C}^{in}_{sf}$ is constructed by using the block Markov coding strategy for the feedback system and
the multiplexing coding scheme for the FSMC with delayed state feedback \cite{vis}, i.e., the messages $W_{1}=(W_{1,1},...,W_{1,n})$ and
$W_{2}=(W_{2,1},...,W_{2,n})$ are transmitted through $n$ blocks,
and in each block $i$ ($1\leq i\leq n$),
the messages $W_{1,i}=(W_{1,i,1},...,W_{1,i,k})$ and $W_{2,i}=(W_{2,i,1},...,W_{2,i,k})$ are divided into $k$ sub-messages,
where $W_{1,i,j}$ and $W_{2,i,j}$ ($1\leq j\leq k$) are with respect to the delayed feedback state $s_{i-d{1}}$ (here note that since $d_{1}\geq d_{2}$,
when transmitter $2$ receives his delayed state $s_{i-d{2}}$, he also knows $s_{i-d{1}}$).

In each block $i$, split the sub-messages $W_{1,i,j}$ and $W_{2,i,j}$ into two part, i.e., $W_{1,i,j}=(W_{1,i,j,0},W_{1,i,j,1})$
and $W_{2,i,j}=(W_{2,i,j,0},W_{2,i,j,1})$.
Here the sub-messages $W_{1,i,j,1}$ and $W_{2,i,j,1}$ will be
encrypted by keys produced from the delayed channel output feedback, and similar to the random binning technique used in Wyner's wiretap channel \cite{Wy},
the sub-messages $W_{1,i,j,0}$ and $W_{2,i,j,0}$ will be respectively protected by
the randomly produced dummy messages $W^{*}_{1,i,j}$ and $W^{*}_{2,i,j}$. In each block, the sub-messages $W_{1,i,j,0}$, $W_{1,i,j,1}$ and
the dummy message $W^{*}_{1,i,j}$ will be encoded as a part of the codeword $X_{1}^{N}$, and analogously,
$W_{2,i,j,0}$, $W_{2,i,j,1}$ and
$W^{*}_{2,i,j}$ will be encoded as a part of the codeword $X_{2}^{N}$. Finally, when the encoding for all the sub-messages of $W_{1,i}$ and $W_{2,i}$ are
completed, multiplexing all parts of $X_{1}^{N}$ and $X_{2}^{N}$, the transmitted codewords are chosen to be transmitted.

The auxiliary random variables $\tilde{S}_{1}$ and $\tilde{S}_{2}$ represent the delayed CSI $S_{i-d_{1}}$ and $S_{i-d_{2}}$, respectively.
In each block $i$ and for a fixed $j$, the auxiliary random variable $Q$ represents a sub-sequence of $q^{N}$ encoded by all the sub-messages
$W_{1,i-1,j,0}$, $W_{1,i-1,j,1}$, $W_{2,i-1,j,0}$, $W_{2,i-1,j,1}$ and all the dummy messages $W^{*}_{1,i-1,j}$ and $W^{*}_{2,i-1,j}$ for the previous block $i-1$
(here note that for $i=1$, we define $W_{1,i-1,j,0}=W_{1,i-1,j,1}=W_{2,i-1,j,0}=W_{2,i-1,j,1}=W^{*}_{1,i-1,j}=W^{*}_{2,i-1,j}=1$).
In block $i$, the transmitter $1$ ($2$) has already known the sequence $q^{n}$ for block $i$, and he attempts to decode the transmitter $2$ ($1$)'s codeword by
finding a unique $x_{2}^{N}$ ($x_{1}^{N}$) such that $x_{1}^{N}$, $x_{2}^{N}$, $q^{n}$, $y^{N}$ and $s^{N}$
(here $y^{N}$ and $s^{N}$ are delayed feedback channel output and state, respectively) are jointly typical. When each transmitter successfully
decodes the other one's codeword for block $i$, he extracts the messages in it, and chooses the sequence $q^{n}$ for block $i+1$ with encoded messages
$W_{1,i,j,0}$, $W_{1,i,j,1}$, $W_{2,i,j,0}$, $W_{2,i,j,1}$, $W^{*}_{1,i,j}$ and $W^{*}_{2,i,j}$, where $1\leq j\leq k$.

From the above encoding scheme, we see that in each block, the delayed channel output feedback $y^{N}$ is not only
used to produce secret keys encrypting the sub-message
$W_{1,i,j,1}$ and $W_{2,i,j,1}$, but also used to allow each transmitter to decode the other one's transmitted codeword. In Section \ref{secIII},
we show that this full use of the delayed channel output feedback helps to increase the achievable secrecy rate region
of the FS-MAC-WT with only delayed state feedback.
The detail of the proof of Theorem \ref{T3} is in Appendix \ref{appen4}.
\end{IEEEproof}

\begin{theorem}\label{T4}
An outer bound $\mathcal{C}^{out}_{sf}$ on $\mathcal{C}_{sf}$ is given by
\begin{eqnarray*}
&&\mathcal{C}^{out}_{sf}=\{(R_{1}, R_{2}): 0\leq R_{1}\leq I(V_{1};Y|U,S,\tilde{S}_{1},\tilde{S}_{2}),\\
&&0\leq R_{2}\leq I(V_{2};Y|U,S,\tilde{S}_{1},\tilde{S}_{2}),\\
&&0\leq R_{1}+R_{2}\leq \min\{H(Y|U,S,\tilde{S}_{1},\tilde{S}_{2},Z), I(V_{1},V_{2};Y|U,S,\tilde{S}_{1},\tilde{S}_{2})\}\},
\end{eqnarray*}
where
\begin{eqnarray}\label{dota2}
&&P_{UV_{1}V_{2}S\tilde{S}_{1}\tilde{S}_{2}X_{1}X_{2}YZ}(u,v_{1},v_{2},s,\tilde{s}_{1},\tilde{s}_{2},x_{1},x_{2},y,z)\nonumber\\
&&=P_{YZ|X_{1}X_{2}S}(y,z|x_{1},x_{2},s)P_{UV_{1}V_{2}S\tilde{S}_{1}\tilde{S}_{2}X_{1}X_{2}}(u,v_{1},v_{2},s,\tilde{s}_{1},\tilde{s}_{2},x_{1},x_{2}),
\end{eqnarray}
$U$ may be assumed to be a (deterministic) function of $V_{1}$ and $V_{2}$, and the alphabets of
the auxiliary random variables $U$, $V_{1}$ and $V_{2}$ satisfy $|\mathcal{U}|\leq |\mathcal{X}_{1}||\mathcal{X}_{2}||\mathcal{S}|+2$,
$|\mathcal{V}_{1}|\leq (|\mathcal{X}_{1}||\mathcal{X}_{2}||\mathcal{S}|+1)(|\mathcal{X}_{1}||\mathcal{X}_{2}||\mathcal{S}|+2)$ and
$|\mathcal{V}_{2}|\leq (|\mathcal{X}_{1}||\mathcal{X}_{2}||\mathcal{S}|+1)(|\mathcal{X}_{1}||\mathcal{X}_{2}||\mathcal{S}|+2)$, respectively.
\end{theorem}

\begin{IEEEproof}
See Appendix \ref{appen5}.
\end{IEEEproof}

The following Theorems \ref{T1} and \ref{T2} show the inner and outer bounds on the secrecy capacity region $\mathcal{C}_{s}$ of
the FS-MAC-WT with delayed state feedback.

\begin{theorem}\label{T1}
An inner bound $\mathcal{C}^{in}_{s}$ on $\mathcal{C}_{s}$ is given by
\begin{eqnarray*}
&&\mathcal{C}^{in}_{s}=\{(R_{1}, R_{2}): 0\leq R_{1}\leq I(X_{1};Y|X_{2},S,\tilde{S}_{1},\tilde{S}_{2},Q)-I(X_{1};Z|S,\tilde{S}_{1},\tilde{S}_{2},Q),\\
&&0\leq R_{2}\leq I(X_{2};Y|X_{1},S,\tilde{S}_{1},\tilde{S}_{2},Q)-I(X_{2};Z|S,\tilde{S}_{1},\tilde{S}_{2},Q),\\
&&0\leq R_{1}+R_{2}\leq I(X_{1},X_{2};Y|S,\tilde{S}_{1},\tilde{S}_{2},Q)-I(X_{1},X_{2};Z|S,\tilde{S}_{1},\tilde{S}_{2},Q)\},
\end{eqnarray*}
where the joint probability mass function $P_{QS\tilde{S}_{1}\tilde{S}_{2}X_{1}X_{2}YZ}(q,s,\tilde{s}_{1},\tilde{s}_{2},x_{1},x_{2},y,z)$
is given by (\ref{dota1}),
and the cardinality of the auxiliary random variable $Q$ satisfies $|\mathcal{Q}|\leq 6$.
\end{theorem}

\begin{IEEEproof}
Here $Q$ is a standard time sharing random variable which is used to increase the achievable secrecy rate region
$\mathcal{C}^{in}_{s}$. The proof of $|\mathcal{Q}|\leq 6$ is directly from the support lemma \cite[pp. 631-633]{network}, and thus we omit it here.
The inner bound $\mathcal{C}^{in}_{s}$ is constructed by simply combining Wyner's random binning coding scheme for the wiretap channel \cite{Wy} with
the multiplexing coding scheme for the FSMC with delayed state feedback \cite{vis}, i.e., the transmitted messages $W_{1}=(W_{1,1},...,W_{1,k})$ and
$W_{2}=(W_{2,1},...,W_{2,k})$ are divided into $k$ sub-messages,
where $W_{1,j}$ and $W_{2,j}$ ($1\leq j\leq k$) are with respect to the delayed feedback state $s_{i-d{1}}$.

The sub-messages $W_{1,j}$ and $W_{2,j}$ will be respectively protected by
the randomly produced dummy messages $W^{*}_{1,j}$ and $W^{*}_{2,j}$, i.e.,
the sub-messages $W_{i,j}$ ($i=1,2$) together with
the dummy message $W^{*}_{i,j}$ will be encoded as a part of the codeword $X_{i}^{N}$. Finally, when the encoding for all the
sub-messages of $W_{1}$ and $W_{2}$ are
completed, multiplexing all parts of $X_{i}^{N}$, the entire transmitted codeword $X_{i}^{N}$ is chosen to be transmitted.
The legal receiver attempts to find unique $x_{1}^{N}$ and $x_{2}^{N}$ such that $x_{1}^{N}$, $x_{2}^{N}$, $y^{N}$ and $s^{N}$ are jointly typical.

The achievability proof of Theorems \ref{T1} is similar to that of the multiple-access wiretap channel \cite{TY1},
and hence we omit the proof here.
\end{IEEEproof}

\begin{theorem}\label{T2}
An outer bound $\mathcal{C}^{out}_{s}$ on $\mathcal{C}_{s}$ is given by
\begin{eqnarray*}
&&\mathcal{C}^{out}_{s}=\{(R_{1}, R_{2}): 0\leq R_{1}\leq I(V_{1};Y|S,\tilde{S}_{1},\tilde{S}_{2},U)-I(V_{1};Z|S,\tilde{S}_{1},\tilde{S}_{2},U),\\
&&0\leq R_{2}\leq I(V_{2};Y|S,\tilde{S}_{1},\tilde{S}_{2},U)-I(V_{2};Z|S,\tilde{S}_{1},\tilde{S}_{2},U),\\
&&0\leq R_{1}+R_{2}\leq I(V_{1},V_{2};Y|S,\tilde{S}_{1},\tilde{S}_{2},U)-I(V_{1},V_{2};Z|S,\tilde{S}_{1},\tilde{S}_{2},U)\},
\end{eqnarray*}
where the joint probability $P_{UV_{1}V_{2}S\tilde{S}_{1}\tilde{S}_{2}X_{1}X_{2}YZ}(u,v_{1},v_{2},s,\tilde{s}_{1},\tilde{s}_{2},x_{1},x_{2},y,z)$
is given by (\ref{dota2}),
$U$ may be assumed to be a (deterministic) function of $V_{1}$ and $V_{2}$, and the alphabets of
the auxiliary random variables $U$, $V_{1}$ and $V_{2}$ satisfy $|\mathcal{U}|\leq |\mathcal{X}_{1}||\mathcal{X}_{2}||\mathcal{S}|+1$,
$|\mathcal{V}_{1}|\leq (|\mathcal{X}_{1}||\mathcal{X}_{2}||\mathcal{S}|+2)(|\mathcal{X}_{1}||\mathcal{X}_{2}||\mathcal{S}|+3)$ and
$|\mathcal{V}_{2}|\leq (|\mathcal{X}_{1}||\mathcal{X}_{2}||\mathcal{S}|+2)(|\mathcal{X}_{1}||\mathcal{X}_{2}||\mathcal{S}|+3)$, respectively.

\end{theorem}

\begin{IEEEproof}
First, note that the bounds on the cardinality of the auxiliary random variables $U$, $V_{1}$ and $V_{2}$ are directly from the support lemma \cite[pp. 633-634]{network},
and thus we omit the proof here. Then, the outer bound $\mathcal{C}^{out}_{s}$ is obtained by the following steps:
\begin{itemize}

\item Using the definition (\ref{e202}) (including $\frac{1}{N}H(W_{j}|Z^{N},S^{N})\geq R_{j}-\epsilon$ for $j=1, 2$) and
Fano's inequality, the secrecy transmission rates $R_{1}$, $R_{2}$ and $R_{1}+R_{2}$ are upper bounded by
$\frac{1}{N}(I(W_{1};Y^{N}|S^{N})-I(W_{1};Z^{N}|S^{N}))$, $\frac{1}{N}(I(W_{2};Y^{N}|S^{N})-I(W_{2};Z^{N}|S^{N}))$
and $\frac{1}{N}(I(W_{1},W_{2};Y^{N}|S^{N})-I(W_{1},W_{2};Z^{N}|S^{N}))$, respectively.

\item The definition of the auxiliary random variables in $\mathcal{C}^{out}_{s}$ follows that in \cite{CK}. To be specific,
in \cite{CK}, Csisz$\acute{a}$r and K\"{o}rner define the auxiliary random variable $U_{i}$ as $U_{i}\triangleq (Y^{i-1},Z_{i+1}^{N})$.
In this paper, considering the delayed feedback states $S_{i-d_{1}}$ and $S_{i-d_{2}}$, we slightly modify the definition of $U_{i}$ in \cite{CK},
i.e., we define $U_{i}\triangleq (Y^{i-1},Z_{i+1}^{N},S^{N})$, and here note that $S_{i-d_{1}}$ and $S_{i-d_{2}}$ are included in $S^{N}$.
Then, similar to the definition in \cite{CK}, we let $V_{1,i}\triangleq (U_{i}, W_{1})$ and $V_{2,i}\triangleq (U_{i}, W_{2})$.

\item Applying chain rule and the above definitions of the auxiliary random variables
$U_{i}$, $V_{1,i}$ and $V_{2,i}$ into the upper bounds of $R_{1}$, $R_{2}$ and $R_{1}+R_{2}$, and using Csisz$\acute{a}$r's equality \cite{CK}
to eliminate some identities in these bounds, the outer bound $\mathcal{C}^{out}_{s}$ is obtained.
\end{itemize}

The proof is similar to that of Theorem \ref{T4}, hence we omit the proof here.
\end{IEEEproof}

\section{Degraded Gaussian Fading Example}\label{secIII}

\subsection{Capacity Results on the Degraded Gaussian Fading Case}\label{sub31}

In this subsection, we compute the bounds in Theorems \ref{T3}-\ref{T2} for a degraded Gaussian fading case.
and investigate how the delays $d_{1}$ and $d_{2}$ affect the secrecy rate regions.

For the degraded Gaussian fading case, at time instant $i$ ($1\leq i\leq N$), the channel inputs and outputs satisfy
\begin{equation}\label{e301}
Y_{i}=h_{1}(s_{i})X_{1,i}+h_{2}(s_{i})X_{2,i}+N_{s_{i}}, \,\, Z_{i}=h_{3}(s_{i})Y_{i}+N_{w,i},
\end{equation}
where $s_{i}$ is the $i$-th time channel state which goes through a Markov process, $h_{j}(s_{i})$ ($j=1,2$) is the fading process of the transmitter $j$, and
$h_{3}(s_{i})$ is the fading process of the eavesdropper. In this example, we assume that $h_{1}(s_{i})$, $h_{2}(s_{i})$ and $h_{3}(s_{i})$
are related with the $i$-th time channel state $s_{i}$.
The noise $N_{s_{i}}$ for the legal receiver is Gaussian distributed
with zero mean and variance $\sigma^{2}_{s_{i}}$ depending on the state $s_{i}$.
The noise $N_{w,i}$ for the eavesdropper is also Gaussian distributed with zero mean and constant variance $\sigma^{2}_{w}$, i.e.,
$N_{w,i}\sim \mathcal{N}(0, \sigma^{2}_{w})$ for all $i\in \{1,2,...,N\}$. Let $\mathcal{P}_{1}$ and $\mathcal{P}_{2}$ be the power constraints satisfying
\begin{eqnarray}\label{ex-3}
&&\sum_{\tilde{s}_{1}}\pi(\tilde{s}_{1})E[X_{1}^{2}|\tilde{s}_{1}]\leq \mathcal{P}_{1},
\end{eqnarray}
\begin{eqnarray}\label{ex-4}
&&\sum_{\tilde{s}_{1}}\pi(\tilde{s}_{1})\sum_{\tilde{s}_{2}}P_{\tilde{S}_{2}|\tilde{S}_{1}}(\tilde{s}_{2}|\tilde{s}_{1})
E[X_{2}^{2}|\tilde{s}_{1},\tilde{s}_{2}]\leq \mathcal{P}_{2}.
\end{eqnarray}

At time instant $i$, the legal receiver receives the state $S_{i}$ and his own channel output $Y_{i}$, and then he sends $S_{i}$
(or $S_{i}$ and $Y_{i}$) back to the transmitter $j$ ($j=1,2$) after a delay time $d_{j}$.
The steady probability distribution and the one step transition probability matrix of the state are denoted by $\pi(s)$ and $K$, respectively.
The following Corollaries \ref{T5}-\ref{T6} provide bounds on the secrecy capacity region $\mathcal{C}_{s}^{(dg)}$
of the degraded Gaussian fading FS-MAC-WT with delayed state feedback, and Corollaries \ref{T7}-\ref{T8}
provide bounds on the secrecy capacity region $\mathcal{C}_{sf}^{(dg)}$ of the degraded Gaussian fading FS-MAC-WT with delayed state
and channel output feedback.

\begin{corollary}\label{T5}
An inner bound $\mathcal{C}^{dg-in}_{s}$ on $\mathcal{C}_{s}^{(dg)}$
is given by
\begin{eqnarray*}
&&\mathcal{C}^{dg-in}_{s}\nonumber\\
&&=\bigcup_{\mathcal{P}_{1}(\tilde{s}_{1}),\mathcal{P}_{2}(\tilde{s}_{1},\tilde{s}_{2})}
\left\{
\begin{array}{ll}
(R_{1}, R_{2}): R_{1}\geq 0, R_{2}\geq 0,\\
R_{1}\leq \sum_{\tilde{s}_{1}}\pi(\tilde{s}_{1})\sum_{\tilde{s}_{2}}K^{d_{1}-d_{2}}(\tilde{s}_{2},\tilde{s}_{1})
\sum_{s}K^{d_{2}}(s,\tilde{s}_{2})(\frac{1}{2}\log(1+\frac{h^{2}_{1}(s)\mathcal{P}_{1}(\tilde{s}_{1})}{\sigma^{2}_{s}})\\
-\frac{1}{2}\log(\frac{h^{2}_{3}(s)h^{2}_{1}(s)\mathcal{P}_{1}(\tilde{s}_{1})+h^{2}_{3}(s)h^{2}_{2}(s)\mathcal{P}_{2}(\tilde{s}_{1},\tilde{s}_{2})+
h^{2}_{3}(s)\sigma^{2}_{s}+\sigma^{2}_{w}}
{h^{2}_{3}(s)h^{2}_{2}(s)\mathcal{P}_{2}(\tilde{s}_{1},\tilde{s}_{2})+h^{2}_{3}(s)\sigma^{2}_{s}+\sigma^{2}_{w}})),\\
R_{2}\leq \sum_{\tilde{s}_{1}}\pi(\tilde{s}_{1})\sum_{\tilde{s}_{2}}K^{d_{1}-d_{2}}(\tilde{s}_{2},\tilde{s}_{1})
\sum_{s}K^{d_{2}}(s,\tilde{s}_{2})(\frac{1}{2}\log(1+\frac{h^{2}_{2}(s)\mathcal{P}_{2}(\tilde{s}_{1},\tilde{s}_{2})}{\sigma^{2}_{s}})\\
-\frac{1}{2}\log(\frac{h^{2}_{3}(s)h^{2}_{1}(s)\mathcal{P}_{1}(\tilde{s}_{1})+h^{2}_{3}(s)h^{2}_{2}(s)\mathcal{P}_{2}(\tilde{s}_{1},\tilde{s}_{2})+
h^{2}_{3}(s)\sigma^{2}_{s}+\sigma^{2}_{w}}
{h^{2}_{3}(s)h^{2}_{1}(s)\mathcal{P}_{1}(\tilde{s}_{1})+h^{2}_{3}(s)\sigma^{2}_{s}+\sigma^{2}_{w}})),\\
R_{1}+R_{2}\leq \sum_{\tilde{s}_{1}}\pi(\tilde{s}_{1})\sum_{\tilde{s}_{2}}K^{d_{1}-d_{2}}(\tilde{s}_{2},\tilde{s}_{1})
\sum_{s}K^{d_{2}}(s,\tilde{s}_{2})\cdot\\
(\frac{1}{2}\log(1+\frac{h^{2}_{1}(s)\mathcal{P}_{1}(\tilde{s}_{1})
+h^{2}_{2}(s)\mathcal{P}_{2}(\tilde{s}_{1},\tilde{s}_{2})}{\sigma^{2}_{s}})\\
-\frac{1}{2}\log(1+\frac{h^{2}_{3}(s)h^{2}_{1}(s)\mathcal{P}_{1}(\tilde{s}_{1})
+h^{2}_{3}(s)h^{2}_{2}(s)\mathcal{P}_{2}(\tilde{s}_{1},\tilde{s}_{2})}{h^{2}_{3}(s)\sigma^{2}_{s}+\sigma^{2}_{w}})),
\end{array}
\right\},
\end{eqnarray*}
where $\mathcal{P}_{1}(\tilde{s}_{1})$ is the power allocated to the state $\tilde{s}_{1}$, i.e., $\mathcal{P}_{1}(\tilde{s}_{1})=E[X_{1}^{2}|\tilde{s}_{1}]$,
and $\mathcal{P}_{2}(\tilde{s}_{1},\tilde{s}_{2})$ is the power allocated to the states $\tilde{s}_{1}$ and $\tilde{s}_{2}$,
i.e., $\mathcal{P}_{2}(\tilde{s}_{1},\tilde{s}_{2})=E[X_{2}^{2}|\tilde{s}_{1},\tilde{s}_{2}]$, and they satisfy
\begin{eqnarray}\label{ex-7}
\sum_{\tilde{s}_{1}}\pi(\tilde{s}_{1})\mathcal{P}_{1}(\tilde{s}_{1})\leq \mathcal{P}_{1},
\end{eqnarray}
\begin{eqnarray}\label{ex-8}
\sum_{\tilde{s}_{1}}\pi(\tilde{s}_{1})\sum_{\tilde{s}_{2}}P_{\tilde{S}_{2}|\tilde{S}_{1}}(\tilde{s}_{2}|\tilde{s}_{1})
\mathcal{P}_{2}(\tilde{s}_{1},\tilde{s}_{2})\leq \mathcal{P}_{2}.
\end{eqnarray}
\end{corollary}
\begin{IEEEproof}
The inner bound $\mathcal{C}^{dg-in}_{s}$ is obtained by letting the time sharing random variable $Q$ be a constant, and substituting (\ref{e301}),
$X_{1}(\tilde{s}_{1})\sim \mathcal{N}(0, \mathcal{P}_{1}(\tilde{s}_{1}))$ and
$X_{2}(\tilde{s}_{1},\tilde{s}_{2})\sim \mathcal{N}(0, \mathcal{P}_{2}(\tilde{s}_{1},\tilde{s}_{2}))$ into Theorem \ref{T1}, and thus we omit the proof here.
\end{IEEEproof}

\begin{corollary}\label{T6}
An outer bound $\mathcal{C}^{dg-out}_{s}$ on $\mathcal{C}_{s}^{(dg)}$
is given by
\begin{eqnarray*}
&&\mathcal{C}^{dg-out}_{s}\nonumber\\
&&=\bigcup_{\mathcal{P}_{1}(\tilde{s}_{1}),\mathcal{P}_{2}(\tilde{s}_{1},\tilde{s}_{2})}
\left\{
\begin{array}{ll}
(R_{1}, R_{2}): R_{1}\geq 0, R_{2}\geq 0,\\
R_{1}\leq \sum_{\tilde{s}_{1}}\pi(\tilde{s}_{1})\sum_{\tilde{s}_{2}}K^{d_{1}-d_{2}}(\tilde{s}_{2},\tilde{s}_{1})
\sum_{s}K^{d_{2}}(s,\tilde{s}_{2})(\frac{1}{2}\log(1+\frac{h^{2}_{1}(s)\mathcal{P}_{1}(\tilde{s}_{1})}{\sigma^{2}_{s}})\\
-\frac{1}{2}\log(\frac{h^{2}_{1}(s)\mathcal{P}_{1}(\tilde{s}_{1})+\sigma^{2}_{s}+\sigma^{2}_{w}}
{h^{2}_{3}(s)h^{2}_{2}(s)\mathcal{P}_{2}(\tilde{s}_{1},\tilde{s}_{2})+h^{2}_{3}(s)\sigma^{2}_{s}+\sigma^{2}_{w}})),\\
R_{2}\leq \sum_{\tilde{s}_{1}}\pi(\tilde{s}_{1})\sum_{\tilde{s}_{2}}K^{d_{1}-d_{2}}(\tilde{s}_{2},\tilde{s}_{1})
\sum_{s}K^{d_{2}}(s,\tilde{s}_{2})(\frac{1}{2}\log(1+\frac{h^{2}_{2}(s)\mathcal{P}_{2}(\tilde{s}_{1},\tilde{s}_{2})}{\sigma^{2}_{s}})\\
-\frac{1}{2}\log(\frac{h^{2}_{2}(s)\mathcal{P}_{2}(\tilde{s}_{1},\tilde{s}_{2})+\sigma^{2}_{s}+\sigma^{2}_{w}}
{h^{2}_{3}(s)h^{2}_{1}(s)\mathcal{P}_{1}(\tilde{s}_{1})+h^{2}_{3}(s)\sigma^{2}_{s}+\sigma^{2}_{w}})),\\
R_{1}+R_{2}\leq \sum_{\tilde{s}_{1}}\pi(\tilde{s}_{1})\sum_{\tilde{s}_{2}}K^{d_{1}-d_{2}}(\tilde{s}_{2},\tilde{s}_{1})
\sum_{s}K^{d_{2}}(s,\tilde{s}_{2})\cdot\\
(\frac{1}{2}\log(1+\frac{h^{2}_{1}(s)\mathcal{P}_{1}(\tilde{s}_{1})
+h^{2}_{2}(s)\mathcal{P}_{2}(\tilde{s}_{1},\tilde{s}_{2})}{\sigma^{2}_{s}})\\
-\frac{1}{2}\log(1+\frac{h^{2}_{3}(s)h^{2}_{1}(s)\mathcal{P}_{1}(\tilde{s}_{1})
+h^{2}_{3}(s)h^{2}_{2}(s)\mathcal{P}_{2}(\tilde{s}_{1},\tilde{s}_{2})}{h^{2}_{3}(s)\sigma^{2}_{s}+\sigma^{2}_{w}})),
\end{array}
\right\},
\end{eqnarray*}
where $\mathcal{P}_{1}(\tilde{s}_{1})$ and $\mathcal{P}_{2}(\tilde{s}_{1},\tilde{s}_{2})$ satisfy (\ref{ex-7}) and (\ref{ex-8}), respectively.
\end{corollary}
\begin{IEEEproof}
The outer bound $\mathcal{C}^{dg-out}_{s}$ is obtained by the following two steps:
\begin{itemize}

\item First, note that for the discrete memoryless degraded FS-MAC-WT with delayed state feedback, it is not difficult to show that
the outer bound $\mathcal{C}^{out}_{s}$
on the secrecy capacity region is exactly the same as the inner bound $\mathcal{C}^{in}_{s}$, except that the joint probability distribution is not defined
by (\ref{dota1}), and it is given by
\begin{eqnarray}\label{dota1.rmx}
&&P_{QS\tilde{S}_{1}\tilde{S}_{2}X_{1}X_{2}YZ}(q,s,\tilde{s}_{1},\tilde{s}_{2},x_{1},x_{2},y,z)\nonumber\\
&&=P_{Z|Y}(z|y)P_{Y|X_{1},X_{2},S}(y|x_{1},x_{2},s)P_{X_{1}X_{2}S\tilde{S}_{1}\tilde{S}_{2}Q}(x_{1},x_{2},s,\tilde{s}_{1},\tilde{s}_{2},q).
\end{eqnarray}

\item Then applying the outer bound for the degraded FS-MAC-WT with delayed state feedback, and using the entropy power inequality and the definitions of
$\mathcal{P}_{1}(\tilde{s}_{1})$ and $\mathcal{P}_{2}(\tilde{s}_{1},\tilde{s}_{2})$ (see Corollary \ref{T5}), it is not difficult
to show that $\mathcal{C}^{dg-out}_{s}$ is obtained. The detail of the proof is omitted here.
\end{itemize}
\end{IEEEproof}

The following Corollaries \ref{T7}-\ref{T8} provide bounds on the secrecy capacity region $\mathcal{C}_{sf}^{(dg)}$ of
the degraded Gaussian fading FS-MAC-WT with delayed state and channel output feedback.

\begin{corollary}\label{T7}
An inner bound $\mathcal{C}_{sf}^{(dg-in)}$ on $\mathcal{C}_{sf}^{(dg)}$
is given by
\begin{eqnarray*}
&&\mathcal{C}^{dg-in}_{sf}\nonumber\\
&&=\bigcup_{\mathcal{P}_{1}(\tilde{s}_{1}),\mathcal{P}_{2}(\tilde{s}_{1},\tilde{s}_{2})}
\left\{
\begin{array}{ll}
(R_{1}, R_{2}): R_{1}\geq 0, R_{2}\geq 0,\\
R_{1}\leq \sum_{\tilde{s}_{1}}\pi(\tilde{s}_{1})\sum_{\tilde{s}_{2}}K^{d_{1}-d_{2}}(\tilde{s}_{2},\tilde{s}_{1})
\sum_{s}K^{d_{2}}(s,\tilde{s}_{2})\frac{1}{2}\log(1+\frac{h^{2}_{1}(s)\mathcal{P}_{1}(\tilde{s}_{1})}{\sigma^{2}_{s}}),\\
R_{2}\leq \sum_{\tilde{s}_{1}}\pi(\tilde{s}_{1})\sum_{\tilde{s}_{2}}K^{d_{1}-d_{2}}(\tilde{s}_{2},\tilde{s}_{1})
\sum_{s}K^{d_{2}}(s,\tilde{s}_{2})\frac{1}{2}\log(1+\frac{h^{2}_{2}(s)\mathcal{P}_{2}(\tilde{s}_{1},\tilde{s}_{2})}{\sigma^{2}_{s}}),\\
R_{1}+R_{2}\leq \sum_{\tilde{s}_{1}}\pi(\tilde{s}_{1})\sum_{\tilde{s}_{2}}K^{d_{1}-d_{2}}(\tilde{s}_{2},\tilde{s}_{1})
\sum_{s}K^{d_{2}}(s,\tilde{s}_{2})\cdot\\
(\frac{1}{2}\log(1+\frac{h^{2}_{1}(s)\mathcal{P}_{1}(\tilde{s}_{1})
+h^{2}_{2}(s)\mathcal{P}_{2}(\tilde{s}_{1},\tilde{s}_{2})}{\sigma^{2}_{s}})
-\frac{1}{2}\log(1+\frac{h^{2}_{3}(s)h^{2}_{1}(s)\mathcal{P}_{1}(\tilde{s}_{1})
+h^{2}_{3}(s)h^{2}_{2}(s)\mathcal{P}_{2}(\tilde{s}_{1},\tilde{s}_{2})}{h^{2}_{3}(s)\sigma^{2}_{s}+\sigma^{2}_{w}})\\
+\min\{\frac{1}{2}\log(1+\frac{h^{2}_{3}(s)h^{2}_{1}(s)\mathcal{P}_{1}(\tilde{s}_{1})
+h^{2}_{3}(s)h^{2}_{2}(s)\mathcal{P}_{2}(\tilde{s}_{1},\tilde{s}_{2})}{h^{2}_{3}(s)\sigma^{2}_{s}+\sigma^{2}_{w}}),
\frac{1}{2}\log(2\pi e\sigma^{2}_{w})+\frac{1}{2}\log\frac{\sigma^{2}_{s}}{h^{2}_{3}(s)\sigma^{2}_{s}+\sigma^{2}_{w}}\}),
\end{array}
\right\},
\end{eqnarray*}
where $\mathcal{P}_{1}(\tilde{s}_{1})$ and $\mathcal{P}_{2}(\tilde{s}_{1},\tilde{s}_{2})$ satisfy (\ref{ex-7}) and (\ref{ex-8}), respectively.
\end{corollary}
\begin{IEEEproof}
The inner bound $\mathcal{C}^{dg-in}_{sf}$ is obtained by letting the time sharing random variable $Q$ be a constant, and substituting (\ref{e301}),
$X_{1}(\tilde{s}_{1})\sim \mathcal{N}(0, \mathcal{P}_{1}(\tilde{s}_{1}))$ and
$X_{2}(\tilde{s}_{1},\tilde{s}_{2})\sim \mathcal{N}(0, \mathcal{P}_{2}(\tilde{s}_{1},\tilde{s}_{2}))$ into Theorem \ref{T3}, and thus we omit the proof here.
\end{IEEEproof}

\begin{corollary}\label{T8}
An outer bound $\mathcal{C}^{dg-out}_{sf}$ on $\mathcal{C}_{sf}^{(dg)}$
is given by
\begin{eqnarray*}
&&\mathcal{C}^{dg-out}_{sf}\nonumber\\
&&=\bigcup_{\mathcal{P}_{1}(\tilde{s}_{1}),\mathcal{P}_{2}(\tilde{s}_{1},\tilde{s}_{2})}
\left\{
\begin{array}{ll}
(R_{1}, R_{2}): R_{1}\geq 0, R_{2}\geq 0,\\
R_{1}\leq \sum_{\tilde{s}_{1}}\pi(\tilde{s}_{1})\sum_{\tilde{s}_{2}}K^{d_{1}-d_{2}}(\tilde{s}_{2},\tilde{s}_{1})
\sum_{s}K^{d_{2}}(s,\tilde{s}_{2})\frac{1}{2}\log(1+\frac{h^{2}_{1}(s)\mathcal{P}_{1}(\tilde{s}_{1})
+h^{2}_{2}(s)\mathcal{P}_{2}(\tilde{s}_{1},\tilde{s}_{2})}{\sigma^{2}_{s}}),\\
R_{2}\leq \sum_{\tilde{s}_{1}}\pi(\tilde{s}_{1})\sum_{\tilde{s}_{2}}K^{d_{1}-d_{2}}(\tilde{s}_{2},\tilde{s}_{1})
\sum_{s}K^{d_{2}}(s,\tilde{s}_{2})\frac{1}{2}\log(1+\frac{h^{2}_{1}(s)\mathcal{P}_{1}(\tilde{s}_{1})
+h^{2}_{2}(s)\mathcal{P}_{2}(\tilde{s}_{1},\tilde{s}_{2})}{\sigma^{2}_{s}}),\\
R_{1}+R_{2}\leq \min\{\sum_{\tilde{s}_{1}}\pi(\tilde{s}_{1})\sum_{\tilde{s}_{2}}K^{d_{1}-d_{2}}(\tilde{s}_{2},\tilde{s}_{1})
\sum_{s}K^{d_{2}}(s,\tilde{s}_{2})\cdot\\
\frac{1}{2}\log(1+\frac{h^{2}_{1}(s)\mathcal{P}_{1}(\tilde{s}_{1})
+h^{2}_{2}(s)\mathcal{P}_{2}(\tilde{s}_{1},\tilde{s}_{2})}{\sigma^{2}_{s}}),\\
\sum_{\tilde{s}_{1}}\pi(\tilde{s}_{1})\sum_{\tilde{s}_{2}}K^{d_{1}-d_{2}}(\tilde{s}_{2},\tilde{s}_{1})
\sum_{s}K^{d_{2}}(s,\tilde{s}_{2})\cdot\\
(\frac{1}{2}\log(2\pi e\sigma^{2}_{w})
+\frac{1}{2}\log(\frac{h^{2}_{1}(s)\mathcal{P}_{1}(\tilde{s}_{1})
+h^{2}_{2}(s)\mathcal{P}_{2}(\tilde{s}_{1},\tilde{s}_{2})+\sigma^{2}_{s}}{h^{2}_{3}(s)(h^{2}_{1}(s)\mathcal{P}_{1}(\tilde{s}_{1})
+h^{2}_{2}(s)\mathcal{P}_{2}(\tilde{s}_{1},\tilde{s}_{2})+\sigma^{2}_{s})+\sigma^{2}_{w}}))\},
\end{array}
\right\},
\end{eqnarray*}
where $\mathcal{P}_{1}(\tilde{s}_{1})$ and $\mathcal{P}_{2}(\tilde{s}_{1},\tilde{s}_{2})$ satisfy (\ref{ex-7}) and (\ref{ex-8}), respectively.
\end{corollary}
\begin{IEEEproof}
The outer bound $\mathcal{C}^{dg-out}_{sf}$ is obtained by the following two steps:
\begin{itemize}

\item First, note that the three bounds in Theorem \ref{T4} can be further upper bounded by
\begin{eqnarray}
&&R_{1}\leq I(V_{1};Y|U,S,\tilde{S}_{1},\tilde{S}_{2})\stackrel{(a)}\leq I(X_{1},X_{2};Y|S,\tilde{S}_{1},\tilde{S}_{2}),\label{dsb1}\\
&&R_{2}\leq I(V_{2};Y|U,S,\tilde{S}_{1},\tilde{S}_{2})\stackrel{(b)}\leq I(X_{1},X_{2};Y|S,\tilde{S}_{1},\tilde{S}_{2}),\label{dsb2}\\
&&R_{1}+R_{2}\leq\min\{I(V_{1},V_{2};Y|U,S,\tilde{S}_{1},\tilde{S}_{2}),H(Y|Z,U,S,\tilde{S}_{1},\tilde{S}_{2})\}\nonumber\\
&&\stackrel{(c)}\leq \min\{H(Y|S,\tilde{S}_{1},\tilde{S}_{2})-H(Y|X_{1},X_{2},S,\tilde{S}_{1},\tilde{S}_{2}),H(Y|Z,S,\tilde{S}_{1},\tilde{S}_{2})\}\nonumber\\
&&=\min\{I(X_{1},X_{2};Y|S,\tilde{S}_{1},\tilde{S}_{2}),H(Y|Z,S,\tilde{S}_{1},\tilde{S}_{2})\},\label{dsb3}
\end{eqnarray}
where (a) is from the Markov chain $(V_{1},U)\rightarrow (S,\tilde{S}_{1},\tilde{S}_{2},X_{1},X_{2})\rightarrow Y$,
(b) is from $(V_{2},U)\rightarrow (S,\tilde{S}_{1},\tilde{S}_{2},X_{1},X_{2})\rightarrow Y$, and (c) is from
$(V_{1},V_{2},U)\rightarrow (S,\tilde{S}_{1},\tilde{S}_{2},X_{1},X_{2})\rightarrow Y$.

\item Using (\ref{dsb1}), (\ref{dsb2}), (\ref{dsb3}), the entropy power inequality and the definitions of
$\mathcal{P}_{1}(\tilde{s}_{1})$ and $\mathcal{P}_{2}(\tilde{s}_{1},\tilde{s}_{2})$ (see Corollary \ref{T5}), it is not difficult
to show that $\mathcal{C}^{dg-out}_{sf}$ is obtained. The detail of the proof is omitted here.

\end{itemize}

\end{IEEEproof}

\subsection{Numerical results on the Degraded Gaussian Fading Example}\label{sub32}

\begin{figure}[htb]
\centerline{\includegraphics[scale=0.55]{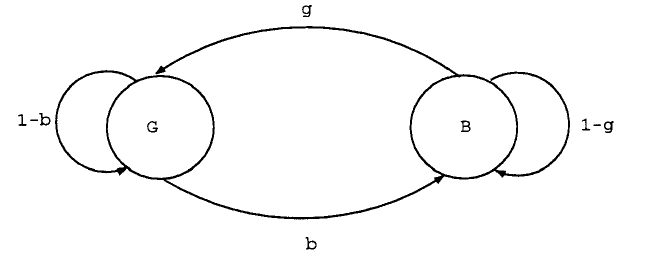}}
\caption{The transition probabilities of the two-state case}
\label{f5}
\end{figure}

To gain some intuition on the bounds shown in Subsection \ref{sub32}, in this subsection, we study a simple two-state case where
the state alphabet $\mathcal{S}$ contains only two elements $G$ (good state) and $B$ (bad state).
The noise variance of the channel in state $G$ is $\sigma^{2}_{G}$, and in state $B$ is $\sigma^{2}_{B}$. Here $\sigma^{2}_{B}>\sigma^{2}_{G}$.
The state process of this two-state case is shown in Figure \ref{f5}, and it is given by
\begin{equation}\label{e307}
P(G|G)=1-b, \,\, P(B|G)=b, \,\, P(B|B)=1-g, \,\, P(G|B)=g.
\end{equation}
Moreover, the steady probabilities of the states $G$ and $B$ are given by
\begin{equation}\label{e308}
\pi(G)=\frac{g}{g+b},\,\, \pi(B)=\frac{b}{g+b}.
\end{equation}

Let $u=1-g-b$ and $c=g/b$. Here note that in \cite{mush}, the authors show that $u$ is with respect to the channel memory, i.e.,
the channel memory is a monotonic increasing function of $u$.
Moreover, from (\ref{e308}),
it is obvious that the steady state distributions depend on $c$.
For the case that $d_{1}=d_{2}=d$ (which implies that the delays for the transmitters are the same) and a fixed $c$ (e.g., $c=1$),
the following Figure \ref{fw1} shows the
effects of the delay $d$ and the channel memory $u$ on the maximum achievable secrecy sum rate
$R^{dg-f}_{sum}$ in $\mathcal{C}_{sf}^{(dg-in)}$ (Theorem \ref{T7}) for
$\mathcal{P}_{1}=\mathcal{P}_{2}=100$, $\sigma^{2}_{G}=1$, $\sigma^{2}_{w}=400$, $h_{1}(g)=1$, $h_{1}(b)=0.5$,
$h_{2}(g)=1$, $h_{2}(b)=0.7$, $h_{3}(g)=0.8$, $h_{3}(b)=0.2$, $c=1$ and different values of $u$ and $\sigma^{2}_{B}$.
\begin{figure}[htb]
\centerline{\includegraphics[scale=0.5]{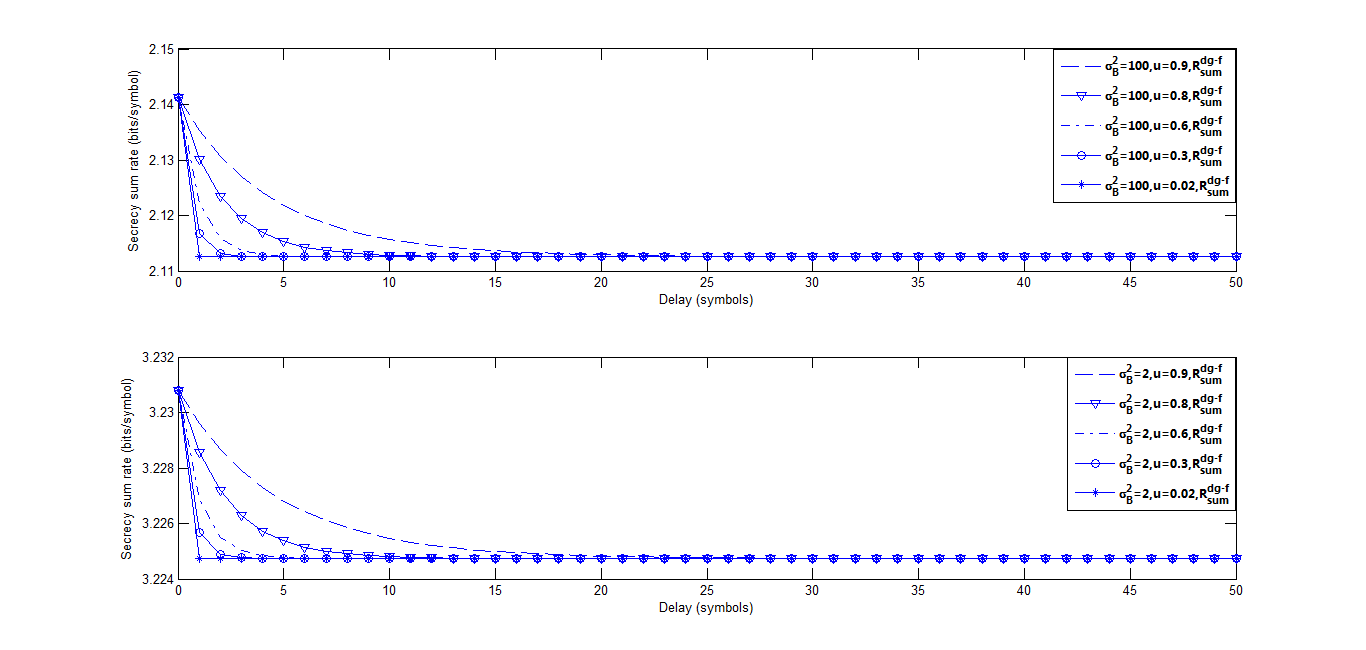}}
\caption{The maximum secrecy sum rates versus delays $d_{1}=d_{2}=d$ for
$\mathcal{P}_{1}=\mathcal{P}_{2}=100$, $\sigma^{2}_{G}=1$, $\sigma^{2}_{w}=400$, $h_{1}(g)=1$, $h_{1}(b)=0.5$,
$h_{2}(g)=1$, $h_{2}(b)=0.7$, $h_{3}(g)=0.8$, $h_{3}(b)=0.2$, $c=1$ and several values of $u$ and $\sigma^{2}_{B}$}
\label{fw1}
\end{figure}
In addition, for the case that $d_{1}=d$ and $d_{2}=0$ (which implies that there is no delay for the transmitter $2$) and a fixed $c=1$,
the following Figure \ref{fw2} shows the
effects of the delay $d$ and the channel memory $u$ on the maximum achievable secrecy sum rate
$R^{dg-f}_{sum}$ in $\mathcal{C}_{sf}^{(dg-in)}$ for
$\mathcal{P}_{1}=\mathcal{P}_{2}=100$, $\sigma^{2}_{G}=1$, $\sigma^{2}_{w}=400$, $h_{1}(g)=1$, $h_{1}(b)=0.5$,
$h_{2}(g)=1$, $h_{2}(b)=0.7$, $h_{3}(g)=0.8$, $h_{3}(b)=0.2$, $c=1$ and different values of $u$ and $\sigma^{2}_{B}$.
\begin{figure}[htb]
\centerline{\includegraphics[scale=0.5]{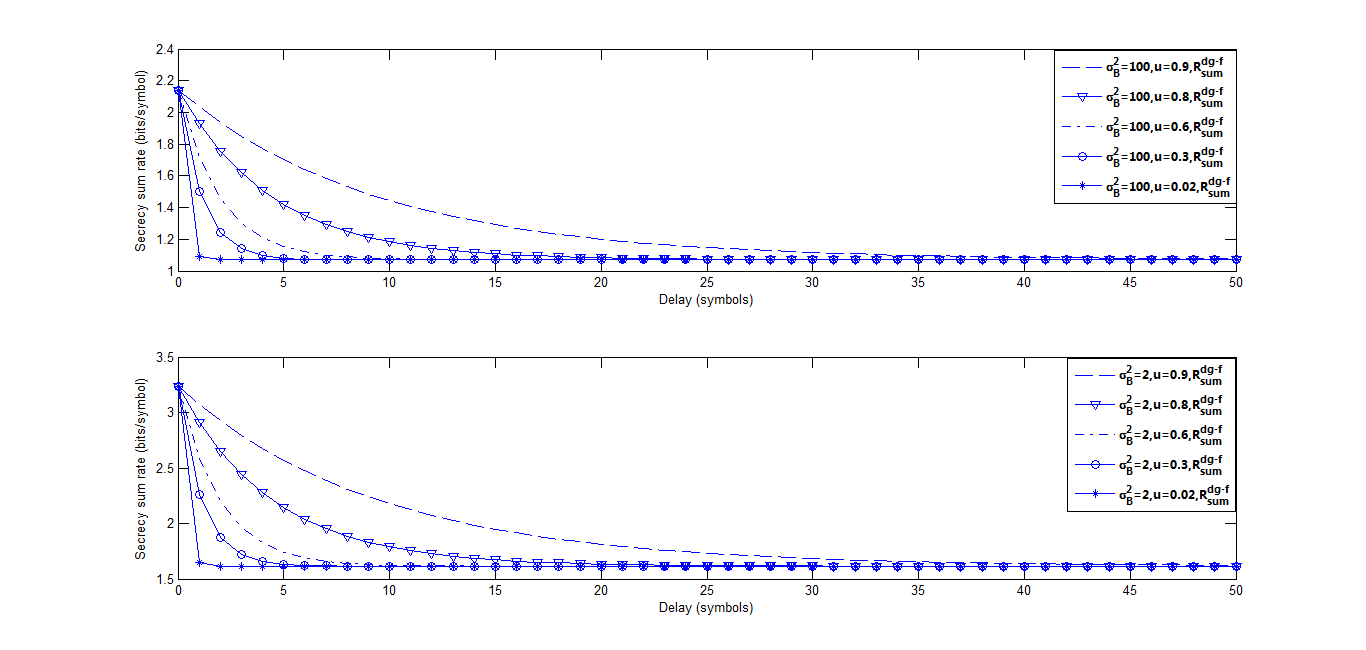}}
\caption{The maximum secrecy sum rates versus delays $d_{1}=d$ and $d_{2}=0$ for
$\mathcal{P}_{1}=\mathcal{P}_{2}=100$, $\sigma^{2}_{G}=1$, $\sigma^{2}_{w}=400$, $h_{1}(g)=1$, $h_{1}(b)=0.5$,
$h_{2}(g)=1$, $h_{2}(b)=0.7$, $h_{3}(g)=0.8$, $h_{3}(b)=0.2$, $c=1$ and several values of $u$ and $\sigma^{2}_{B}$}
\label{fw2}
\end{figure}
From Figs. \ref{fw1} and \ref{fw2}, we see that the maximum achievable secrecy sum rate is approaching the infinite asymptote
while the delay $d$ is increasing, and the secrecy sum rate is changing rapidly while the channel memory $u$ is decreasing.
Moreover, it is easy to see that $R^{dg-f}_{sum}$ is increasing while $\sigma_{B}$ is decreasing,
and this is because for a given $\sigma^{2}_{G}$, the decrease of $\sigma_{B}$ implies
the decrease of the average channel noise.

For $\mathcal{P}_{1}=\mathcal{P}_{2}=100$, $\sigma^{2}_{G}=1$,
$g=b=0.05$, $h_{1}(g)=1$, $h_{1}(b)=0.5$,
$h_{2}(g)=1$, $h_{2}(b)=0.7$, $h_{3}(g)=1$, $h_{3}(b)=0.9$, $d_{1}=100$, $d_{2}=10$ and several values of $\sigma^{2}_{w}$ and $\sigma_{B}$, the following
Figs. \ref{fw3} and \ref{fw4} show the inner and outer bounds on the secrecy capacity
regions of the degraded Gaussian fading case of Figure \ref{f2} with or without channel output feedback, and the capacity region $\mathcal{C}^{(dg*)}$
of the FS-MAC with only delayed state feedback (see \cite[pp.3442-3443(equations 57,60,62)]{bash}). From Figure \ref{fw3}, we see that
$\mathcal{C}_{sf}^{(dg-in)}$ is larger than $\mathcal{C}_{s}^{(dg-in)}$ (even as large as $\mathcal{C}^{(dg*)}$,
which indicates that the Shannon capacity is achieved). Comparing Figure \ref{fw3} with Figure \ref{fw4}, we see that
the gap between $\mathcal{C}_{s}^{(dg-in)}$ and $\mathcal{C}_{sf}^{(dg-in)}$ is increasing while $\sigma^{2}_{w}$ is decreasing. From Figure \ref{fw4}, we see that
the inner bounds $\mathcal{C}_{s}^{(dg-in)}$ and $\mathcal{C}_{sf}^{(dg-in)}$ respectively meet the outer bounds
$\mathcal{C}_{s}^{(dg-out)}$ and $\mathcal{C}_{sf}^{(dg-out)}$ when $\sigma^{2}_{w}$ is small, and this is because when $\sigma^{2}_{w}$ is small enough,
the bounds on the individual rates are larger than the bound on the sum rate,
and the sum rate bounds of $\mathcal{C}_{s}^{(dg-in)}$ and $\mathcal{C}_{sf}^{(dg-in)}$
are respectively approaching those of $\mathcal{C}_{s}^{(dg-out)}$ and $\mathcal{C}_{sf}^{(dg-out)}$.
Moreover, from Figs. \ref{fw3}-\ref{fw4}, we see that for fixed
$\sigma_{G}$, $\sigma^{2}_{w}$, $h_{1}(g)$, $h_{1}(b)$,
$h_{2}(g)$, $h_{2}(b)$, $h_{3}(g)$, $h_{3}(b)$, $d_{1}$ and $d_{2}$,
all the bounds ($\mathcal{C}_{s}^{(dg-in)}$, $\mathcal{C}_{s}^{(dg-out)}$, $\mathcal{C}_{sf}^{(dg-in)}$, $\mathcal{C}_{sf}^{(dg-out)}$,
$\mathcal{C}^{(dg*)}$) are enlarging
while $\sigma_{B}$ is decreasing, which is due to the fact that the decrease of $\sigma_{B}$ indicates
the decrease of the average channel noise.

\begin{figure}[htb]
\centering
\includegraphics[scale=0.5]{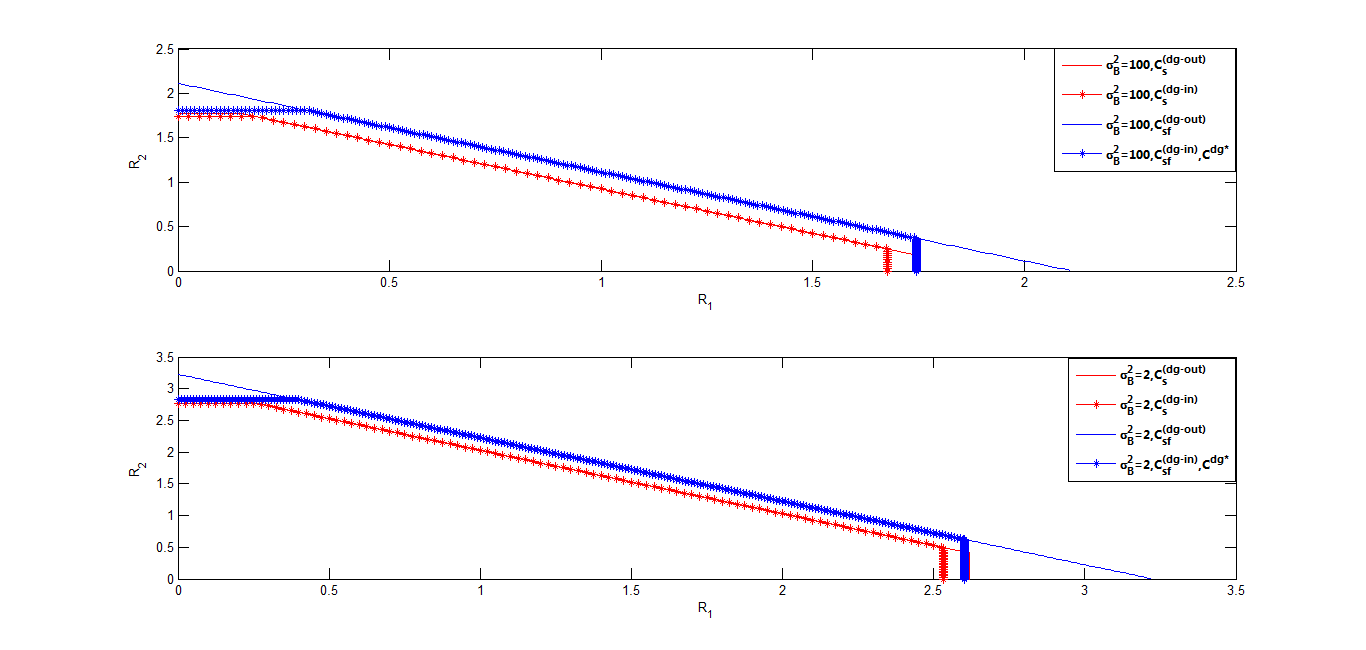}
\caption{The comparison of $\mathcal{C}^{dg-in}_{sf}$, $\mathcal{C}^{dg-out}_{sf}$, $\mathcal{C}^{dg-in}_{s}$, $\mathcal{C}^{dg-out}_{s}$
and $\mathcal{C}^{dg*}$ for
$\mathcal{P}_{1}=\mathcal{P}_{2}=100$, $\sigma^{2}_{G}=1$, $g=b=0.05$, $h_{1}(g)=1$, $h_{1}(b)=0.5$,
$h_{2}(g)=1$, $h_{2}(b)=0.7$, $h_{3}(g)=1$, $h_{3}(b)=0.9$, $d_{1}=100$, $d_{2}=10$, $\sigma^{2}_{w}=400$ and several values of $\sigma^{2}_{B}$}
\label{fw3}
\end{figure}

\begin{figure}[htb]
\centering
\includegraphics[scale=0.5]{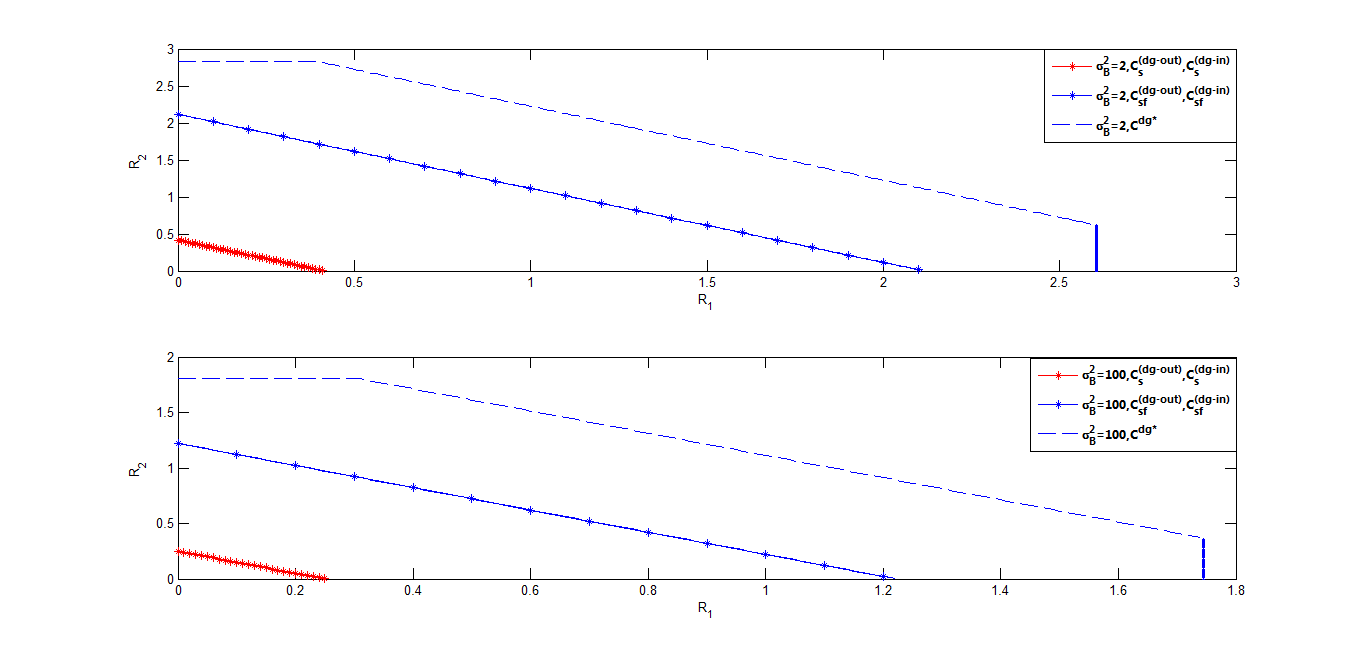}
\caption{The comparison of $\mathcal{C}^{dg-in}_{sf}$, $\mathcal{C}^{dg-out}_{sf}$, $\mathcal{C}^{dg-in}_{s}$, $\mathcal{C}^{dg-out}_{s}$
and $\mathcal{C}^{dg*}$ for
$\mathcal{P}_{1}=\mathcal{P}_{2}=100$, $\sigma^{2}_{G}=1$, $g=b=0.05$, $h_{1}(g)=1$, $h_{1}(b)=0.5$,
$h_{2}(g)=1$, $h_{2}(b)=0.7$, $h_{3}(g)=1$, $h_{3}(b)=0.9$, $d_{1}=100$, $d_{2}=10$, $\sigma^{2}_{w}=1$ and several values of $\sigma^{2}_{B}$}
\label{fw4}
\end{figure}

\section{Summary\label{secIV}}

In this paper, we investigate the FS-MAC-WT with delayed feedback. Bounds on the secrecy capacity region of this model are provided,
and the achievability of the inner bound implies that the legal receiver's delayed channel output
feedback can be not only used to allow the transmitters to cooperate with each other, but also used to
produce secret keys encrypting the transmitted messages.
The capacity results are further explained via a degraded Gaussian fading example. Numerical result of this example
shows that the maximum achievable secrecy sum rate is approaching the infinite asymptote
while the delays are increasing, and the secrecy sum rate is changing rapidly while the channel memory is decreasing. Moreover, from this example, we
see that feeding back the legal receiver's channel output
greatly enhances the achievable secrecy rate region of the FS-MAC-WT with only delayed state feedback.
The result of this paper is an intermediate step toward understanding the secure transmission in wireless communication networks with
delayed feedback.

\renewcommand{\theequation}{A\arabic{equation}}
\appendices\section{Proof of Theorem \ref{T3}\label{appen4}}
\setcounter{equation}{0}

Several already existing coding strategies, such as block Markov coding strategy for the feedback systems,
generating secret keys from the legal receiver's channel output feedback \cite{AC} and
the decode-and-forward (DF) strategy for the MAC with noiseless feedback \cite{coverz}, Wyner's random binning technique \cite{Wy}
have been combined with the
multiplexing coding scheme for the FSMC with delayed state feedback \cite{vis}
to show the achievability of Theorem \ref{T3}.
Now the remainder of this section is organized as follows. Basic notations and definitions are introduced in Subsection \ref{sub-x1},
the coding scheme is shown in Subsection \ref{sub-x2}, and the equivocation analysis is given in Subsection \ref{sub-x3}.

\subsection{Basic notations and definitions}\label{sub-x1}

\begin{itemize}

\item The messages are transmitted over $n$ blocks, and the codeword length in each block is $N$.
Without loss of generality, denote the state alphabet $\mathcal{S}$ by $\mathcal{S}=\{1,2,...,k\}$, and note that
the steady state probability $\pi(l)>0$ for all $l\in \mathcal{S}$. In addition, denote $N_{\tilde{s}_{1}}$ ($1\leq \tilde{s}_{1}\leq k$) by
\begin{eqnarray}\label{c3.q2}
&&N_{\tilde{s}_{1}}=NP_{\tilde{S}_{1}}(\tilde{s}_{1})-\epsilon_{1},
\end{eqnarray}
and $N_{\tilde{s}_{1},\tilde{s}_{2}}$ ($1\leq \tilde{s}_{1},\tilde{s}_{2}\leq k$) by
\begin{eqnarray}\label{c3.q2.koudai}
&&N_{\tilde{s}_{1},\tilde{s}_{2}}=NP_{\tilde{S}_{1}\tilde{S}_{2}}(\tilde{s}_{1},\tilde{s}_{2})-\frac{\epsilon_{1}}{k},
\end{eqnarray}
where $\epsilon_{1}>0$ and $\epsilon_{1}\rightarrow 0$ as $N\rightarrow \infty$. Here note that from (\ref{c3.q2}) and (\ref{c3.q2.koudai}),
we have
\begin{eqnarray}\label{daqin-1}
&&\sum_{\tilde{s}_{2}=1}^{k}N_{\tilde{s}_{1},\tilde{s}_{2}}=N_{\tilde{s}_{1}}.
\end{eqnarray}

\item The messages $W_{1}=(W_{1,1},...,W_{1,n})$ and $W_{2}=(W_{2,1},...,W_{2,n})$ are transmitted through $n$ blocks.
In block $i$ ($1\leq i\leq n$),
the transmitted message $w_{1,i}$ is denoted by $w_{1,i}=(w_{1,i,0},w_{1,i,1})$, where $w_{1,i,0}\in\{1,2,...,2^{NR_{10}}\}$
and $w_{1,i,1}\in\{1,2,...,2^{NR_{11}}\}$.  For a given delayed feedback state $\tilde{s}_{1}$ ($1\leq \tilde{s}_{1}\leq k$),
we further divide the messages $w_{1,i,0}$ and $w_{1,i,1}$ into $k$ sub-messages, i.e.,
$w_{1,i,0}=(w_{1,i,0,1},...,w_{1,i,0,k})$ and $w_{1,i,1}=(w_{1,i,1,1},...,w_{1,i,1,k})$, where for each $\tilde{s}_{1}$, the messages
$w_{1,i,0,\tilde{s}_{1}}$ and $w_{1,i,1,\tilde{s}_{1}}$ take values in the sets
$\{1,2,...,2^{N_{\tilde{s}_{1}}R_{10}(\tilde{s}_{1})}\}$ and
$\{1,2,...,2^{N_{\tilde{s}_{1}}R_{11}(\tilde{s}_{1})}\}$, respectively. Here note that
\begin{eqnarray}\label{burning-1}
&&\sum_{\tilde{s}_{1}=1}^{k}P_{\tilde{S}_{1}}(\tilde{s}_{1})R_{10}(\tilde{s}_{1})=R_{10},
\end{eqnarray}
\begin{eqnarray}\label{burning-2}
&&\sum_{\tilde{s}_{1}=1}^{k}P_{\tilde{S}_{1}}(\tilde{s}_{1})R_{11}(\tilde{s}_{1})=R_{11}.
\end{eqnarray}
Analogously, the message $w_{2,i}$ is denoted by $w_{2,i}=(w_{2,i,0},w_{2,i,1})$,
where $w_{2,i,0}\in\{1,2,...,2^{NR_{20}}\}$
and $w_{2,i,1}\in\{1,2,...,2^{NR_{21}}\}$. For a given $\tilde{s}_{1}$ ($1\leq \tilde{s}_{1}\leq k$),
define $w_{2,i,0}=(w_{2,i,0,1},...,w_{2,i,0,k})$ and $w_{2,i,1}=(w_{2,i,1,1},...,w_{2,i,1,k})$, where for each $1\leq \tilde{s}_{1}\leq k$,
the messages
$w_{2,i,0,\tilde{s}_{1}}$ and $w_{2,i,1,\tilde{s}_{1}}$ take values in the sets
$\{1,2,...,2^{N_{\tilde{s}_{1}}R_{20}(\tilde{s}_{1})}\}$ and
$\{1,2,...,2^{N_{\tilde{s}_{1}}R_{21}(\tilde{s}_{1})}\}$, respectively. Moreover, the messages
$w_{2,i,0,\tilde{s}_{1}}$ and $w_{2,i,1,\tilde{s}_{1}}$ can be further divided by a given delayed state $\tilde{s}_{2}$ ($1\leq \tilde{s}_{2}\leq k$),
i.e., $w_{2,i,0,\tilde{s}_{1}}=(w_{2,i,0,\tilde{s}_{1},1},...,w_{2,i,0,\tilde{s}_{1},k})$,
$w_{2,i,1,\tilde{s}_{1}}=(w_{2,i,1,\tilde{s}_{1},1},...,w_{2,i,1,\tilde{s}_{1},k})$, where
$w_{2,i,0,\tilde{s}_{1},\tilde{s}_{2}}$ and $w_{2,i,1,\tilde{s}_{1},\tilde{s}_{2}}$ take values in the sets
$\{1,2,...,\\2^{N_{\tilde{s}_{1},\tilde{s}_{2}}R_{20}(\tilde{s}_{1},\tilde{s}_{2})}\}$ and
$\{1,2,...,2^{N_{\tilde{s}_{1},\tilde{s}_{2}}R_{21}(\tilde{s}_{1},\tilde{s}_{2})}\}$, respectively. From the above definitions, it is easy to see that
\begin{eqnarray}\label{burning-3.1}
&&\sum_{\tilde{s}_{1}=1}^{k}P_{\tilde{S}_{1}}(\tilde{s}_{1})R_{20}(\tilde{s}_{1})=R_{20},
\end{eqnarray}
\begin{eqnarray}\label{burning-4.1}
&&\sum_{\tilde{s}_{1}=1}^{k}P_{\tilde{S}_{1}}(\tilde{s}_{1})R_{21}(\tilde{s}_{1})=R_{21}.
\end{eqnarray}
Moreover, we have
\begin{eqnarray}\label{burning-3}
&&R_{20}(\tilde{s}_{1})=\sum_{\tilde{s}_{2}=1}^{k}P_{\tilde{S}_{2}|\tilde{S}_{1}}(\tilde{s}_{2}|\tilde{s}_{1})R_{20}(\tilde{s}_{1},\tilde{s}_{2})-\epsilon^{*}_{1},
\end{eqnarray}
\begin{eqnarray}\label{burning-4}
&&R_{21}(\tilde{s}_{1})=\sum_{\tilde{s}_{2}=1}^{k}P_{\tilde{S}_{2}|\tilde{S}_{1}}(\tilde{s}_{2}|\tilde{s}_{1})R_{21}(\tilde{s}_{1},\tilde{s}_{2})-\epsilon^{*}_{2},
\end{eqnarray}
where $\epsilon^{*}_{1}$ and $\epsilon^{*}_{2}$ tend to zero while $N$ tends to infinity.

\item For block $i$ ($1\leq i\leq n$), let $W_{1,i}^{*}$ and $W_{2,i}^{*}$ be the dummy messages taking values in $\{1,2,...,2^{NR^{*}_{1}}\}$
and $\{1,2,...,2^{NR^{*}_{2}}\}$, respectively. Further divide $W_{1,i}^{*}$ into $k$ sub-messages, i.e.,
$w^{*}_{1,i}=(w^{*}_{1,i,1},...,w^{*}_{1,i,k})$ and for each $1\leq \tilde{s}_{1}\leq k$, the message
$w^{*}_{1,i,\tilde{s}_{1}}$ takes values in the set
$\{1,2,...,2^{N_{\tilde{s}_{1}}R^{*}_{1}(\tilde{s}_{1})}\}$. Similarly, let $w^{*}_{2,i}=(w^{*}_{2,i,1},...,w^{*}_{2,i,k})$, where
for each $1\leq \tilde{s}_{1}\leq k$, the message
$w^{*}_{2,i,\tilde{s}_{1}}$ takes values in the set
$\{1,2,...,2^{N_{\tilde{s}_{1}}R^{*}_{2}(\tilde{s}_{1})}\}$. Moreover, the message $w^{*}_{2,i,\tilde{s}_{1}}$ can be further divided by
$w^{*}_{2,i,\tilde{s}_{1}}=(w^{*}_{2,i,\tilde{s}_{1},1},...,w^{*}_{2,i,\tilde{s}_{1},k})$, where for each $1\leq \tilde{s}_{2}\leq k$, the message
$w^{*}_{2,i,\tilde{s}_{1},\tilde{s}_{2}}$ takes values in the set
$\{1,2,...,2^{N_{\tilde{s}_{1},\tilde{s}_{2}}R^{*}_{2}(\tilde{s}_{1},\tilde{s}_{2})}\}$.
Here note that
\begin{eqnarray}\label{burning-5}
&&\sum_{\tilde{s}_{1}=1}^{k}P_{\tilde{S}_{1}}(\tilde{s}_{1})R^{*}_{1}(\tilde{s}_{1})=R^{*}_{1},
\end{eqnarray}
\begin{eqnarray}\label{burning-6.1}
&&\sum_{\tilde{s}_{1}=1}^{k}P_{\tilde{S}_{1}}(\tilde{s}_{1})R^{*}_{2}(\tilde{s}_{1})=R^{*}_{2},
\end{eqnarray}
\begin{eqnarray}\label{burning-6}
&&R^{*}_{2}(\tilde{s}_{1})=\sum_{\tilde{s}_{2}=1}^{k}P_{\tilde{S}_{2}|\tilde{S}_{1}}(\tilde{s}_{2}|\tilde{s}_{1})R^{*}_{2}(\tilde{s}_{1},\tilde{s}_{2})
-\epsilon^{*}_{3},
\end{eqnarray}
where $\epsilon^{*}_{3}\rightarrow 0$ as $N\rightarrow \infty$.

\item Let $\widetilde{X}_{j,i}$ ($j=1,2$), $\widetilde{Q}_{i}$, $\widetilde{S}_{i}$, $\widetilde{Y}_{i}$
and $\widetilde{Z}_{i}$ be the random vectors for block $i$, and let
$X_{j}^{n}=(\widetilde{X}_{j,1},...,\widetilde{X}_{j,n})$ ($j=1,2$),
$Q^{n}=(\widetilde{Q}_{1},...,\widetilde{Q}_{n})$, $S^{n}=(\widetilde{S}_{1},...,\widetilde{S}_{n})$,
$Y^{n}=(\widetilde{Y}_{1},...,\widetilde{Y}_{n})$ and $Z^{n}=(\widetilde{Z}_{1},...,\widetilde{Z}_{n})$. Moreover, for given $\tilde{s}_{1}$,
the sub-vectors of $\widetilde{X}_{1,i}$, $\widetilde{X}_{2,i}$, $\widetilde{Q}_{i}$, $\widetilde{S}_{i}$, $\widetilde{Y}_{i}$
and $\widetilde{Z}_{i}$ are denoted by $\widetilde{X}^{N_{\tilde{s}_{1}}}_{1,i}$, $\widetilde{X}^{N_{\tilde{s}_{1}}}_{2,i}$,
$\widetilde{Q}^{N_{\tilde{s}_{1}}}_{i}$, $\widetilde{S}^{N_{\tilde{s}_{1}}}_{i}$, $\widetilde{Y}^{N_{\tilde{s}_{1}}}_{i}$
and $\widetilde{Z}^{N_{\tilde{s}_{1}}}_{i}$, respectively.
The real values of the above random vectors are denoted by lower case letters.

\end{itemize}

\subsection{Encoding and decoding schemes}\label{sub-x2}

\subsubsection*{1). Codebooks construction}

\begin{itemize}

\item First, fix the probability $P_{X_{1}|\tilde{S}_{1},Q}(x_{1}|\tilde{s}_{1},q)P_{X_{2}|\tilde{S}_{1},\tilde{S}_{2},Q}(x_{2}|\tilde{s}_{1},\tilde{s}_{2},q)
P_{Q|\tilde{S}_{1}}(q|\tilde{s}_{1})$.
Then, in block $i$ ($1\leq i\leq n$), for a given $\tilde{s}_{1}$ ($1\leq \tilde{s}_{1}\leq k$), randomly produce
$2^{N_{\tilde{s}_{1}}(R_{10}(\tilde{s}_{1})+R_{11}(\tilde{s}_{1})+R^{*}_{1}(\tilde{s}_{1})+R_{20}(\tilde{s}_{1})+R_{21}(\tilde{s}_{1})+R^{*}_{2}(\tilde{s}_{1}))}$
i.i.d. sequences $\widetilde{q}^{N_{\tilde{s}_{1}}}_{i}$ according to $P_{Q|\tilde{S}_{1}}(q|\tilde{s}_{1})$, and index
these sequences as $\widetilde{q}^{N_{\tilde{s}_{1}}}_{i}(w^{'}_{0,i,\tilde{s}_{1}})$, where
$1\leq w^{'}_{0,i,\tilde{s}_{1}}\leq 2^{N_{\tilde{s}_{1}}(R_{10}(\tilde{s}_{1})+R_{11}(\tilde{s}_{1})+R^{*}_{1}(\tilde{s}_{1})+R_{20}(\tilde{s}_{1})
+R_{21}(\tilde{s}_{1})+R^{*}_{2}(\tilde{s}_{1}))}$.

\item For each $\widetilde{q}^{N_{\tilde{s}_{1}}}_{i}(w^{'}_{0,i,\tilde{s}_{1}})$, randomly produce
$2^{N_{\tilde{s}_{1}}(R_{10}(\tilde{s}_{1})+R_{11}(\tilde{s}_{1})+R^{*}_{1}(\tilde{s}_{1}))}$ i.i.d. sequences
$\widetilde{x}^{N_{\tilde{s}_{1}}}_{1,i}$ according to \\$P_{X_{1}|\tilde{S}_{1},Q}(x_{1}|\tilde{s}_{1},q)$, and index these sequences as
$\widetilde{x}^{N_{\tilde{s}_{1}}}_{1,i}(w_{1,i,\tilde{s}_{1}}^{'})$, where
$1\leq w_{1,i,\tilde{s}_{1}}^{'}\leq 2^{N(R_{10}(\tilde{s}_{1})+R_{11}(\tilde{s}_{1})+R^{*}_{1}(\tilde{s}_{1}))}$.

\item For each $\widetilde{q}^{N_{\tilde{s}_{1}}}_{i}(w^{'}_{0,i,\tilde{s}_{1}})$, divide it into $k$ sub-sequences, i.e.,
$\widetilde{q}^{N_{\tilde{s}_{1}}}_{i}(w^{'}_{0,i,\tilde{s}_{1}})
=(\widetilde{q}^{N_{\tilde{s}_{1},1}}_{i}(w^{'}_{0,i,\tilde{s}_{1},1}),
\widetilde{q}^{N_{\tilde{s}_{1},2}}_{i}(w^{'}_{0,i,\tilde{s}_{1},2}),...,\\
\widetilde{q}^{N_{\tilde{s}_{1},k}}_{i}(w^{'}_{0,i,\tilde{s}_{1},k}))$, where for each $1\leq \tilde{s}_{2}\leq k$, the message
$w^{'}_{0,i,\tilde{s}_{1},\tilde{s}_{2}}$ takes values in the set \\$\{1,2,...,
2^{N_{\tilde{s}_{1},\tilde{s}_{2}}(R_{10}(\tilde{s}_{1})+R_{11}(\tilde{s}_{1})+R^{*}_{1}(\tilde{s}_{1})+
R_{20}(\tilde{s}_{1},\tilde{s}_{2})+R_{21}(\tilde{s}_{1},\tilde{s}_{2})+R^{*}_{2}(\tilde{s}_{1},\tilde{s}_{2}))}\}$.
For each $\widetilde{q}^{N_{\tilde{s}_{1},\tilde{s}_{2}}}_{i}(w^{'}_{0,i,\tilde{s}_{1},\tilde{s}_{2}})$,
randomly produce
$2^{N_{\tilde{s}_{1},\tilde{s}_{2}}(R_{20}(\tilde{s}_{1},\tilde{s}_{2})+R_{21}(\tilde{s}_{1},\tilde{s}_{2})+R^{*}_{2}(\tilde{s}_{1},\tilde{s}_{2}))}$ i.i.d. sequences
$\widetilde{x}^{N_{\tilde{s}_{1},\tilde{s}_{2}}}_{2,i}$ according to $P_{X_{2}|Q,\tilde{S}_{1},\tilde{S}_{2}}(x_{2}|q,\tilde{s}_{1},\tilde{s}_{2})$,
and index these sequences as
$\widetilde{x}^{N_{\tilde{s}_{1},\tilde{s}_{2}}}_{2,i}(w_{2,i,\tilde{s}_{1},\tilde{s}_{2}}^{'})$, where
$1\leq w_{2,i,\tilde{s}_{1},\tilde{s}_{2}}^{'}\leq 2^{N_{\tilde{s}_{1},\tilde{s}_{2}}(R_{20}(\tilde{s}_{1},\tilde{s}_{2})
+R_{21}(\tilde{s}_{1},\tilde{s}_{2})+R^{*}_{2}(\tilde{s}_{1},\tilde{s}_{2}))}$.

\end{itemize}

\subsubsection*{2). Encoding scheme}

\begin{itemize}
\item \textbf{Encoding scheme for $\widetilde{Q}_{i}$ ($1\leq i\leq n$):}

\begin{itemize}
\item \textbf{Transmitter $1$'s encoding scheme of $\widetilde{Q}_{i}$:}
In block $1\leq i\leq 2d_{1}$, for each $\tilde{s}_{1}$, the transmitter $1$ chooses $w^{'}_{0,i,\tilde{s}_{1}}=1$ as the index of the transmitted
$\widetilde{q}^{N_{\tilde{s}_{1}}}_{i}$.
In block $i$ ($2d_{1}+1\leq i\leq n$), for each $\tilde{s}_{1}$, the transmitter $1$ has already known the delayed state sequence
$\widetilde{s}^{N_{\tilde{s}_{1}}}_{i-2d_{1}}$, $w^{'}_{0,i-d_{1},\tilde{s}_{1}}$ and $w_{1,i-d_{1},\tilde{s}_{1}}^{'}=(w_{1,i-d_{1},0,\tilde{s}_{1}},
w_{1,i-d_{1},1,\tilde{s}_{1}},w_{1,i-d_{1},\tilde{s}_{1}}^{*})$, where $\widetilde{s}^{N_{\tilde{s}_{1}}}_{i-2d_{1}}$
is the delayed feedback state used to de-multiplex $\widetilde{y}_{i-d_{1}}$ into the sub-sequences
$\widetilde{y}^{N_{1}}_{i-d_{1}}$, ..., $\widetilde{y}^{N_{k}}_{i-d_{1}}$.
Once the transmitter $1$ receives the feedback $\widetilde{y}^{N_{\tilde{s}_{1}}}_{i-d_{1}}$, he attempts to find a unique
sequence $\widetilde{x}^{N_{\tilde{s}_{1}}}_{2,i-d_{1}}(\check{w}_{2,i-d_{1},\tilde{s}_{1}}^{'},w^{'}_{0,i-d_{1},\tilde{s}_{1}})$
such that $(\widetilde{x}^{N_{\tilde{s}_{1}}}_{2,i-d_{1}}(\check{w}_{2,i-d_{1},\tilde{s}_{1}}^{'},w^{'}_{0,i-d_{1},\tilde{s}_{1}}),\\
\widetilde{x}^{N_{\tilde{s}_{1}}}_{1,i-d_{1}}(w_{1,i-d_{1},\tilde{s}_{1}}^{'},w^{'}_{0,i-d_{1},\tilde{s}_{1}}),
\widetilde{q}^{N_{\tilde{s}_{1}}}_{i-d_{1}}(w^{'}_{0,i-d_{1},\tilde{s}_{1}}),\widetilde{s}^{N_{\tilde{s}_{1}}}_{i-d_{1}},
\widetilde{y}^{N_{\tilde{s}_{1}}}_{i-d_{1}})$ are jointly typical sequences,
where $\check{w}_{2,i-d_{1},\tilde{s}_{1}}^{'}\\=(\check{w}_{2,i-d_{1},\tilde{s}_{1},1}^{'},...,\check{w}_{2,i-d_{1},\tilde{s}_{1},k}^{'})$, and
$\check{w}_{2,i-d_{1},\tilde{s}_{1},\tilde{s}_{2}}^{'}$ ($1\leq \tilde{s}_{2}\leq k$) is the transmitter $1$'s estimation of
$w_{2,i-d_{1},\tilde{s}_{1},\tilde{s}_{2}}^{'}$.
From the packing lemma \cite{network}, the error probability $Pr\{\check{w}_{2,i-d_{1},\tilde{s}_{1}}^{'}\neq w_{2,i-d_{1},\tilde{s}_{1}}^{'}\}$ approaches to $0$ if
\begin{eqnarray}\label{bgod-1}
&&R_{20}(\tilde{s}_{1})+R_{21}(\tilde{s}_{1})+R^{*}_{2}(\tilde{s}_{1})\leq I(X_{2};Y|X_{1},Q,S,\tilde{S}_{1}=\tilde{s}_{1}).
\end{eqnarray}
Here note that (\ref{bgod-1}) indicates that
\begin{eqnarray}\label{bgod-1.x}
&&R_{20}+R_{21}+R^{*}_{2}\nonumber\\
&&=\sum_{\tilde{s}_{1}=1}^{k}P_{\tilde{S}_{1}}(\tilde{s}_{1})(R_{20}(\tilde{s}_{1})+R_{21}(\tilde{s}_{1})+R^{*}_{2}(\tilde{s}_{1}))\nonumber\\
&&\leq\sum_{\tilde{s}_{1}=1}^{k}P_{\tilde{S}_{1}}(\tilde{s}_{1})I(X_{2};Y|X_{1},Q,S,\tilde{S}_{1}=\tilde{s}_{1})\nonumber\\
&&=I(X_{2};Y|X_{1},Q,S,\tilde{S}_{1})\stackrel{(1)}=I(X_{2};Y|X_{1},Q,S,\tilde{S}_{1},\tilde{S}_{2}),
\end{eqnarray}
where (1) follows from the Markov chains $\tilde{S}_{2}\rightarrow (X_{1},Q,S,\tilde{S}_{1})\rightarrow Y$ and
$\tilde{S}_{2}\rightarrow (X_{1},Q,S,\tilde{S}_{1},X_{2})\rightarrow Y$.
Thus in block $i$ ($2d_{1}+1\leq i\leq n$) and given $\tilde{s}_{1}$, the transmitter $1$ chooses $\widetilde{q}^{N_{\tilde{s}_{1}}}_{i}$
with the index $w^{'}_{0,i,\tilde{s}_{1}}=(w_{1,i-d_{1},\tilde{s}_{1}}^{'},\check{w}_{2,i-d_{1},\tilde{s}_{1}}^{'})$.
Finally, the transmitter $1$ sends $\widetilde{q}_{i}$ by multiplexing the different sub-codewords $\widetilde{q}^{N_{\tilde{s}_{1}}}_{i}$.

\item \textbf{Transmitter $2$'s encoding scheme of $\widetilde{Q}_{i}$:}
Analogously, in block $1\leq i\leq 2d_{1}$ and for each $\tilde{s}_{1}$,
the transmitter $2$ also chooses $w^{'}_{0,i,\tilde{s}_{1}}=1$ as the index of the transmitted
$\widetilde{q}^{N_{\tilde{s}_{1}}}_{i}$. In block $i$ ($2d_{1}+1\leq i\leq n$), for each $\tilde{s}_{1}$, the transmitter $2$ has
already known the delayed state sequence
$\widetilde{s}^{N_{\tilde{s}_{1}}}_{i-2d_{1}}$, $w^{'}_{0,i-d_{1},\tilde{s}_{1}}$ and $w_{2,i-d_{1},\tilde{s}_{1}}^{'}=(w_{2,i-d_{1},0,\tilde{s}_{1}},
w_{2,i-d_{1},1,\tilde{s}_{1}},w_{2,i-d_{1},\tilde{s}_{1}}^{*})$.
Once the transmitter $2$ receives the feedback $\widetilde{y}^{N_{\tilde{s}_{1}}}_{i-d_{1}}$, he attempts to find a unique
sequence $\widetilde{x}^{N_{\tilde{s}_{1}}}_{1,i-d_{1}}(\tilde{w}_{1,i-d_{1},\tilde{s}_{1}}^{'},w^{'}_{0,i-d_{1},\tilde{s}_{1}})$
such that $(\widetilde{x}^{N_{\tilde{s}_{1}}}_{2,i-d_{1}}(w_{2,i-d_{1},\tilde{s}_{1}}^{'},w^{'}_{0,i-d_{1},\tilde{s}_{1}}),\\
\widetilde{x}^{N_{\tilde{s}_{1}}}_{1,i-d_{1}}(\tilde{w}_{1,i-d_{1},\tilde{s}_{1}}^{'},w^{'}_{0,i-d_{1},\tilde{s}_{1}}),
\widetilde{q}^{N_{\tilde{s}_{1}}}_{i-d_{1}}(w^{'}_{0,i-d_{1},\tilde{s}_{1}}),\widetilde{s}^{N_{\tilde{s}_{1}}}_{i-d_{1}},
\widetilde{y}^{N_{\tilde{s}_{1}}}_{i-d_{1}})$ are jointly typical sequences,
where
$\tilde{w}_{1,i-d_{1},\tilde{s}_{1}}^{'}$ is the transmitter $2$'s estimation of $w_{1,i-d_{1},\tilde{s}_{1}}^{'}$.
From the packing lemma \cite{network}, the error probability $Pr\{\tilde{w}_{1,i-1,\tilde{s}_{1}}^{'}\neq w_{1,i-1,\tilde{s}_{1}}^{'}\}$ approaches to $0$ if
\begin{eqnarray}\label{bgod-2}
&&R_{10}(\tilde{s}_{1})+R_{11}(\tilde{s}_{1})+R^{*}_{1}(\tilde{s}_{1})\leq I(X_{1};Y|X_{2},Q,S,\tilde{S}_{1}=\tilde{s}_{1}).
\end{eqnarray}
Here note that (\ref{bgod-2}) implies that
\begin{eqnarray}\label{bgod-2.x}
&&R_{10}+R_{11}+R^{*}_{1}\nonumber\\
&&=\sum_{\tilde{s}_{1}=1}^{k}P_{\tilde{S}_{1}}(\tilde{s}_{1})(R_{10}(\tilde{s}_{1})+R_{11}(\tilde{s}_{1})+R^{*}_{1}(\tilde{s}_{1}))\nonumber\\
&&\leq\sum_{\tilde{s}_{1}=1}^{k}P_{\tilde{S}_{1}}(\tilde{s}_{1})I(X_{1};Y|X_{2},Q,S,\tilde{S}_{1}=\tilde{s}_{1})\nonumber\\
&&=I(X_{1};Y|X_{2},Q,S,\tilde{S}_{1})\stackrel{(2)}=I(X_{1};Y|X_{2},Q,S,\tilde{S}_{1},\tilde{S}_{2}),
\end{eqnarray}
where (2) follows from the Markov chains $\tilde{S}_{2}\rightarrow (X_{2},Q,S,\tilde{S}_{1})\rightarrow Y$ and
$\tilde{S}_{2}\rightarrow (X_{2},Q,S,\tilde{S}_{1},X_{1})\rightarrow Y$.
Thus in block $i$ and given $\tilde{s}_{1}$, the transmitter $2$ chooses $\widetilde{q}^{N_{\tilde{s}_{1}}}_{i}$
with the index $w^{'}_{0,i,\tilde{s}_{1}}=(\tilde{w}_{1,i-d_{1},\tilde{s}_{1}}^{'},w_{2,i-d_{1},\tilde{s}_{1}}^{'})$.
Finally, the transmitter $2$ sends $\widetilde{q}_{i}$ by multiplexing the different sub-codewords $\widetilde{q}^{N_{\tilde{s}_{1}}}_{i}$.
\end{itemize}

\item \textbf{Encoding schemes for $\widetilde{X}_{1,i}$ and $\widetilde{X}_{2,i}$ ($1\leq i\leq n$):}
\begin{itemize}

\item In block $1\leq i\leq 2d_{1}$ and for each $\tilde{s}_{1}$, the transmitter $j$ ($j=1,2$) chooses
$w_{j,i,\tilde{s}_{1}}^{'}=(w_{j,i,0,\tilde{s}_{1}},w_{j,i,1,\tilde{s}_{1}}=const,w_{j,i,\tilde{s}_{1}}^{*})$
as the index of the transmitted codeword $\widetilde{x}^{N_{\tilde{s}_{1}}}_{j,i}$. The codeword $\widetilde{x}_{j,i}$ is
chosen by multiplexing the different sub-codewords $\widetilde{x}^{N_{\tilde{s}_{1}}}_{j,i}$.

\item In block $2d_{1}+1\leq i\leq n$, the transmitters have already received the delayed state sequence
$\widetilde{s}^{N_{\tilde{s}_{1}}}_{i-2d_{1}}$, which is the delayed feedback state used to de-multiplex $\widetilde{y}_{i-d_{1}}$ into the sub-sequences
$\widetilde{y}^{N_{1}}_{i-d_{1}}$, ..., $\widetilde{y}^{N_{k}}_{i-d_{1}}$.
Once the transmitters obtain the delayed channel output feedback $\widetilde{y}_{i-d_{1}}$,
they first demultiplex them into sub-sequences
$\widetilde{y}^{N_{1}}_{i-d_{1}}$, $\widetilde{y}^{N_{2}}_{i-d_{1}}$,...,
$\widetilde{y}^{N_{k}}_{i-d_{1}}$.
Then, for the sub-sequence $\widetilde{y}^{N_{\tilde{s}_{1}}}_{i-d_{1}}$,
produce a mapping
$g_{i,\tilde{s}_{1}}: \widetilde{y}^{N_{\tilde{s}_{1}}}_{i-d_{1}}\rightarrow \{1,2,...,2^{N_{\tilde{s}_{1}}(R_{11}(\tilde{s}_{1})+R_{21}(\tilde{s}_{1}))}\}$.
Furthermore, define $K_{i,\tilde{s}_{1}}^{*}=(K_{1,i,\tilde{s}_{1}}^{*},K_{2,i,\tilde{s}_{1}}^{*})
=g_{i,\tilde{s}_{1}}(\widetilde{Y}^{N_{\tilde{s}_{1}}}_{i-d_{1}})$
as a random variable uniformly distributed over $\{1,2,...,2^{N_{\tilde{s}_{1}}(R_{11}(\tilde{s}_{1})+R_{21}(\tilde{s}_{1}))}\}$, and it is
independent of $\widetilde{X}^{N_{\tilde{s}_{1}}}_{1,i}$, $\widetilde{X}^{N_{\tilde{s}_{1}}}_{2,i}$, $\widetilde{S}^{N_{\tilde{s}_{1}}}_{i}$,
$\widetilde{Y}^{N_{\tilde{s}_{1}}}_{i}$, $\widetilde{Z}^{N_{\tilde{s}_{1}}}_{i}$, $W_{1,i}$, $W_{2,i}$, $W_{1,i}^{*}$ and $W_{2,i}^{*}$.
Here note that $K_{j,i,\tilde{s}_{1}}^{*}$ ($j=1,2$) is used as a secret key of
the $i$-th block shared by the transmitter $j$ and the legal receiver, and
$k_{j,i,\tilde{s}_{1}}^{*}\in \{1,2,...,2^{N_{\tilde{s}_{1}}R_{j1}(\tilde{s}_{1})}\}$ is a specific value of $K_{j,i,\tilde{s}_{1}}^{*}$.
Moreover, note that $k_{2,i,\tilde{s}_{1}}^{*}$ can be further divided by the delayed state $\tilde{s}_{2}$, i.e.,
$k_{2,i,\tilde{s}_{1}}^{*}=(k_{2,i,\tilde{s}_{1},1}^{*},...,k_{2,i,\tilde{s}_{1},k}^{*})$, where $k_{2,i,\tilde{s}_{1},\tilde{s}_{2}}^{*}$
($1\leq \tilde{s}_{2}\leq k$) takes values in $\{1,2,...,2^{N_{\tilde{s}_{1},\tilde{s}_{2}}R_{21}(\tilde{s}_{1},\tilde{s}_{2})}\}$.

Reveal the mapping $g_{i,\tilde{s}_{1}}$ to the transmitters, legal receiver and the eavesdropper.
After the generation of the secret key, the transmitter $1$ chooses $\widetilde{x}^{N_{\tilde{s}_{1}}}_{1,i}$ with the index
$w_{1,i,\tilde{s}_{1}}^{'}=(w_{1,i,0,\tilde{s}_{1}},w_{1,i,1,\tilde{s}_{1}}\oplus k_{1,i,\tilde{s}_{1}}^{*},w_{1,i,\tilde{s}_{1}}^{*})$.
The codeword $\widetilde{x}_{1,i}$ is
chosen by multiplexing the different sub-codewords $\widetilde{x}^{N_{\tilde{s}_{1}}}_{1,i}$.

Similarly, for given $\tilde{s}_{1}$ and $\tilde{s}_{2}$, the transmitter $2$ chooses $\widetilde{x}^{N_{\tilde{s}_{1},\tilde{s}_{2}}}_{2,i}$ with the index
$w_{2,i,\tilde{s}_{1},\tilde{s}_{2}}^{'}=(w_{2,i,0,\tilde{s}_{1},\tilde{s}_{2}},\\
w_{2,i,1,\tilde{s}_{1},\tilde{s}_{2}}\oplus k_{2,i,\tilde{s}_{1},\tilde{s}_{2}}^{*},w_{2,i,\tilde{s}_{1},\tilde{s}_{2}}^{*})$.
The codeword $\widetilde{x}_{2,i}$ is
chosen by multiplexing the different sub-codewords $\widetilde{x}^{N_{\tilde{s}_{1},\tilde{s}_{2}}}_{1,i}$.

\end{itemize}

\end{itemize}

\subsubsection*{3). Decoding scheme}

Once the legal receiver receives all $n$ blocks
$y^{n}=(\widetilde{y}_{1},...,\widetilde{y}_{n})$ and $s^{n}=(\widetilde{s}_{1},...,\widetilde{s}_{n})$, first, he demultiplexes them into
sub-sequences
$\widetilde{y}^{N_{1}}_{1}$, $\widetilde{y}^{N_{2}}_{1}$,...,
$\widetilde{y}^{N_{k}}_{1}$,...,$\widetilde{y}^{N_{1}}_{n}$,...,$\widetilde{y}^{N_{k}}_{n}$, $\widetilde{s}^{N_{1}}_{1}$, $\widetilde{s}^{N_{2}}_{1}$,...,
$\widetilde{s}^{N_{k}}_{1}$,...,$\widetilde{s}^{N_{1}}_{n}$,...,$\widetilde{s}^{N_{k}}_{n}$. Then, since the legal receiver also knows the secret key
produced by the delayed channel output feedback,
he does backward decoding
which is exactly the same as that of the classical MAC with noiseless feedback, see \cite{coverz}.
Following similar steps of error probability analysis for MAC with noiseless feedback (see \cite[pp. 295-296]{coverz}), we can conclude that
the legal receiver can decode the transmitted messages and the dummy messages with decoding error probability less than any $\epsilon>0$
if
\begin{eqnarray}\label{bgod-3}
&&R_{10}+R_{11}+R^{*}_{1}+R_{20}+R_{21}+R^{*}_{2}\leq I(X_{1},X_{2};Y|S,\widetilde{S}_{1})\nonumber\\
&&\stackrel{(a)}=H(Y|S,\widetilde{S}_{1},\widetilde{S}_{2})-H(Y|X_{1},X_{2},S,\widetilde{S}_{1},\widetilde{S}_{2})\nonumber\\
&&=I(X_{1},X_{2};Y|S,\widetilde{S}_{1},\widetilde{S}_{2}),
\end{eqnarray}
where (a) is from the Markov chains $\widetilde{S}_{2}\rightarrow (S,\widetilde{S}_{1})\rightarrow Y$
and $\widetilde{S}_{2}\rightarrow (X_{1},X_{2},S,\widetilde{S}_{1})\rightarrow Y$.

\subsection{Equivocation analysis}\label{sub-x3}

First, we give a lower bound
on $H(K_{i,\tilde{s}_{1}}^{*}|\widetilde{X}^{N_{\tilde{s}_{1}}}_{1,i-d_{1}},\widetilde{X}^{N_{\tilde{s}_{1}}}_{2,i-d_{1}},
\widetilde{S}^{N_{\tilde{s}_{1}}}_{i-d_{1}},\widetilde{Z}^{N_{\tilde{s}_{1}}}_{i-d_{1}},\tilde{S}_{1}=\tilde{s}_{1})$, which will be used in the analysis of
the eavesdropper's equivocation about the transmitted messages $W_{1}$ and $W_{2}$.

In block $i-d_{1}$ ($2d_{1}+1\leq i\leq n$) and for a given $\tilde{S}_{1}=\tilde{s}_{1}$, suppose that the eavesdropper
knows not only $\widetilde{S}^{N_{\tilde{s}_{1}}}_{i-d_{1}}$ and $\widetilde{Z}^{N_{\tilde{s}_{1}}}_{i-d_{1}}$, but also
$\widetilde{X}^{N_{\tilde{s}_{1}}}_{1,i-d_{1}}$, $\widetilde{X}^{N_{\tilde{s}_{1}}}_{2,i-d_{1}}$, the eavesdropper's equivocation
about the secret key $K_{i,\tilde{s}_{1}}^{*}$ can be bounded by Ahlswede and Cai's balanced coloring lemma \cite[p. 260]{AC}, see the followings.
\begin{lemma}\label{Lx}
\textbf{(Balanced coloring lemma)} Given $\tilde{S}_{1}=\tilde{s}_{1}$, for
arbitrary $\epsilon, \delta>0$, sufficiently large $N_{\tilde{s}_{1}}$, all $N_{\tilde{s}_{1}}$-type
$P_{X_{1}X_{2}S\tilde{S}_{1}Y}(x_{1},x_{2},s,\tilde{s}_{1},y)$ and all
$\widetilde{x}^{N_{\tilde{s}_{1}}}_{1,i-d_{1}}, \widetilde{x}^{N_{\tilde{s}_{1}}}_{2,i-d_{2}}, \widetilde{s}^{N_{\tilde{s}_{1}}}_{i-d_{1}}
\in T_{X_{1}X_{2}S|\tilde{S}_{1}}^{N_{\tilde{s}_{1}}}(\tilde{s}_{1})$ (where $2d_{1}+1\leq i\leq n$),
there exists a $\gamma$-coloring $c: T_{Y|X_{1},X_{2},S,\tilde{S}_{1}}^{N_{\tilde{s}_{1}}}(\widetilde{x}^{N_{\tilde{s}_{1}}}_{1,i-d_{1}},
\widetilde{x}^{N_{\tilde{s}_{1}}}_{2,i-d_{1}},\widetilde{s}^{N_{\tilde{s}_{1}}}_{i-d_{1}},\tilde{s}_{1})\rightarrow \{1,2,..,\gamma\}$
such that for all joint $N_{\tilde{s}_{1}}$-type $P_{X_{1}X_{2}S\tilde{S}_{1}YZ}(x_{1},x_{2},s,\tilde{s}_{1},y,z)$
with marginal distribution $P_{X_{1}X_{2}S\tilde{S}_{1}Z}(x_{1},x_{2},s,\tilde{s}_{1},z)$,
$$\frac{|T_{Y|X_{1},X_{2},S,\tilde{S}_{1},Z}^{N_{\tilde{s}_{1}}}(\widetilde{x}^{N_{\tilde{s}_{1}}}_{1,i-d_{1}},
\widetilde{x}^{N_{\tilde{s}_{1}}}_{2,i-d_{1}},\widetilde{s}^{N_{\tilde{s}_{1}}}_{i-d_{1}},\widetilde{z}^{N_{\tilde{s}_{1}}}_{i-d_{1}},\tilde{s}_{1})|}{\gamma}
\geq 2^{N_{\tilde{s}_{1}}\epsilon},$$
and
$\widetilde{x}^{N_{\tilde{s}_{1}}}_{1,i-d_{1}},
\widetilde{x}^{N_{\tilde{s}_{1}}}_{2,i-d_{1}},\widetilde{s}^{N_{\tilde{s}_{1}}}_{i-d_{1}},\widetilde{z}^{N_{\tilde{s}_{1}}}_{i-d_{1}}\in
T^{N_{\tilde{s}_{1}}}_{X_{1}X_{2}SZ|\tilde{S}_{1}}$,
\begin{equation}\label{bgod-3.xxx}
|c^{-1}(k)|\leq \frac{|T_{Y|X_{1},X_{2},S,\tilde{S}_{1},Z}^{N_{\tilde{s}_{1}}}(\widetilde{x}^{N_{\tilde{s}_{1}}}_{1,i-d_{1}},
\widetilde{x}^{N_{\tilde{s}_{1}}}_{2,i-d_{1}},\widetilde{s}^{N_{\tilde{s}_{1}}}_{i-d_{1}},\widetilde{z}^{N_{\tilde{s}_{1}}}_{i-d_{1}},\tilde{s}_{1})|
(1+\delta)}{\gamma},
\end{equation}
for $k=1,2,...,\gamma$, where $c^{-1}$ is the inverse image of $c$.
\end{lemma}
From Lemma \ref{Lx}, we see that the typical set
$T_{Y|X_{1},X_{2},S,\tilde{S}_{1},Z}^{N_{\tilde{s}_{1}}}(\widetilde{x}^{N_{\tilde{s}_{1}}}_{1,i-d_{1}},
\widetilde{x}^{N_{\tilde{s}_{1}}}_{2,i-d_{1}},\widetilde{s}^{N_{\tilde{s}_{1}}}_{i-d_{1}},\widetilde{z}^{N_{\tilde{s}_{1}}}_{i-d_{1}},\tilde{s}_{1})$
maps into at least
\begin{eqnarray}\label{bgod-4}
&&\frac{|T_{Y|X_{1},X_{2},S,\tilde{S}_{1},Z}^{N_{\tilde{s}_{1}}}(\widetilde{x}^{N_{\tilde{s}_{1}}}_{1,i-d_{1}},
\widetilde{x}^{N_{\tilde{s}_{1}}}_{2,i-d_{1}},\widetilde{s}^{N_{\tilde{s}_{1}}}_{i-d_{1}},\widetilde{z}^{N_{\tilde{s}_{1}}}_{i-d_{1}},\tilde{s}_{1})|}
{\frac{|T_{Y|X_{1},X_{2},S,\tilde{S}_{1},Z}^{N_{\tilde{s}_{1}}}(\widetilde{x}^{N_{\tilde{s}_{1}}}_{1,i-d_{1}},
\widetilde{x}^{N_{\tilde{s}_{1}}}_{2,i-d_{1}},\widetilde{s}^{N_{\tilde{s}_{1}}}_{i-d_{1}},\widetilde{z}^{N_{\tilde{s}_{1}}}_{i-d_{1}},\tilde{s}_{1})|
(1+\delta)}{\gamma}}=\frac{\gamma}{1+\delta}
\end{eqnarray}
colors. On the other hand, the typical set $T_{Y|X_{1},X_{2},S,\tilde{S}_{1},Z}^{N_{\tilde{s}_{1}}}(\widetilde{x}^{N_{\tilde{s}_{1}}}_{1,i-d_{1}},
\widetilde{x}^{N_{\tilde{s}_{1}}}_{2,i-d_{1}},\widetilde{s}^{N_{\tilde{s}_{1}}}_{i-d_{1}},\widetilde{z}^{N_{\tilde{s}_{1}}}_{i-d_{1}},\tilde{s}_{1})$
maps into at most $\gamma$ colors.
From (\ref{bgod-4}), we can conclude that
\begin{eqnarray}\label{bgod-5}
&&H(K_{i,\tilde{s}_{1}}^{*}|\widetilde{X}^{N_{\tilde{s}_{1}}}_{1,i-d_{1}},\widetilde{X}^{N_{\tilde{s}_{1}}}_{2,i-d_{1}},
\widetilde{S}^{N_{\tilde{s}_{1}}}_{i-d_{1}},\widetilde{Z}^{N_{\tilde{s}_{1}}}_{i-d_{1}},\tilde{S}_{1}=\tilde{s}_{1})\geq \log\frac{\gamma}{1+\delta}.
\end{eqnarray}
Here note that $$\frac{|T_{Y|X_{1},X_{2},S,\tilde{S}_{1},Z}^{N_{\tilde{s}_{1}}}(\widetilde{x}^{N_{\tilde{s}_{1}}}_{1,i-d_{1}},
\widetilde{x}^{N_{\tilde{s}_{1}}}_{2,i-d_{1}},\widetilde{s}^{N_{\tilde{s}_{1}}}_{i-d_{1}},\widetilde{z}^{N_{\tilde{s}_{1}}}_{i-d_{1}},\tilde{s}_{1})|}
{\gamma}\geq 2^{N_{\tilde{s}_{1}}\epsilon}$$ implies that
$$\gamma\leq |T_{Y|X_{1},X_{2},S,\tilde{S}_{1},Z}^{N_{\tilde{s}_{1}}}(\widetilde{x}^{N_{\tilde{s}_{1}}}_{1,i-d_{1}},
\widetilde{x}^{N_{\tilde{s}_{1}}}_{2,i-d_{1}},\widetilde{s}^{N_{\tilde{s}_{1}}}_{i-d_{1}},\widetilde{z}^{N_{\tilde{s}_{1}}}_{i-d_{1}},\tilde{s}_{1})|.$$
Choosing $\gamma=|T_{Y|X_{1},X_{2},S,\tilde{S}_{1},Z}^{N_{\tilde{s}_{1}}}(\widetilde{x}^{N_{\tilde{s}_{1}}}_{1,i-d_{1}},
\widetilde{x}^{N_{\tilde{s}_{1}}}_{2,i-d_{1}},\widetilde{s}^{N_{\tilde{s}_{1}}}_{i-d_{1}},\widetilde{z}^{N_{\tilde{s}_{1}}}_{i-d_{1}},\tilde{s}_{1})|$
and noticing that
\begin{eqnarray}\label{bgod-6}
&&|T_{Y|X_{1},X_{2},S,\tilde{S}_{1},Z}^{N_{\tilde{s}_{1}}}(\widetilde{x}^{N_{\tilde{s}_{1}}}_{1,i-d_{1}},
\widetilde{x}^{N_{\tilde{s}_{1}}}_{2,i-d_{1}},\widetilde{s}^{N_{\tilde{s}_{1}}}_{i-d_{1}},\widetilde{z}^{N_{\tilde{s}_{1}}}_{i-d_{1}},\tilde{s}_{1})|\nonumber\\
&&\geq (1-\epsilon_{1})2^{N_{\tilde{s}_{1}}(1-\epsilon_{2})H(Y|X_{1},X_{2},S,Z,\tilde{S}_{1}=\tilde{s}_{1})},
\end{eqnarray}
where $\epsilon_{1}$ and $\epsilon_{2}$ tend to $0$ as $N$ tends to infinity, (\ref{bgod-5}) can be further bounded by
\begin{eqnarray}\label{bgod-7}
&&H(K_{i,\tilde{s}_{1}}^{*}|\widetilde{X}^{N_{\tilde{s}_{1}}}_{1,i-d_{1}},\widetilde{X}^{N_{\tilde{s}_{1}}}_{2,i-d_{1}},
\widetilde{S}^{N_{\tilde{s}_{1}}}_{i-d_{1}},\widetilde{Z}^{N_{\tilde{s}_{1}}}_{i-d_{1}},\tilde{S}_{1}=\tilde{s}_{1})\nonumber\\
&&\geq \log\frac{1-\epsilon_{1}}{1+\delta}+N_{\tilde{s}_{1}}(1-\epsilon_{2})H(Y|X_{1},X_{2},S,Z,\tilde{S}_{1}=\tilde{s}_{1}).
\end{eqnarray}
Now we show the bound on the eavesdropper's equivocation $\Delta$ to the transmitted messages, see the followings.
The overall equivocation $\Delta$, which is denoted by $\Delta=\frac{1}{nN}H(W_{1},W_{2}|Z^{n},S^{n})$, can be expressed as
\begin{eqnarray}\label{bgod-8}
&&\Delta=\frac{1}{nN}H(W_{1},W_{2}|Z^{n},S^{n})\nonumber\\
&&=\frac{1}{nN}\sum_{i=1}^{n}H(W_{1,i},W_{2,i}|Z^{n},S^{n},W_{1,1},W_{2,1},...,W_{1,i-1},W_{2,i-1})\nonumber\\
&&=\frac{1}{nN}\sum_{i=1}^{n}H(W_{1,i,0},W_{1,i,1},W_{2,i,0},W_{2,i,1}|Z^{n},S^{n},W_{1,1},W_{2,1},...,W_{1,i-1},W_{2,i-1})\nonumber\\
&&=\frac{1}{nN}\sum_{i=1}^{n}(H(W_{1,i,0},W_{2,i,0}|Z^{n},S^{n},W_{1,1},W_{2,1},...,W_{1,i-1},W_{2,i-1})\nonumber\\
&&+H(W_{1,i,1},W_{2,i,1}|Z^{n},S^{n},W_{1,1},W_{2,1},...,W_{1,i-1},W_{2,i-1},W_{1,i,0},W_{2,i,0}))\nonumber\\
&&\stackrel{(a)}=\frac{1}{nN}\sum_{i=1}^{n}H(W_{1,i,0},W_{2,i,0}|Z^{n},S^{n},W_{1,1},W_{2,1},...,W_{1,i-1},W_{2,i-1})\nonumber\\
&&+\frac{1}{nN}\sum_{i=2d_{1}+1}^{n}
H(W_{1,i,1},W_{2,i,1}|Z^{n},S^{n},W_{1,1},W_{2,1},...,W_{1,i-1},W_{2,i-1},W_{1,i,0},W_{2,i,0}),
\end{eqnarray}
where (a) is from the fact that $W_{1,i,1}$ and $W_{2,i,1}$ are constants when $1\leq i\leq 2d_{1}$.

The first conditional entropy $H(W_{1,i,0},W_{2,i,0}|Z^{n},S^{n},W_{1,1},W_{2,1},...,W_{1,i-1},W_{2,i-1})$ of (\ref{bgod-8}) is bounded by
\begin{eqnarray}\label{bgod-9}
&&H(W_{1,i,0},W_{2,i,0}|Z^{n},S^{n},W_{1,1},W_{2,1},...,W_{1,i-1},W_{2,i-1})\nonumber\\
&&=\sum_{\tilde{s}_{1}=1}^{k}H(W_{1,i,0,\tilde{s}_{1}},W_{2,i,0,\tilde{s}_{1}}|
W_{1,i,0,1},W_{2,i,0,1},...,W_{1,i,0,\tilde{s}_{1}-1},W_{2,i,0,\tilde{s}_{1}-1},Z^{n},S^{n},W_{1,1},W_{2,1},...,W_{1,i-1},W_{2,i-1})\nonumber\\
&&\geq \sum_{\tilde{s}_{1}=1}^{k}H(W_{1,i,0,\tilde{s}_{1}},W_{2,i,0,\tilde{s}_{1}}|
W_{1,i,0,1},W_{2,i,0,1},...,W_{1,i,0,\tilde{s}_{1}-1},W_{2,i,0,\tilde{s}_{1}-1},Z^{n},S^{n},W_{1,1},W_{2,1},...,W_{1,i-1},W_{2,i-1},\nonumber\\
&&\tilde{S}_{1}=\tilde{s}_{1})\nonumber\\
&&\stackrel{(b)}=\sum_{\tilde{s}_{1}=1}^{k}H(W_{1,i,0,\tilde{s}_{1}},W_{2,i,0,\tilde{s}_{1}}|
\widetilde{Z}_{i}^{N_{\tilde{s}_{1}}},\widetilde{S}_{i}^{N_{\tilde{s}_{1}}},\tilde{S}_{1}=\tilde{s}_{1})\nonumber\\
&&=\sum_{\tilde{s}_{1}=1}^{k}(H(W_{1,i,0,\tilde{s}_{1}},W_{2,i,0,\tilde{s}_{1}},\widetilde{Z}_{i}^{N_{\tilde{s}_{1}}},\widetilde{S}_{i}^{N_{\tilde{s}_{1}}},\tilde{S}_{1}=\tilde{s}_{1})
-H(\widetilde{Z}_{i}^{N_{\tilde{s}_{1}}},\widetilde{S}_{i}^{N_{\tilde{s}_{1}}},\tilde{S}_{1}=\tilde{s}_{1}))\nonumber\\
&&=\sum_{\tilde{s}_{1}=1}^{k}(H(\widetilde{X}_{1,i}^{N_{\tilde{s}_{1}}},\widetilde{X}_{2,i}^{N_{\tilde{s}_{1}}},W_{1,i,0,\tilde{s}_{1}},W_{2,i,0,\tilde{s}_{1}},
\widetilde{Z}_{i}^{N_{\tilde{s}_{1}}},\widetilde{S}_{i}^{N_{\tilde{s}_{1}}},\tilde{S}_{1}=\tilde{s}_{1})\nonumber\\
&&-H(\widetilde{X}_{1,i}^{N_{\tilde{s}_{1}}},\widetilde{X}_{2,i}^{N_{\tilde{s}_{1}}}|W_{1,i,0,\tilde{s}_{1}},
W_{2,i,0,\tilde{s}_{1}},\widetilde{Z}_{i}^{N_{\tilde{s}_{1}}},\widetilde{S}_{i}^{N_{\tilde{s}_{1}}},\tilde{S}_{1}=\tilde{s}_{1})
-H(\widetilde{Z}_{i}^{N_{\tilde{s}_{1}}},\widetilde{S}_{i}^{N_{\tilde{s}_{1}}},\tilde{S}_{1}=\tilde{s}_{1}))\nonumber\\
&&\stackrel{(c)}=\sum_{\tilde{s}_{1}=1}^{k}(H(\widetilde{Z}_{i}^{N_{\tilde{s}_{1}}}|\widetilde{X}_{1,i}^{N_{\tilde{s}_{1}}},
\widetilde{X}_{2,i}^{N_{\tilde{s}_{1}}},\widetilde{S}_{i}^{N_{\tilde{s}_{1}}},\tilde{S}_{1}=\tilde{s}_{1})
+H(\widetilde{X}_{1,i}^{N_{\tilde{s}_{1}}},\widetilde{X}_{2,i}^{N_{\tilde{s}_{1}}}|\widetilde{S}_{i}^{N_{\tilde{s}_{1}}},\tilde{S}_{1}=\tilde{s}_{1})\nonumber\\
&&-H(\widetilde{X}_{1,i}^{N_{\tilde{s}_{1}}},\widetilde{X}_{2,i}^{N_{\tilde{s}_{1}}}|W_{1,i,0,\tilde{s}_{1}},
W_{2,i,0,\tilde{s}_{1}},\widetilde{Z}_{i}^{N_{\tilde{s}_{1}}},\widetilde{S}_{i}^{N_{\tilde{s}_{1}}},\tilde{S}_{1}=\tilde{s}_{1})
-H(\widetilde{Z}_{i}^{N_{\tilde{s}_{1}}}|\widetilde{S}_{i}^{N_{\tilde{s}_{1}}},\tilde{S}_{1}=\tilde{s}_{1}))\nonumber\\
&&\stackrel{(d)}=\sum_{\tilde{s}_{1}=1}^{k}(H(\widetilde{Z}_{i}^{N_{\tilde{s}_{1}}}|\widetilde{X}_{1,i}^{N_{\tilde{s}_{1}}},
\widetilde{X}_{2,i}^{N_{\tilde{s}_{1}}},\widetilde{S}_{i}^{N_{\tilde{s}_{1}}},\tilde{S}_{1}=\tilde{s}_{1})
+H(\widetilde{X}_{1,i}^{N_{\tilde{s}_{1}}}|\tilde{S}_{1}=\tilde{s}_{1})
+H(\widetilde{X}_{2,i}^{N_{\tilde{s}_{1}}}|\tilde{S}_{1}=\tilde{s}_{1})\nonumber\\
&&-H(\widetilde{X}_{1,i}^{N_{\tilde{s}_{1}}},\widetilde{X}_{2,i}^{N_{\tilde{s}_{1}}}|W_{1,i,0,\tilde{s}_{1}},
W_{2,i,0,\tilde{s}_{1}},\widetilde{Z}_{i}^{N_{\tilde{s}_{1}}},\widetilde{S}_{i}^{N_{\tilde{s}_{1}}},\tilde{S}_{1}=\tilde{s}_{1})
-H(\widetilde{Z}_{i}^{N_{\tilde{s}_{1}}}|\widetilde{S}_{i}^{N_{\tilde{s}_{1}}},\tilde{S}_{1}=\tilde{s}_{1}))\nonumber\\
&&\stackrel{(e)}\geq\sum_{\tilde{s}_{1}=1}^{k}(H(\widetilde{X}_{1,i}^{N_{\tilde{s}_{1}}}|\tilde{S}_{1}=\tilde{s}_{1})
+H(\widetilde{X}_{2,i}^{N_{\tilde{s}_{1}}}|\tilde{S}_{1}=\tilde{s}_{1})\nonumber\\
&&-H(\widetilde{X}_{1,i}^{N_{\tilde{s}_{1}}},\widetilde{X}_{2,i}^{N_{\tilde{s}_{1}}}|W_{1,i,0,\tilde{s}_{1}},
W_{2,i,0,\tilde{s}_{1}},\widetilde{Z}_{i}^{N_{\tilde{s}_{1}}},\widetilde{S}_{i}^{N_{\tilde{s}_{1}}},\tilde{S}_{1}=\tilde{s}_{1})
-N_{\tilde{s}_{1}}I(X_{1},X_{2};Z|S,\tilde{S}_{1}=\tilde{s}_{1}))\nonumber\\
&&\stackrel{(f)}=\sum_{\tilde{s}_{1}=1}^{k}(N_{\tilde{s}_{1}}(R_{10}(\tilde{s}_{1})+R_{11}(\tilde{s}_{1})+R^{*}_{1}(\tilde{s}_{1}))
+N_{\tilde{s}_{1}}(R_{20}(\tilde{s}_{1})+R_{21}(\tilde{s}_{1})+R^{*}_{2}(\tilde{s}_{1}))\nonumber\\
&&-H(\widetilde{X}_{1,i}^{N_{\tilde{s}_{1}}},\widetilde{X}_{2,i}^{N_{\tilde{s}_{1}}}|W_{1,i,0,\tilde{s}_{1}},
W_{2,i,0,\tilde{s}_{1}},\widetilde{Z}_{i}^{N_{\tilde{s}_{1}}},\widetilde{S}_{i}^{N_{\tilde{s}_{1}}},\tilde{S}_{1}=\tilde{s}_{1})
-N_{\tilde{s}_{1}}I(X_{1},X_{2};Z|S,\tilde{S}_{1}=\tilde{s}_{1}))\nonumber\\
&&\stackrel{(g)}\geq\sum_{\tilde{s}_{1}=1}^{k}(N_{\tilde{s}_{1}}(R_{10}(\tilde{s}_{1})+R_{11}(\tilde{s}_{1})+R^{*}_{1}(\tilde{s}_{1}))
+N_{\tilde{s}_{1}}(R_{20}(\tilde{s}_{1})+R_{21}(\tilde{s}_{1})+R^{*}_{2}(\tilde{s}_{1}))\nonumber\\
&&-N_{\tilde{s}_{1}}\epsilon_{3}-N_{\tilde{s}_{1}}I(X_{1},X_{2};Z|S,\tilde{S}_{1}=\tilde{s}_{1})),
\end{eqnarray}
where (b) follows from the Markov chain $(W_{1,i,0,1},W_{2,i,0,1},...,W_{1,i,0,\tilde{s}_{1}-1},W_{2,i,0,\tilde{s}_{1}-1},
W_{1,1},W_{2,1},...,W_{1,i-1},\\W_{2,i-1},\widetilde{Z}_{i}^{N_{1}},\widetilde{S}_{i}^{N_{1}},...,\widetilde{Z}_{i}^{N_{\tilde{s}_{1}-1}},
\widetilde{S}_{i}^{N_{\tilde{s}_{1}-1}},\widetilde{Z}_{i}^{N_{\tilde{s}_{1}+1}},
\widetilde{S}_{i}^{N_{\tilde{s}_{1}+1}}...,\widetilde{Z}_{i}^{N_{k}},\widetilde{S}_{i}^{N_{k}},\widetilde{Z}_{1},\widetilde{S}_{1},...,
\widetilde{Z}_{i-1},\widetilde{S}_{i-1},\widetilde{Z}_{i+1},\widetilde{S}_{i+1},...,\widetilde{Z}_{n},\widetilde{S}_{n}
)\rightarrow (\widetilde{Z}_{i}^{N_{\tilde{s}_{1}}},\widetilde{S}_{i}^{N_{\tilde{s}_{1}}},\tilde{S}_{1}=\tilde{s}_{1})\rightarrow
(W_{1,i,0,\tilde{s}_{1}},W_{2,i,0,\tilde{s}_{1}})$, (c) follows from the fact that
$H(W_{1,i,0,\tilde{s}_{1}},W_{2,i,0,\tilde{s}_{1}}|\widetilde{X}_{1,i}^{N_{\tilde{s}_{1}}},\widetilde{X}_{2,i}^{N_{\tilde{s}_{1}}})=0$,
(d) follows from the fact that given $\tilde{S}_{1}=\tilde{s}_{1}$,
$\widetilde{X}_{1,i}^{N_{\tilde{s}_{1}}}$ is independent of $\widetilde{S}_{i}^{N_{\tilde{s}_{1}}}$ and $\widetilde{X}_{2,i}^{N_{\tilde{s}_{1}}}$,
and given $\tilde{S}_{1}=\tilde{s}_{1}$, $\widetilde{X}_{2,i}^{N_{\tilde{s}_{1}}}$ is independent of $\widetilde{S}_{i}^{N_{\tilde{s}_{1}}}$,
(e) follows from the construction of the codebooks and the
fact that the channel is memoryless, (f) follows from the fact that given $\tilde{S}_{1}=\tilde{s}_{1}$, there are
$2^{N_{\tilde{s}_{1}}(R_{10}(\tilde{s}_{1})+R_{11}(\tilde{s}_{1})+R^{*}_{1}(\tilde{s}_{1}))}$ codewords $\widetilde{X}_{1,i}^{N_{\tilde{s}_{1}}}$,
and there are $2^{N_{\tilde{s}_{1}}(R_{20}(\tilde{s}_{1})+R_{21}(\tilde{s}_{1})+R^{*}_{2}(\tilde{s}_{1}))}$ codewords
$\widetilde{X}_{2,i}^{N_{\tilde{s}_{1}}}$, and (g) follows from the fact that given $W_{1,i,0,\tilde{s}_{1}}$,
$W_{2,i,0,\tilde{s}_{1}}$, $\widetilde{Z}_{i}^{N_{\tilde{s}_{1}}}$, $\widetilde{S}_{i}^{N_{\tilde{s}_{1}}}$ and $\tilde{S}_{1}=\tilde{s}_{1}$,
the eavesdropper's decoding error probability of $\widetilde{X}_{1,i}^{N_{\tilde{s}_{1}}}$ and $\widetilde{X}_{2,i}^{N_{\tilde{s}_{1}}}$
tends to $0$ if
\begin{eqnarray}\label{bgod-10}
&&R_{11}(\tilde{s}_{1})+R^{*}_{1}(\tilde{s}_{1})+R_{21}(\tilde{s}_{1})+R^{*}_{2}(\tilde{s}_{1})\leq
I(X_{1},X_{2};Z|S,\tilde{S}_{1}=\tilde{s}_{1}),
\end{eqnarray}
then by using Fano's inequality, we have $\frac{1}{N_{\tilde{s}_{1}}}H(\widetilde{X}_{1,i}^{N_{\tilde{s}_{1}}},\widetilde{X}_{2,i}^{N_{\tilde{s}_{1}}}
|W_{1,i,0,\tilde{s}_{1}},
W_{2,i,0,\tilde{s}_{1}},\widetilde{Z}_{i}^{N_{\tilde{s}_{1}}},\widetilde{S}_{i}^{N_{\tilde{s}_{1}}},\tilde{S}_{1}=\tilde{s}_{1})\leq \epsilon_{3}$,
where $\epsilon_{3}\rightarrow 0$
as $N_{\tilde{s}_{1}}\rightarrow \infty$. Here note that (\ref{bgod-10}) implies that
\begin{eqnarray}\label{bgod-10.x}
&&R_{11}+R^{*}_{1}+R_{21}+R^{*}_{2}\nonumber\\
&&=\sum_{\tilde{s}_{1}=1}^{k}P_{\tilde{S}_{1}}(\tilde{s}_{1})(R_{11}(\tilde{s}_{1})+R^{*}_{1}(\tilde{s}_{1})+R_{21}(\tilde{s}_{1})+R^{*}_{2}(\tilde{s}_{1}))\nonumber\\
&&\leq\sum_{\tilde{s}_{1}=1}^{k}P_{\tilde{S}_{1}}(\tilde{s}_{1})I(X_{1},X_{2};Z|S,\tilde{S}_{1}=\tilde{s}_{1})\nonumber\\
&&=I(X_{1},X_{2};Z|S,\tilde{S}_{1})\stackrel{(1)}=I(X_{1},X_{2};Z|S,\tilde{S}_{1},\tilde{S}_{2}),
\end{eqnarray}
where (1) follows from the Markov chains $\tilde{S}_{2}\rightarrow (S,\tilde{S}_{1})\rightarrow Z$ and
$\tilde{S}_{2}\rightarrow (S,\tilde{S}_{1},X_{1},X_{2})\rightarrow Z$.

For $2d_{1}+1\leq i\leq n$, the second conditional entropy $H(W_{1,i,1},W_{2,i,1}|Z^{n},S^{n},W_{1,1},W_{2,1},...,W_{1,i-1},W_{2,i-1},\\W_{1,i,0},W_{2,i,0})$
of (\ref{bgod-8}) is bounded by
\begin{eqnarray}\label{bgod-11}
&&H(W_{1,i,1},W_{2,i,1}|Z^{n},S^{n},W_{1,1},W_{2,1},...,W_{1,i-1},W_{2,i-1},W_{1,i,0},W_{2,i,0})\nonumber\\
&&\geq H(W_{1,i,1},W_{2,i,1}|Z^{n},S^{n},W_{1,1},W_{2,1},...,W_{1,i-1},W_{2,i-1},W_{1,i,0},W_{2,i,0},X_{1}^{n},X_{2}^{n})\nonumber\\
&&\stackrel{(h)}=H(W_{1,i,1},W_{2,i,1}|\widetilde{Z}_{i},\widetilde{S}_{i},\widetilde{X}_{1,i},\widetilde{X}_{2,i},
\widetilde{Z}_{i-d_{1}},\widetilde{S}_{i-d_{1}},\widetilde{X}_{1,i-d_{1}},\widetilde{X}_{2,i-d_{1}})\nonumber\\
&&=\sum_{\tilde{s}_{1}=1}^{k}H(W_{1,i,1,\tilde{s}_{1}},W_{2,i,1,\tilde{s}_{1}}|W_{1,i,1,1},W_{2,i,1,1},...,W_{1,i,1,\tilde{s}_{1}-1},W_{2,i,1,\tilde{s}_{1}-1},\nonumber\\
&&\widetilde{Z}_{i},\widetilde{S}_{i},\widetilde{X}_{1,i},\widetilde{X}_{2,i},
\widetilde{Z}_{i-d_{1}},\widetilde{S}_{i-d_{1}},\widetilde{X}_{1,i-d_{1}},\widetilde{X}_{2,i-d_{1}})\nonumber\\
&&\geq \sum_{\tilde{s}_{1}=1}^{k}H(W_{1,i,1,\tilde{s}_{1}},W_{2,i,1,\tilde{s}_{1}}|W_{1,i,1,1},W_{2,i,1,1},...,W_{1,i,1,\tilde{s}_{1}-1},W_{2,i,1,\tilde{s}_{1}-1},\nonumber\\
&&\widetilde{Z}_{i},\widetilde{S}_{i},\widetilde{X}_{1,i},\widetilde{X}_{2,i},
\widetilde{Z}_{i-d_{1}},\widetilde{S}_{i-d_{1}},\widetilde{X}_{1,i-d_{1}},\widetilde{X}_{2,i-d_{1}},\tilde{S}_{1}=\tilde{s}_{1})\nonumber\\
&&\stackrel{(i)}=\sum_{\tilde{s}_{1}=1}^{k}H(W_{1,i,1,\tilde{s}_{1}},W_{2,i,1,\tilde{s}_{1}}|
\widetilde{Z}^{N_{\tilde{s}_{1}}}_{i},\widetilde{S}^{N_{\tilde{s}_{1}}}_{i},\widetilde{X}^{N_{\tilde{s}_{1}}}_{1,i},\widetilde{X}^{N_{\tilde{s}_{1}}}_{2,i},
\widetilde{Z}^{N_{\tilde{s}_{1}}}_{i-d_{1}},\widetilde{S}^{N_{\tilde{s}_{1}}}_{i-d_{1}},\widetilde{X}^{N_{\tilde{s}_{1}}}_{1,i-d_{1}},
\widetilde{X}^{N_{\tilde{s}_{1}}}_{2,i-d_{1}},\tilde{S}_{1}=\tilde{s}_{1})\nonumber\\
&&\geq \sum_{\tilde{s}_{1}=1}^{k}H(W_{1,i,1,\tilde{s}_{1}},W_{2,i,1,\tilde{s}_{1}}|
\widetilde{Z}^{N_{\tilde{s}_{1}}}_{i},\widetilde{S}^{N_{\tilde{s}_{1}}}_{i},\widetilde{X}^{N_{\tilde{s}_{1}}}_{1,i},\widetilde{X}^{N_{\tilde{s}_{1}}}_{2,i},
\widetilde{Z}^{N_{\tilde{s}_{1}}}_{i-d_{1}},\widetilde{S}^{N_{\tilde{s}_{1}}}_{i-d_{1}},\widetilde{X}^{N_{\tilde{s}_{1}}}_{1,i-d_{1}},
\widetilde{X}^{N_{\tilde{s}_{1}}}_{2,i-d_{1}},\tilde{S}_{1}=\tilde{s}_{1},\nonumber\\
&&W_{1,i,1,\tilde{s}_{1}}\oplus K_{1,i,\tilde{s}_{1}}^{*},
W_{2,i,1,\tilde{s}_{1}}\oplus K_{2,i,\tilde{s}_{1}}^{*})\nonumber\\
&&\stackrel{(j)}=\sum_{\tilde{s}_{1}=1}^{k}H(W_{1,i,1,\tilde{s}_{1}},W_{2,i,1,\tilde{s}_{1}}|
\widetilde{Z}^{N_{\tilde{s}_{1}}}_{i-d_{1}},\widetilde{S}^{N_{\tilde{s}_{1}}}_{i-d_{1}},\widetilde{X}^{N_{\tilde{s}_{1}}}_{1,i-d_{1}},
\widetilde{X}^{N_{\tilde{s}_{1}}}_{2,i-d_{1}},\tilde{S}_{1}=\tilde{s}_{1},W_{1,i,1,\tilde{s}_{1}}\oplus K_{1,i,\tilde{s}_{1}}^{*},
W_{2,i,1,\tilde{s}_{1}}\oplus K_{2,i,\tilde{s}_{1}}^{*})\nonumber\\
&&=\sum_{\tilde{s}_{1}=1}^{k}H(K_{1,i,\tilde{s}_{1}}^{*},K_{2,i,\tilde{s}_{1}}^{*}|
\widetilde{Z}^{N_{\tilde{s}_{1}}}_{i-d_{1}},\widetilde{S}^{N_{\tilde{s}_{1}}}_{i-d_{1}},\widetilde{X}^{N_{\tilde{s}_{1}}}_{1,i-d_{1}},
\widetilde{X}^{N_{\tilde{s}_{1}}}_{2,i-d_{1}},\tilde{S}_{1}=\tilde{s}_{1},W_{1,i,1,\tilde{s}_{1}}\oplus K_{1,i,\tilde{s}_{1}}^{*},
W_{2,i,1,\tilde{s}_{1}}\oplus K_{2,i,\tilde{s}_{1}}^{*})\nonumber\\
&&\stackrel{(k)}=\sum_{\tilde{s}_{1}=1}^{k}H(K_{1,i,\tilde{s}_{1}}^{*},K_{2,i,\tilde{s}_{1}}^{*}|
\widetilde{Z}^{N_{\tilde{s}_{1}}}_{i-d_{1}},\widetilde{S}^{N_{\tilde{s}_{1}}}_{i-d_{1}},\widetilde{X}^{N_{\tilde{s}_{1}}}_{1,i-d_{1}},
\widetilde{X}^{N_{\tilde{s}_{1}}}_{2,i-d_{1}},\tilde{S}_{1}=\tilde{s}_{1})\nonumber\\
&&\stackrel{(l)}\geq \sum_{\tilde{s}_{1}=1}^{k}(\log\frac{1-\epsilon_{1}}{1+\delta}+N_{\tilde{s}_{1}}(1-\epsilon_{2})
H(Y|X_{1},X_{2},S,Z,\tilde{S}_{1}=\tilde{s}_{1})),
\end{eqnarray}
where (h) follows from the fact that given the random vectors of the $i$-th block and the $i-d_{1}$-th block, the messages
$W_{1,i,1}$ and $W_{2,i,1}$ are independent of the random vectors of the other blocks, and from the fact that
$H(W_{1,i,0},W_{2,i,0}|\widetilde{X}_{1,i},\widetilde{X}_{2,i})=0$, (i) follows from the fact that given the $\tilde{s}_{1}$-th part of
the random vectors $\widetilde{Z}_{i}$, $\widetilde{S}_{i}$, $\widetilde{X}_{1,i}$, $\widetilde{X}_{2,i}$,
$\widetilde{Z}_{i-d_{1}}$, $\widetilde{S}_{i-d_{1}}$, $\widetilde{X}_{1,i-d_{1}}$, $\widetilde{X}_{2,i-d_{1}}$,
the messages $W_{1,i,1,\tilde{s}_{1}}$ and $W_{2,i,1,\tilde{s}_{1}}$ are independent of
the other parts of these random vectors, (j) follows from the Markov chain
$(\widetilde{Z}^{N_{\tilde{s}_{1}}}_{i},\widetilde{S}^{N_{\tilde{s}_{1}}}_{i},\widetilde{X}^{N_{\tilde{s}_{1}}}_{1,i},\widetilde{X}^{N_{\tilde{s}_{1}}}_{2,i})
\rightarrow (\widetilde{Z}^{N_{\tilde{s}_{1}}}_{i-d_{1}},\widetilde{S}^{N_{\tilde{s}_{1}}}_{i-d_{1}},\widetilde{X}^{N_{\tilde{s}_{1}}}_{1,i-d_{1}},\\
\widetilde{X}^{N_{\tilde{s}_{1}}}_{2,i-d_{1}},\tilde{S}_{1}=\tilde{s}_{1},W_{1,i,1,\tilde{s}_{1}}\oplus K_{1,i,\tilde{s}_{1}}^{*},
W_{2,i,1,\tilde{s}_{1}}\oplus K_{2,i,\tilde{s}_{1}}^{*})\rightarrow (W_{1,i,1,\tilde{s}_{1}},W_{2,i,1,\tilde{s}_{1}})$,
(k) follows from the Markov chain $(W_{1,i,1,\tilde{s}_{1}}\oplus K_{1,i,\tilde{s}_{1}}^{*},
W_{2,i,1,\tilde{s}_{1}}\oplus K_{2,i,\tilde{s}_{1}}^{*})\rightarrow (\widetilde{Z}^{N_{\tilde{s}_{1}}}_{i-d_{1}},\widetilde{S}^{N_{\tilde{s}_{1}}}_{i-d_{1}},\widetilde{X}^{N_{\tilde{s}_{1}}}_{1,i-d_{1}},
\widetilde{X}^{N_{\tilde{s}_{1}}}_{2,i-d_{1}},\tilde{S}_{1}=\tilde{s}_{1})\rightarrow (K_{1,i,\tilde{s}_{1}}^{*},K_{2,i,\tilde{s}_{1}}^{*})$,
and (l) follows from (\ref{bgod-7}).

Substituting (\ref{bgod-9}) and (\ref{bgod-11}) into (\ref{bgod-8}), the equivocation $\Delta$ is lower bounded by
\begin{eqnarray}\label{bgod-12}
&&\Delta\geq\frac{1}{nN}\sum_{i=1}^{n}\sum_{\tilde{s}_{1}=1}^{k}(N_{\tilde{s}_{1}}(R_{10}(\tilde{s}_{1})+R_{11}(\tilde{s}_{1})+R^{*}_{1}(\tilde{s}_{1}))
+N_{\tilde{s}_{1}}(R_{20}(\tilde{s}_{1})+R_{21}(\tilde{s}_{1})+R^{*}_{2}(\tilde{s}_{1}))\nonumber\\
&&-N_{\tilde{s}_{1}}\epsilon_{3}-N_{\tilde{s}_{1}}I(X_{1},X_{2};Z|S,\tilde{S}_{1}=\tilde{s}_{1}))+\frac{1}{nN}\sum_{i=2d_{1}+1}^{n}
\sum_{\tilde{s}_{1}=1}^{k}(\log\frac{1-\epsilon_{1}}{1+\delta}+N_{\tilde{s}_{1}}(1-\epsilon_{2})
H(Y|X_{1},X_{2},S,Z,\tilde{S}_{1}=\tilde{s}_{1}))\nonumber\\
&&\stackrel{(m)}=R_{10}+R_{11}+R^{*}_{1}+R_{20}+R_{21}+R^{*}_{2}-\epsilon_{3}-I(X_{1},X_{2};Z|S,\tilde{S}_{1})\nonumber\\
&&-\frac{\epsilon_{1}\sum_{\tilde{s}_{1}=1}^{k}(R_{10}(\tilde{s}_{1})+R_{11}(\tilde{s}_{1})+R^{*}_{1}(\tilde{s}_{1})+
R_{20}(\tilde{s}_{1})+R_{21}(\tilde{s}_{1})+R^{*}_{2}(\tilde{s}_{1})-\epsilon_{3}-I(X_{1},X_{2};Z|S,\tilde{S}_{1}=\tilde{s}_{1}))}{N}\nonumber\\
&&+\frac{n-2d_{1}}{nN}k\log\frac{1-\epsilon_{1}}{1+\delta}+\frac{n-2d_{1}}{n}(1-\epsilon_{2})H(Y|X_{1},X_{2},S,Z,\tilde{S}_{1})\nonumber\\
&&-\frac{n-2d_{1}}{nN}\epsilon_{1}(1-\epsilon_{2})\sum_{\tilde{s}_{1}=1}^{k}H(Y|X_{1},X_{2},S,Z,\tilde{S}_{1}=\tilde{s}_{1}),
\end{eqnarray}
where (m) follows from the definitions in (\ref{c3.q2}) and (\ref{c3.q2.koudai}).
The lower bound (\ref{bgod-12}) implies that if we choose
sufficiently large $N$ and $n$, we have
\begin{eqnarray}\label{bgod-13}
&&\Delta\geq R_{10}+R_{11}+R^{*}_{1}+R_{20}+R_{21}+R^{*}_{2}-I(X_{1},X_{2};Z|S,\tilde{S}_{1})+H(Y|X_{1},X_{2},S,Z,\tilde{S}_{1})-\epsilon\nonumber\\
&&\stackrel{(n)}=R_{10}+R_{11}+R^{*}_{1}+R_{20}+R_{21}+R^{*}_{2}
-I(X_{1},X_{2};Z|S,\tilde{S}_{1},\tilde{S}_{2})+H(Y|X_{1},X_{2},S,Z,\tilde{S}_{1},\tilde{S}_{2})-\epsilon,\nonumber\\
\end{eqnarray}
where (n) follows from the Markov chains $\tilde{S}_{2}\rightarrow (S,\tilde{S}_{1})\rightarrow Z$,
$\tilde{S}_{2}\rightarrow (X_{1},X_{2},S,\tilde{S}_{1})\rightarrow Z$ and
$\tilde{S}_{2}\rightarrow (X_{1},X_{2},S,\tilde{S}_{1},Z)\rightarrow Y$.
From (\ref{bgod-13}), we see that $\Delta\geq R_{10}+R_{11}+R_{20}+R_{21}-\epsilon$ is achieved if
\begin{eqnarray}\label{bgod-14}
&&R^{*}_{1}+R^{*}_{2}\geq I(X_{1},X_{2};Z|S,\tilde{S}_{1},\tilde{S}_{2})-H(Y|X_{1},X_{2},S,Z,\tilde{S}_{1},\tilde{S}_{2}).
\end{eqnarray}
Finally, combining (\ref{bgod-1.x}), (\ref{bgod-2.x}), (\ref{bgod-3}) and (\ref{bgod-10.x}) with (\ref{bgod-14}), and
applying Fourier-Motzkin elimination (see, e.g., \cite{lall}) to eliminate $R^{*}_{1}$, $R^{*}_{2}$, $R_{10}$,
$R_{11}$, $R_{20}$ and $R_{21}$ (here note that $R_{1}=R_{10}+R_{11}$ and $R_{2}=R_{20}+R_{21}$),
Theorem \ref{T3} is obtained.

The proof of Theorem \ref{T3} is completed.

\section{Proof of Theorem \ref{T4}\label{appen5}}

The bounds on the cardinality of the auxiliary random variables $U$, $V_{1}$ and $V_{2}$ are directly from the support lemma \cite[pp. 633-634]{network},
and thus we omit the proof here.
Theorem \ref{T4} is proved by showing that for any achievable secrecy rate pair $(R_{1}, R_{2})$, the inequalities
$R_{1}\leq I(V_{1};Y|U,S,\tilde{S}_{1},\tilde{S}_{2})$,
$R_{2}\leq I(V_{2};Y|U,S,\tilde{S}_{1},\tilde{S}_{2})$ and
$R_{1}+R_{2}\leq \min\{H(Y|U,S,\tilde{S}_{1},\tilde{S}_{2},Z), I(V_{1},V_{2};Y|U,S,\tilde{S}_{1},\tilde{S}_{2})\}$ hold. Here
the random variables $U$, $V_{1}$, $V_{2}$, $S$, $\tilde{S}_{1}$, $\tilde{S}_{2}$, $Y$ and $Z$ are denoted by
\begin{eqnarray}\label{jmds1}
&&U\triangleq (Y^{J-1}, Z_{J+1}^{N}, S^{N}, J),\,\, V_{1}\triangleq (U, W_{1}), \,\, V_{2}\triangleq (U, W_{2}),\,\,Y\triangleq Y_{J},\,\,Z\triangleq Z_{J}\nonumber\\
&&S\triangleq S_{J},\,\, \tilde{S}_{1}\triangleq S_{J-d_{1}}, \,\, \tilde{S}_{2}\triangleq S_{J-d_{2}},
\end{eqnarray}
where the uniformly distributed random variable $J$ takes values in the set $\{1, 2, ,...,N\}$, and it is independent of
$Y^{N}$, $Z^{N}$, $W_{1}$, $W_{2}$ and $S^{N}$.

\textbf{Proof of $R_{1}\leq I(V_{1};Y|U,S,\tilde{S}_{1},\tilde{S}_{2})$:}

First, note that the joint secrecy ensures the individual secrecy, and thus we have
\begin{eqnarray}\label{jmds2}
R_{1}-\epsilon&\leq&\frac{1}{N}H(W_{1}|Z^{N},S^{N})\nonumber\\
&=&\frac{1}{N}(H(W_{1})-I(W_{1};Z^{N},S^{N}))\stackrel{(a)}=\frac{1}{N}(H(W_{1})-I(W_{1};Z^{N}|S^{N}))\nonumber\\
&\stackrel{(b)}=&\frac{1}{N}(H(W_{1}|S^{N})-H(W_{1}|Y^{N},S^{N})+H(W_{1}|Y^{N},S^{N})-I(W_{1};Z^{N}|S^{N}))\nonumber\\
&\stackrel{(c)}\leq&\frac{1}{N}(I(W_{1};Y^{N}|S^{N})+\delta(P_{e})-I(W_{1};Z^{N}|S^{N}))\nonumber\\
&=&\frac{1}{N}\sum_{i=1}^{N}(I(W_{1};Y_{i}|Y^{i-1},S^{N})-I(W_{1};Z_{i}|Z_{i+1}^{N},S^{N}))+\frac{\delta(P_{e})}{N}\nonumber\\
&=&\frac{1}{N}\sum_{i=1}^{N}(I(W_{1};Y_{i}|Y^{i-1},S^{N},Z_{i+1}^{N})-I(W_{1};Z_{i}|Y^{i-1},S^{N},Z_{i+1}^{N})
+I(Y_{i};Z_{i+1}^{N}|Y^{i-1},S^{N})\nonumber\\
&&-I(Z_{i};Y^{i-1}|Z_{i+1}^{N},S^{N})
-I(Y_{i};Z_{i+1}^{N}|W_{1},Y^{i-1},S^{N})+I(Z_{i};Y^{i-1}|W_{1},Z_{i+1}^{N},S^{N}))+\frac{\delta(P_{e})}{N}\nonumber\\
&\stackrel{(d)}=&\frac{1}{N}\sum_{i=1}^{N}(I(W_{1};Y_{i}|Y^{i-1},S^{N},Z_{i+1}^{N})-I(W_{1};Z_{i}|Y^{i-1},S^{N},Z_{i+1}^{N}))+\frac{\delta(P_{e})}{N}\nonumber\\
&\stackrel{(e)}=&\frac{1}{N}\sum_{i=1}^{N}(I(W_{1};Y_{i}|Y^{i-1},S^{N},Z_{i+1}^{N},S_{i},S_{i-d_{1}},S_{i-d_{2}})\nonumber\\
&&-I(W_{1};Z_{i}|Y^{i-1},S^{N},Z_{i+1}^{N},S_{i},S_{i-d_{1}},S_{i-d_{2}}))+\frac{\delta(P_{e})}{N}\nonumber\\
&\stackrel{(f)}=&\frac{1}{N}\sum_{i=1}^{N}(I(V_{1,i};Y_{i}|U_{i},S_{i},S_{i-d_{1}},S_{i-d_{2}})
-I(V_{1,i};Z_{i}|U_{i},S_{i},S_{i-d_{1}},S_{i-d_{2}}))+\frac{\delta(P_{e})}{N}\nonumber\\
&\leq&\frac{1}{N}\sum_{i=1}^{N}I(V_{1,i};Y_{i}|U_{i},S_{i},S_{i-d_{1}},S_{i-d_{2}})+\frac{\delta(P_{e})}{N}\nonumber\\
&\stackrel{(g)}=&\frac{1}{N}\sum_{i=1}^{N}I(V_{1,i};Y_{i}|U_{i},S_{i},S_{i-d_{1}},S_{i-d_{2}},J=i)+\frac{\delta(P_{e})}{N}\nonumber\\
&=&I(V_{1,J};Y_{J}|U_{J},S_{J},S_{J-d_{1}},S_{J-d_{2}},J)+\frac{\delta(P_{e})}{N}\nonumber\\
&\stackrel{(h)}=&I(V_{1};Y|U,S,\tilde{S}_{1},\tilde{S}_{2})+\frac{\delta(P_{e})}{N}\nonumber\\
&\stackrel{(i)}\leq&I(V_{1};Y|U,S,\tilde{S}_{1},\tilde{S}_{2})+\frac{\delta(\epsilon)}{N},
\end{eqnarray}
where (a) and (b) are deduced from $W_{1}$ is independent of $S^{N}$, (c) is deduced from Fano's inequality,
(d) is deduced from Csisz$\acute{a}$r's equality \cite{CK}, i.e.,
\begin{eqnarray}\label{jmds3.xxs}
&&I(Y_{i};Z_{i+1}^{N}|Y^{i-1},S^{N})=I(Z_{i};Y^{i-1}|Z_{i+1}^{N},S^{N}),
\end{eqnarray}
\begin{eqnarray}\label{jmds3.xxs1}
&&I(Y_{i};Z_{i+1}^{N}|W_{1},Y^{i-1},S^{N})=I(Z_{i};Y^{i-1}|W_{1},Z_{i+1}^{N},S^{N}),
\end{eqnarray}
(e) is deduced from $S_{i}$, $S_{i-d_{1}}$ and $S_{i-d_{2}}$ are included in $S^{N}$,
hence we have $H(S_{i},S_{i-d_{1}},S_{i-d_{2}}|S^{N})=0$, and here
note that $S_{i-d_{1}}=const$ (or $S_{i-d_{2}}=const$) when $i\leq d_{1}$ (or $i\leq d_{2}$),
(f) is deduced from the definitions $U_{i}=(Y^{i-1},S^{N},Z_{i+1}^{N})$ and $V_{1,i}=(W_{1},Y^{i-1},S^{N},Z_{i+1}^{N})$,
(g) is deduced from $J$ is a uniformly distributed random variable  which takes values in the set $\{1,2,...,N\}$, and it is
independent of $Y^{N}$, $Z^{N}$, $W_{1}$, $W_{2}$ and $S^{N}$, (h) is from the definitions in (\ref{jmds1}), and (i) follows from the fact that
$\delta(P_{e})$ is a monotonic increasing function of $P_{e}$ and $P_{e}\leq \epsilon$.
Then, letting $\epsilon\rightarrow 0$, the bound $R_{1}\leq I(V_{1};Y|U,S,\tilde{S}_{1},\tilde{S}_{2})$ is obtained.

\textbf{Proof of $R_{1}\leq I(V_{1};Y|U,S,\tilde{S}_{1},\tilde{S}_{2})$:}

The proof of $R_{2}\leq I(V_{2};Y|U,S,\tilde{S}_{1},\tilde{S}_{2})$ is analogous to the proof of
$R_{1}\leq I(V_{1};Y|U,S,\tilde{S}_{1},\tilde{S}_{2})$, and thus we omit the proof here.

\textbf{Proof of $R_{1}+R_{2}\leq \min\{H(Y|U,S,\tilde{S}_{1},\tilde{S}_{2},Z), I(V_{1},V_{2};Y|U,S,\tilde{S}_{1},\tilde{S}_{2})\}$:}

From (\ref{e202}), we know that
\begin{eqnarray}\label{jmds-xss1}
R_{1}+R_{2}-\epsilon&\leq&\frac{1}{N}H(W_{1},W_{2}|Z^{N},S^{N})\nonumber\\
&=&\frac{1}{N}(H(W_{1},W_{2})-I(W_{1},W_{2};Z^{N},S^{N}))\stackrel{(a)}=\frac{1}{N}(H(W_{1},W_{2})-I(W_{1},W_{2};Z^{N}|S^{N}))\nonumber\\
&\stackrel{(b)}=&\frac{1}{N}(H(W_{1},W_{2}|S^{N})-H(W_{1},W_{2}|Y^{N},S^{N})+H(W_{1},W_{2}|Y^{N},S^{N})-I(W_{1},W_{2};Z^{N}|S^{N}))\nonumber\\
&\stackrel{(c)}\leq&\frac{1}{N}(I(W_{1},W_{2};Y^{N}|S^{N})+\delta(P_{e})-I(W_{1},W_{2};Z^{N}|S^{N}))\nonumber\\
&=&\frac{1}{N}\sum_{i=1}^{N}(I(W_{1},W_{2};Y_{i}|Y^{i-1},S^{N})-I(W_{1},W_{2};Z_{i}|Z_{i+1}^{N},S^{N}))+\frac{\delta(P_{e})}{N}\nonumber\\
&=&\frac{1}{N}\sum_{i=1}^{N}(I(W_{1},W_{2};Y_{i}|Y^{i-1},S^{N},Z_{i+1}^{N})-I(W_{1},W_{2};Z_{i}|Y^{i-1},S^{N},Z_{i+1}^{N})\nonumber\\
&&+I(Y_{i};Z_{i+1}^{N}|Y^{i-1},S^{N})
-I(Z_{i};Y^{i-1}|Z_{i+1}^{N},S^{N})\nonumber\\
&&-I(Y_{i};Z_{i+1}^{N}|W_{1},W_{2},Y^{i-1},S^{N})+I(Z_{i};Y^{i-1}|W_{1},W_{2},Z_{i+1}^{N},S^{N}))+\frac{\delta(P_{e})}{N}\nonumber\\
&\stackrel{(d)}=&\frac{1}{N}\sum_{i=1}^{N}(I(W_{1},W_{2};Y_{i}|Y^{i-1},S^{N},Z_{i+1}^{N})-I(W_{1},W_{2};Z_{i}|Y^{i-1},S^{N},Z_{i+1}^{N}))+\frac{\delta(P_{e})}{N}\nonumber\\
&\stackrel{(e)}=&\frac{1}{N}\sum_{i=1}^{N}(I(W_{1},W_{2};Y_{i}|Y^{i-1},S^{N},Z_{i+1}^{N},S_{i},S_{i-d_{1}},S_{i-d_{2}})\nonumber\\
&&-I(W_{1},W_{2};Z_{i}|Y^{i-1},S^{N},Z_{i+1}^{N},S_{i},S_{i-d_{1}},S_{i-d_{2}}))+\frac{\delta(P_{e})}{N}\nonumber\\
&\stackrel{(f)}=&\frac{1}{N}\sum_{i=1}^{N}(I(V_{1,i},V_{2,i};Y_{i}|U_{i},S_{i},S_{i-d_{1}},S_{i-d_{2}})
-I(V_{1,i},V_{2,i};Z_{i}|U_{i},S_{i},S_{i-d_{1}},S_{i-d_{2}}))+\frac{\delta(P_{e})}{N}\nonumber\\
&\leq&\frac{1}{N}\sum_{i=1}^{N}I(V_{1,i},V_{2,i};Y_{i}|U_{i},S_{i},S_{i-d_{1}},S_{i-d_{2}})+\frac{\delta(P_{e})}{N}\nonumber\\
&\stackrel{(g)}=&I(V_{1,J},V_{2,J};Y_{J}|U_{J},S_{J},S_{J-d_{1}},S_{J-d_{2}},J)+\frac{\delta(P_{e})}{N}\nonumber\\
&\stackrel{(h)}=&I(V_{1},V_{2};Y|U,S,\tilde{S}_{1},\tilde{S}_{2})+\frac{\delta(P_{e})}{N}\nonumber\\
&\stackrel{(i)}\leq&I(V_{1},V_{2};Y|U,S,\tilde{S}_{1},\tilde{S}_{2})+\frac{\delta(\epsilon)}{N},
\end{eqnarray}
where (a) and (b) are deduced from $S^{N}$ is independent of $W_{1}$ and $W_{2}$, (c) is deduced from Fano's inequality,
(d) is deduced from Csisz$\acute{a}$r's equality \cite{CK}, i.e.,
\begin{eqnarray}\label{jmds3.xxs333}
&&I(Y_{i};Z_{i+1}^{N}|Y^{i-1},S^{N})=I(Z_{i};Y^{i-1}|Z_{i+1}^{N},S^{N}),
\end{eqnarray}
\begin{eqnarray}\label{jmds3.xxs1345}
&&I(Y_{i};Z_{i+1}^{N}|W_{1},W_{2},Y^{i-1},S^{N})=I(Z_{i};Y^{i-1}|W_{1},W_{2},Z_{i+1}^{N},S^{N}),
\end{eqnarray}
(e) is deduced from $S_{i}$, $S_{i-d_{1}}$ and $S_{i-d_{2}}$ are included in $S^{N}$,
(f) is deduced from the definitions $U_{i}=(Y^{i-1},S^{N},Z_{i+1}^{N})$, $V_{1,i}=(W_{1},Y^{i-1},S^{N},Z_{i+1}^{N})$ and $V_{2,i}=(W_{2},Y^{i-1},S^{N},Z_{i+1}^{N})$,
(g) is deduced from $J$ is a uniformly distributed random variable which takes values in $\{1,2,...,N\}$, and it is
independent of $Y^{N}$, $Z^{N}$, $W_{1}$, $W_{2}$ and $S^{N}$, (h) is from the definitions in (\ref{jmds1}), and (i) follows from the fact that
$\delta(P_{e})$ is a monotonic increasing function of $P_{e}$ and $P_{e}\leq \epsilon$.
Then, letting $\epsilon\rightarrow 0$, the bound
$R_{1}+R_{2}\leq I(V_{1},V_{2};Y|S,\tilde{S}_{1},\tilde{S}_{2},U)$ is obtained.

Moreover, note that
\begin{eqnarray}\label{jmds-xss1.tank3}
R_{1}+R_{2}-\epsilon&\leq&\frac{1}{N}H(W_{1},W_{2}|Z^{N},S^{N})\nonumber\\
&=&\frac{1}{N}(H(W_{1},W_{2}|Z^{N},S^{N})-H(W_{1},W_{2}|Z^{N},S^{N},Y^{N})+H(W_{1},W_{2}|Z^{N},S^{N},Y^{N}))\nonumber\\
&=&\frac{1}{N}(I(W_{1},W_{2};Y^{N}|Z^{N},S^{N})+H(W_{1},W_{2}|Z^{N},S^{N},Y^{N}))\nonumber\\
&\stackrel{(1)}\leq&\frac{1}{N}H(Y^{N}|Z^{N},S^{N})+\frac{\delta(P_{e})}{N}\nonumber\\
&=&\frac{1}{N}\sum_{i=1}^{N}H(Y_{i}|Y^{i-1},Z_{i+1}^{N},Z_{i},Z^{i-1},S^{N})+\frac{\delta(P_{e})}{N}\nonumber\\
&\stackrel{(2)}=&\frac{1}{N}\sum_{i=1}^{N}H(Y_{i}|Y^{i-1},Z_{i+1}^{N},Z_{i},Z^{i-1},S^{N},S_{i},S_{i-d_{1}},S_{i-d_{2}})+\frac{\delta(P_{e})}{N}\nonumber\\
&\stackrel{(3)}\leq&\frac{1}{N}\sum_{i=1}^{N}H(Y_{i}|Y^{i-1},Z_{i+1}^{N},Z_{i},S^{N},S_{i},S_{i-d_{1}},S_{i-d_{2}})+\frac{\delta(\epsilon)}{N}\nonumber\\
&\stackrel{(4)}=&\frac{1}{N}\sum_{i=1}^{N}H(Y_{i}|U_{i},Z_{i},S_{i},S_{i-d_{1}},S_{i-d_{2}},J=i)+\frac{\delta(\epsilon)}{N}\nonumber\\
&\stackrel{(5)}=&H(Y_{J}|U_{J},Z_{J},S_{J},S_{J-d_{1}},S_{J-d_{2}},J)+\frac{\delta(\epsilon)}{N}\nonumber\\
&\stackrel{(6)}=&H(Y|U,Z,S,\tilde{S}_{1},\tilde{S}_{2})+\frac{\delta(\epsilon)}{N},
\end{eqnarray}
where (1) is deduced from Fano's inequality, (2) is deduced from $S_{i}$, $S_{i-d_{1}}$ and $S_{i-d_{2}}$ are included in $S^{N}$,
hence we have $H(S_{i},S_{i-d_{1}},S_{i-d_{2}}|S^{N})=0$, and here
note that $S_{i-d_{1}}=const$ (or $S_{i-d_{2}}=const$) when $i\leq d_{1}$ (or $i\leq d_{2}$), (3) is deduced from
$\delta(P_{e})$ is a monotonic increasing function of $P_{e}$ and $P_{e}\leq \epsilon$, (4) and (5) are deduced from
the definitions $U_{i}=(Y^{i-1},S^{N},Z_{i+1}^{N})$ and $J$ is a uniformly distributed random variable  which takes values in the set $\{1,2,...,N\}$, and it is
independent of $Y^{N}$, $Z^{N}$, $W_{1}$, $W_{2}$ and $S^{N}$, and (6) is deduced from the definitions in (\ref{jmds1}).
Letting $\epsilon\rightarrow 0$, the bound $R_{1}+R_{2}\leq H(Y|U,S,\tilde{S}_{1},\tilde{S}_{2},Z)$ is obtained.
Thus the proof of $R_{1}+R_{2}\leq \min\{H(Y|U,S,\tilde{S}_{1},\tilde{S}_{2},Z), I(V_{1},V_{2};Y|U,S,\tilde{S}_{1},\tilde{S}_{2})\}$ and the proof
of Theorem \ref{T4} are completed.

\section{Proof of Theorem 3}
\setcounter{equation}{0}

To prove Theorem 3, we first show that the region $\mathcal{R}^{*}$
\begin{eqnarray*}
&&\mathcal{R}^{*}=\{(R_{1}, R_{2}): 0\leq R_{1}\leq I(X_{1};Y|X_{2},S,\tilde{S}_{1},\tilde{S}_{2})-I(X_{1};Z|S,\tilde{S}_{1},\tilde{S}_{2}),\\
&&0\leq R_{2}\leq I(X_{2};Y|X_{1},S,\tilde{S}_{1},\tilde{S}_{2})-I(X_{2};Z|S,\tilde{S}_{1},\tilde{S}_{2}),\\
&&0\leq R_{1}+R_{2}\leq I(X_{1},X_{2};Y|S,\tilde{S}_{1},\tilde{S}_{2})-I(X_{1},X_{2};Z|S,\tilde{S}_{1},\tilde{S}_{2})\}
\end{eqnarray*}
is achievable. Then using a standard time sharing technique \cite[p.3438]{bash},
Theorem 3 is directly obtained. Now the remainder of this section is organized as follows. Some basic definitions used in the code construction
are introduced
in Subsection \ref{fm-1}, the encoding and decoding schemes are shown in Subsection \ref{fm-2}, and the equivocation analysis is given in
Subsection \ref{fm-3}.

\subsection{Basic definitions}\label{fm-1}

\begin{itemize}

\item Without loss of generality, denote the state alphabet $\mathcal{S}$ by $\mathcal{S}=\{1,2,...,k\}$, and note that
the steady state probability $\pi(l)>0$ for all $l\in \mathcal{S}$. In addition, denote $N_{\tilde{s}_{1}}$ ($1\leq \tilde{s}_{1}\leq k$) by
\begin{eqnarray}\label{c3.q2}
&&N_{\tilde{s}_{1}}=NP_{\tilde{S}_{1}}(\tilde{s}_{1})-\epsilon_{1},
\end{eqnarray}
and $N_{\tilde{s}_{1},\tilde{s}_{2}}$ ($1\leq \tilde{s}_{1},\tilde{s}_{2}\leq k$) by
\begin{eqnarray}\label{c3.q2.koudai}
&&N_{\tilde{s}_{1},\tilde{s}_{2}}=NP_{\tilde{S}_{1}\tilde{S}_{2}}(\tilde{s}_{1},\tilde{s}_{2})-\frac{\epsilon_{1}}{k},
\end{eqnarray}
where $\epsilon_{1}>0$ and $\epsilon_{1}\rightarrow 0$ as $N\rightarrow \infty$. Here note that from (\ref{c3.q2}) and (\ref{c3.q2.koudai}),
we have
\begin{eqnarray}\label{daqin-1}
&&\sum_{\tilde{s}_{2}=1}^{k}N_{\tilde{s}_{1},\tilde{s}_{2}}=N_{\tilde{s}_{1}}.
\end{eqnarray}

\item Let $W^{*}_{1}$ and $W^{*}_{2}$ be the dummy messages taking values in $\{1,2,...,2^{NR^{*}_{1}}\}$ and $\{1,2,...,2^{NR^{*}_{2}}\}$, respectively.
In addition, let $R_{1}(\tilde{s}_{1})$, $R_{2}(\tilde{s}_{1},\tilde{s}_{2})$, $R^{*}_{1}(\tilde{s}_{1})$
and $R^{*}_{2}(\tilde{s}_{1},\tilde{s}_{2})$ be the transmission rates $R_{1}$, $R_{2}$, $R^{*}_{1}$ and $R^{*}_{2}$
for given $\tilde{s}_{1}$ and $\tilde{s}_{2}$, respectively, and they satisfy
\begin{eqnarray}\label{c1.q1.rmb1}
&&\sum_{\tilde{s}_{1}=1}^{k}P_{\tilde{S}_{1}}(\tilde{s}_{1})R_{1}(\tilde{s}_{1})=R_{1},
\end{eqnarray}
\begin{eqnarray}\label{c1.q1}
&&\sum_{\tilde{s}_{1}=1}^{k}\sum_{\tilde{s}_{2}=1}^{k}P_{\tilde{S}_{1}\tilde{S}_{2}}(\tilde{s}_{1},\tilde{s}_{2})R_{2}(\tilde{s}_{1},\tilde{s}_{2})=R_{2},
\end{eqnarray}
\begin{eqnarray}\label{c1.q1.rmb11}
&&\sum_{\tilde{s}_{1}=1}^{k}P_{\tilde{S}_{1}}(\tilde{s}_{1})R^{*}_{1}(\tilde{s}_{1})=R^{*}_{1},
\end{eqnarray}
\begin{eqnarray}\label{c1.q11}
&&\sum_{\tilde{s}_{1}=1}^{k}\sum_{\tilde{s}_{2}=1}^{k}P_{\tilde{S}_{1}\tilde{S}_{2}}(\tilde{s}_{1},\tilde{s}_{2})R^{*}_{2}(\tilde{s}_{1},\tilde{s}_{2})
=R^{*}_{2}.
\end{eqnarray}

\item The messages $W_{1}$ and $W^{*}_{1}$ are respectively denoted by $W_{1}=(W_{1,1},...,W_{1,k})$ and $W^{*}_{1}=(W^{*}_{1,1},...,W^{*}_{1,k})$, where
the sub-messages $W_{1,\tilde{s}_{1}}$ and $W^{*}_{1,\tilde{s}_{1}}$ ($\tilde{s}_{1}\in\{1,2,...,k\}$) take values in the sets
$\mathcal{W}_{1,\tilde{s}_{1}}=\{1,2,...,2^{N_{\tilde{s}_{1}}R_{1}(\tilde{s}_{1})}\}$ and
$\mathcal{W}^{*}_{1,\tilde{s}_{1}}=\{1,2,...,2^{N_{\tilde{s}_{1}}R^{*}_{1}(\tilde{s}_{1})}\}$, respectively.
Similarly, denote $W_{2}$ and $W^{*}_{2}$ by $W_{2}=(W_{2,1,1},W_{2,1,2},...,W_{2,k,k})$ and
$W^{*}_{2}=(W^{*}_{2,1,1},W^{*}_{2,1,2},...,W^{*}_{2,k,k})$, respectively, and the sub-messages
$W_{2,\tilde{s}_{1},\tilde{s}_{2}}$ and $W^{*}_{2,\tilde{s}_{1},\tilde{s}_{2}}$ ($\tilde{s}_{1}\in\{1,2,...,k\}$ and $\tilde{s}_{2}\in\{1,2,...,k\}$) respectively
take values in the sets
$\mathcal{W}_{2,\tilde{s}_{1},\tilde{s}_{2}}=\{1,2,...,2^{N_{\tilde{s}_{1},\tilde{s}_{2}}R_{2}(\tilde{s}_{1},\tilde{s}_{2})}\}$ and
$\mathcal{W}^{*}_{2,\tilde{s}_{1},\tilde{s}_{2}}=\{1,2,...,2^{N_{\tilde{s}_{1},\tilde{s}_{2}}R^{*}_{2}(\tilde{s}_{1},\tilde{s}_{2})}\}$.
\end{itemize}

\subsection{Encoding and decoding schemes}\label{fm-2}

\subsubsection*{1). Codebooks construction}

Fix the probabilities $P_{X_{1}|\tilde{S}_{1}}(x_{1}|\tilde{s}_{1})$ and $P_{X_{2}|\tilde{S}_{1},\tilde{S}_{2}}(x_{2}|\tilde{s}_{1},\tilde{s}_{2})$, and then the
construction of the code-book is as follows.

\begin{itemize}

\item \textbf{Codebook construction of $X_{1}^{N}$}: Generating $k$ sub-codebooks $\mathcal{C}_{1}^{\tilde{s}_{1}}$ of $X_{1}^{N}$ for all
$\tilde{s}_{1}\in \mathcal{S}$.
In each sub-codebook $\mathcal{C}_{1}^{\tilde{s}_{1}}$, randomly generate $2^{N_{\tilde{s}_{1}}(R_{1}(\tilde{s}_{1})+R^{*}_{1}(\tilde{s}_{1}))}$
i.i.d. codewords
$x_{1}^{N_{\tilde{s}}}$ according to $P_{X_{1}|\tilde{S}_{1}}(x_{1}|\tilde{s}_{1})$, and index these codewords as $x_{1}^{N_{\tilde{s}_{1}}}(i)$, where
$1\leq i\leq 2^{N_{\tilde{s}_{1}}(R_{1}(\tilde{s}_{1})+R^{*}_{1}(\tilde{s}_{1}))}$.
Here for a fixed block length $N$, we define $L_{\tilde{s}_{1}}$ as the number of times during the $N$ symbols for which
the delayed state at the transmitter $1$ is $\tilde{S}_{1}=\tilde{s}_{1}$. Every time that
the transmitter $1$ receives a delayed state $\tilde{S}_{1}=\tilde{s}_{1}$,
he chooses the next symbol from the sub-codebook $\mathcal{C}_{1}^{\tilde{s}_{1}}$.
Since $L_{\tilde{s}_{1}}$ is not necessarily equal to
$N_{\tilde{s}_{1}}$, an error occurs if $L_{\tilde{s}_{1}}< N_{\tilde{s}_{1}}$, and the code is filled with zero if $L_{\tilde{s}_{1}}> N_{\tilde{s}_{1}}$.
Here note that the state process is stationary and ergodic, thus we have
\begin{eqnarray}\label{miyue0.xc}
&&\lim_{N\rightarrow \infty}\frac{L_{\tilde{s}_{1}}}{N}=Pr\{\tilde{S}_{1}=\tilde{s}_{1}\}.
\end{eqnarray}
Combining (\ref{miyue0.xc}) with (\ref{c3.q2}), we know that
\begin{eqnarray}\label{miyue0}
&&Pr\{L_{\tilde{s}_{1}}< N_{\tilde{s}_{1}}\}\rightarrow 0, \,\, \mbox{as}\,\, N\rightarrow \infty.
\end{eqnarray}

\item \textbf{Codebook construction of $X_{2}^{N}$}: Generating $k\times k$ sub-codebooks $\mathcal{C}_{2}^{\tilde{s}_{1},\tilde{s}_{2}}$ of $X_{2}^{N}$
for all $\tilde{s}_{1}\in \mathcal{S}$ and $\tilde{s}_{2}\in \mathcal{S}$.
In each sub-codebook $\mathcal{C}_{2}^{\tilde{s}_{1},\tilde{s}_{2}}$, randomly generate
$2^{N_{\tilde{s}_{1},\tilde{s}_{2}}(R_{2}(\tilde{s}_{1},\tilde{s}_{2})+R^{*}_{2}(\tilde{s}_{1},\tilde{s}_{2}))}$ i.i.d. codewords
$x_{2}^{N_{\tilde{s}_{1},\tilde{s}_{2}}}$ according to $P_{X_{2}|\tilde{S}_{1},\tilde{S}_{2}}(x_{2}|\tilde{s}_{1},\tilde{s}_{2})$.
Index the codewords of the sub-codebook $\mathcal{C}_{2}^{\tilde{s}_{1},\tilde{s}_{2}}$
as $x_{2}^{N_{\tilde{s}_{1},\tilde{s}_{2}}}(j)$, where
$1\leq j\leq 2^{N_{\tilde{s}_{1},\tilde{s}_{2}}(R_{2}(\tilde{s}_{1},\tilde{s}_{2})+R^{*}_{2}(\tilde{s}_{1},\tilde{s}_{2}))}$.
For a fixed block length $N$, we define $L_{\tilde{s}_{1},\tilde{s}_{2}}$ as the number of times during the $N$ symbols for which
the delayed state at the transmitter $2$ is $(\tilde{S}_{1},\tilde{S}_{2})=(\tilde{s}_{1},\tilde{s}_{2})$.
Every time that the transmitter $2$ receives the delayed state $(\tilde{S}_{1},\tilde{S}_{2})=(\tilde{s}_{1},\tilde{s}_{2})$,
he chooses the next symbol from the sub-codebook $\mathcal{C}_{2}^{\tilde{s}_{1},\tilde{s}_{2}}$.
Since $L_{\tilde{s}_{1},\tilde{s}_{2}}$ is not necessarily equal to
$N_{\tilde{s}_{1},\tilde{s}_{2}}$, an error occurs if $L_{\tilde{s}_{1},\tilde{s}_{2}}< N_{\tilde{s}_{1},\tilde{s}_{2}}$,
and the code is filled with zero if $L_{\tilde{s}_{1},\tilde{s}_{2}}> N_{\tilde{s}_{1},\tilde{s}_{2}}$.
Here note that
\begin{eqnarray}\label{miyue0.xc11}
&&\lim_{N\rightarrow \infty}\frac{L_{\tilde{s}_{1},\tilde{s}_{2}}}{N}=Pr\{\tilde{S}_{1}=\tilde{s}_{1},\tilde{S}_{2}=\tilde{s}_{2}\}.
\end{eqnarray}
Combining (\ref{miyue0.xc11}) with (\ref{c3.q2.koudai}), we know that
\begin{eqnarray}\label{miyue0.xcc}
&&Pr\{L_{\tilde{s}_{1},\tilde{s}_{2}}< N_{\tilde{s}_{1},\tilde{s}_{2}}\}\rightarrow 0, \,\, \mbox{as}\,\, N\rightarrow \infty.
\end{eqnarray}
\end{itemize}

\subsubsection*{2). Encoding scheme}

For the transmitter $1$, suppose that a message $w_{1}=(w_{1,1},...,w_{1,k})$ and a randomly generated dummy message $w_{1}^{*}=(w^{*}_{1,1},...,w^{*}_{1,k})$
are chosen to be transmitted.
In each
sub-codebook $\mathcal{C}_{1}^{\tilde{s}_{1}}$ ($1\leq \tilde{s}_{1}\leq k$), the transmitter $1$ chooses
$x_{1}^{N_{\tilde{s}_{1}}}(w_{1,\tilde{s}_{1}},w^{*}_{1,\tilde{s}_{1}})$ as the $\tilde{s}_{1}$-th component codeword of the transmitted $x_{1}^{N}$.
The transmitted codeword $x_{1}^{N}$ is obtained by multiplexing the different component codewords chosen in the different sub-codebooks.

Similarly, for the transmitter $2$, suppose that a message $w_{2}=(w_{2,1,1},w_{2,1,2},...,w_{2,k,k})$ and
a randomly generated dummy message $w_{2}^{*}=(w^{*}_{2,1,1},w^{*}_{2,1,2},...,w^{*}_{2,k,k})$
are chosen to be transmitted. In each
sub-codebook $\mathcal{C}_{2}^{\tilde{s}_{1},\tilde{s}_{2}}$ ($1\leq \tilde{s}_{1},\tilde{s}_{2}\leq k$), the transmitter $2$ chooses
$x_{2}^{N_{\tilde{s}_{1},\tilde{s}_{2}}}(w_{2,\tilde{s}_{1},\tilde{s}_{2}},w^{*}_{2,\tilde{s}_{1},\tilde{s}_{2}})$ as the
$(\tilde{s}_{1},\tilde{s}_{2})$-th component codeword of the transmitted $x_{2}^{N}$.
The transmitted codeword $x_{2}^{N}$ is obtained by multiplexing the different component codewords chosen in the different sub-codebooks.

\subsubsection*{3). Decoding scheme for the legitimate receiver}

Since the legitimate receiver knows the delayed feedback state $\tilde{S}_{1}$, he uses it to
demultiplex his received channel output $y^{N}$ and the state sequence $s^{N}$ into outputs with respect to the
sub-codebooks of the transmitters. From (\ref{daqin-1}), we know that for each $\tilde{s}_{1}$ ($1\leq \tilde{s}_{1}$),
$x_{2}^{N_{\tilde{s}_{1}}}(w_{2,\tilde{s}_{1}},w^{*}_{2,\tilde{s}_{1}})$ can be re-written as
$x_{2}^{N_{\tilde{s}_{1}}}(w_{2,\tilde{s}_{1}},w^{*}_{2,\tilde{s}_{1}})=(x_{2}^{N_{\tilde{s}_{1},1}}(w_{2,\tilde{s}_{1},1},w^{*}_{2,\tilde{s}_{1},1}),
x_{2}^{N_{\tilde{s}_{1},2}}(w_{2,\tilde{s}_{1},2},w^{*}_{2,\tilde{s}_{1},2})
,...,x_{2}^{N_{\tilde{s}_{1},k}}(w_{2,\tilde{s}_{1},k},w^{*}_{2,\tilde{s}_{1},k}))$, where
\begin{eqnarray}\label{daqindiguo1}
&&w_{2,\tilde{s}_{1}}=(w_{2,\tilde{s}_{1},1},w_{2,\tilde{s}_{1},2},...,w_{2,\tilde{s}_{1},k}),
\end{eqnarray}
and
\begin{eqnarray}\label{daqindiguo2}
&&w^{*}_{2,\tilde{s}_{1}}=(w^{*}_{2,\tilde{s}_{1},1},w^{*}_{2,\tilde{s}_{1},2},...,w^{*}_{2,\tilde{s}_{1},k}).
\end{eqnarray}

Once the legitimate receiver receives $y^{N_{\tilde{s}_{1}}}$ and $s^{N_{\tilde{s}_{1}}}$, he tries to find a unique
quadruple $(\hat{w}_{1,\tilde{s}_{1}},\hat{w}^{*}_{1,\tilde{s}_{1}},\hat{w}_{2,\tilde{s}_{1}},\hat{w}^{*}_{2,\tilde{s}_{1}})$
such that $(x_{1}^{N_{\tilde{s}_{1}}}(\hat{w}_{1,\tilde{s}_{1}},\hat{w}^{*}_{1,\tilde{s}_{1}}),
x_{2}^{N_{\tilde{s}_{1}}}(\hat{w}_{2,\tilde{s}_{1}},\hat{w}^{*}_{2,\tilde{s}_{1}}), y^{N_{\tilde{s}_{1}}}, s^{N_{\tilde{s}_{1}}})$
are strongly jointly typical sequences \cite{coverx}, i.e.,
\begin{eqnarray}\label{miyue4.1}
&&(x_{1}^{N_{\tilde{s}_{1}}}(\hat{w}_{1,\tilde{s}_{1}},\hat{w}^{*}_{1,\tilde{s}_{1}}),
x_{2}^{N_{\tilde{s}_{1}}}(\hat{w}_{2,\tilde{s}_{1}},\hat{w}^{*}_{2,\tilde{s}_{1}}),
y^{N_{\tilde{s}_{1}}}, s^{N_{\tilde{s}_{1}}})\in T^{N_{\tilde{s}_{1}}}_{X_{1},X_{2},S,Y|\tilde{S}_{1}=\tilde{s}_{1}}(\epsilon).
\end{eqnarray}
If there exists such a unique quadruple,
the legitimate receiver declares that
$(\hat{w}_{1,\tilde{s}_{1}},\hat{w}^{*}_{1,\tilde{s}_{1}},\hat{w}_{2,\tilde{s}_{1}},
\hat{w}^{*}_{2,\tilde{s}_{1}})$
is sent, otherwise he declares an error.
Using the Law of Large Numbers, it is easy to see that the ergodic state sequence $S^{N_{\tilde{s}_{1}}}$ satisfies
\begin{eqnarray}\label{hengda-1}
&&Pr\{S^{N_{\tilde{s}_{1}}}\in T^{N_{\tilde{s}_{1}}}_{S|\tilde{s}_{1}}(\epsilon)\}\rightarrow 1
\end{eqnarray}
as $N_{\tilde{s}_{1}}\rightarrow \infty$. Based on the AEP, the construction of the codebooks, (\ref{hengda-1}), (\ref{miyue0}),
(\ref{miyue0.xcc}) and (\ref{miyue4.1}), the legitimate receiver's decoding error probability
$Pr\{(\hat{w}_{1,\tilde{s}_{1}},\hat{w}^{*}_{1,\tilde{s}_{1}},\hat{w}_{2,\tilde{s}_{1}},\hat{w}^{*}_{2,\tilde{s}_{1}})\neq
(w_{1,\tilde{s}_{1}},w^{*}_{1,\tilde{s}_{1}},w_{2,\tilde{s}_{1}},w^{*}_{2,\tilde{s}_{1}})\}$ tends to $0$ if
$N_{\tilde{s}_{1}}\rightarrow \infty$, and
\begin{eqnarray}\label{hengda-2}
&&R_{1}(\tilde{s}_{1})+R^{*}_{1}(\tilde{s}_{1})\leq I(X_{1};Y|X_{2},S,\tilde{S}_{1}=\tilde{s}_{1}),
\end{eqnarray}
\begin{eqnarray}\label{hengda-3}
&&\tilde{R}_{2}(\tilde{s}_{1})+\tilde{R}^{*}_{2}(\tilde{s}_{1})\leq I(X_{2};Y|X_{1},S,\tilde{S}_{1}=\tilde{s}_{1}),
\end{eqnarray}
\begin{eqnarray}\label{hengda-4}
&&R_{1}(\tilde{s}_{1})+R^{*}_{1}(\tilde{s}_{1})+\tilde{R}_{2}(\tilde{s}_{1})+\tilde{R}^{*}_{2}(\tilde{s}_{1})
\leq I(X_{1},X_{2};Y|S,\tilde{S}_{1}=\tilde{s}_{1}).
\end{eqnarray}
Here note that $\tilde{R}_{2}(\tilde{s}_{1})$ and $\tilde{R}^{*}_{2}(\tilde{s}_{1})$ are the rates of the messages $w_{2,\tilde{s}_{1}}$ and
$w^{*}_{2,\tilde{s}_{1}}$, respectively, and they are given by
\begin{eqnarray}\label{hengda-5}
&&\tilde{R}_{2}(\tilde{s}_{1})=\frac{\sum_{\tilde{s}_{2}=1}^{k}N_{\tilde{s}_{1},\tilde{s}_{2}}R_{2}(\tilde{s}_{1},\tilde{s}_{2})}{N_{\tilde{s}_{1}}}\nonumber\\
&&\stackrel{(a)}=\sum_{\tilde{s}_{2}=1}^{k}P_{\tilde{S}_{2}|\tilde{S}_{1}}(\tilde{s}_{2}|\tilde{s}_{1})R_{2}(\tilde{s}_{1},\tilde{s}_{2})-
\sum_{\tilde{s}_{2}=1}^{k}R_{2}(\tilde{s}_{1},\tilde{s}_{2})
\frac{\epsilon_{1}(1-kP_{\tilde{S}_{2}|\tilde{S}_{1}}(\tilde{s}_{2}|\tilde{s}_{1}))}{kN_{\tilde{s}_{1}}},
\end{eqnarray}
\begin{eqnarray}\label{hengda-6}
&&\tilde{R}^{*}_{2}(\tilde{s}_{1})=\frac{\sum_{\tilde{s}_{2}=1}^{k}N_{\tilde{s}_{1},\tilde{s}_{2}}R^{*}_{2}(\tilde{s}_{1},\tilde{s}_{2})}{N_{\tilde{s}_{1}}}\nonumber\\
&&\stackrel{(b)}=\sum_{\tilde{s}_{2}=1}^{k}P_{\tilde{S}_{2}|\tilde{S}_{1}}(\tilde{s}_{2}|\tilde{s}_{1})R^{*}_{2}(\tilde{s}_{1},\tilde{s}_{2})-
\sum_{\tilde{s}_{2}=1}^{k}R^{*}_{2}(\tilde{s}_{1},\tilde{s}_{2})
\frac{\epsilon_{1}(1-kP_{\tilde{S}_{2}|\tilde{S}_{1}}(\tilde{s}_{2}|\tilde{s}_{1}))}{kN_{\tilde{s}_{1}}},
\end{eqnarray}
where (a) and (b) are from (\ref{c3.q2}) and (\ref{c3.q2.koudai}). Hence substituting (\ref{hengda-5}) and (\ref{hengda-6}) into
(\ref{hengda-3}) and (\ref{hengda-4}) and letting $N_{\tilde{s}_{1}}\rightarrow \infty$, we have
\begin{eqnarray}\label{hengda-3.11}
&&\sum_{\tilde{s}_{2}=1}^{k}P_{\tilde{S}_{2}|\tilde{S}_{1}}(\tilde{s}_{2}|\tilde{s}_{1})
(R_{2}(\tilde{s}_{1},\tilde{s}_{2})+R^{*}_{2}(\tilde{s}_{1},\tilde{s}_{2}))\leq I(X_{2};Y|X_{1},S,\tilde{S}_{1}=\tilde{s}_{1}),
\end{eqnarray}
and
\begin{eqnarray}\label{hengda-4.11}
&&R_{1}(\tilde{s}_{1})+R^{*}_{1}(\tilde{s}_{1})+\sum_{\tilde{s}_{2}=1}^{k}P_{\tilde{S}_{2}|\tilde{S}_{1}}(\tilde{s}_{2}|\tilde{s}_{1})
(R_{2}(\tilde{s}_{1},\tilde{s}_{2})+R^{*}_{2}(\tilde{s}_{1},\tilde{s}_{2}))
\leq I(X_{1},X_{2};Y|S,\tilde{S}_{1}=\tilde{s}_{1}).
\end{eqnarray}
Further combining (\ref{hengda-2}), (\ref{hengda-3.11}), (\ref{hengda-4.11}) with (\ref{c1.q1.rmb1}), (\ref{c1.q1}), (\ref{c1.q1.rmb11}) and (\ref{c1.q11}),
we have
\begin{eqnarray}\label{hengda-2.1}
&&R_{1}+R^{*}_{1}=\sum_{\tilde{s}_{1}=1}^{k}P_{\tilde{S}_{1}}(\tilde{s}_{1})(R_{1}(\tilde{s}_{1})+R^{*}_{1}(\tilde{s}_{1}))\nonumber\\
&&\leq \sum_{\tilde{s}_{1}=1}^{k}P_{\tilde{S}_{1}}(\tilde{s}_{1})I(X_{1};Y|X_{2},S,\tilde{S}_{1}=\tilde{s}_{1})\nonumber\\
&&=I(X_{1};Y|X_{2},S,\tilde{S}_{1})=H(Y|X_{2},S,\tilde{S}_{1})-H(Y|X_{1},X_{2},S,\tilde{S}_{1})\nonumber\\
&&\stackrel{(1)}=H(Y|X_{2},S,\tilde{S}_{1},\tilde{S}_{2})-H(Y|X_{1},X_{2},S,\tilde{S}_{1},\tilde{S}_{2})=I(X_{1};Y|X_{2},S,\tilde{S}_{1},\tilde{S}_{2}),
\end{eqnarray}
\begin{eqnarray}\label{hengda-3.1}
&&R_{2}+R^{*}_{2}=\sum_{\tilde{s}_{1}=1}^{k}P_{\tilde{S}_{1}}(\tilde{s}_{1})(\sum_{\tilde{s}_{2}=1}^{k}P_{\tilde{S}_{2}|\tilde{S}_{1}}(\tilde{s}_{2}|\tilde{s}_{1})
(R_{2}(\tilde{s}_{1},\tilde{s}_{2})+R^{*}_{2}(\tilde{s}_{1},\tilde{s}_{2})))\nonumber\\
&&\leq \sum_{\tilde{s}_{1}=1}^{k}P_{\tilde{S}_{1}}(\tilde{s}_{1})I(X_{2};Y|X_{1},S,\tilde{S}_{1}=\tilde{s}_{1})\nonumber\\
&&=I(X_{2};Y|X_{1},S,\tilde{S}_{1})=H(Y|X_{1},S,\tilde{S}_{1})-H(Y|X_{1},X_{2},S,\tilde{S}_{1})\nonumber\\
&&\stackrel{(2)}=H(Y|X_{1},S,\tilde{S}_{1},\tilde{S}_{2})-H(Y|X_{1},X_{2},S,\tilde{S}_{1},\tilde{S}_{2})=I(X_{2};Y|X_{1},S,\tilde{S}_{1},\tilde{S}_{2}),
\end{eqnarray}
\begin{eqnarray}\label{hengda-4.1}
&&R_{1}+R^{*}_{1}+R_{2}+R^{*}_{2}=\sum_{\tilde{s}_{1}=1}^{k}P_{\tilde{S}_{1}}(\tilde{s}_{1})(R_{1}(\tilde{s}_{1})+R^{*}_{1}(\tilde{s}_{1})+
\sum_{\tilde{s}_{2}=1}^{k}P_{\tilde{S}_{2}|\tilde{S}_{1}}(\tilde{s}_{2}|\tilde{s}_{1})
(R_{2}(\tilde{s}_{1},\tilde{s}_{2})+R^{*}_{2}(\tilde{s}_{1},\tilde{s}_{2})))\nonumber\\
&&\leq \sum_{\tilde{s}_{1}=1}^{k}P_{\tilde{S}_{1}}(\tilde{s}_{1})I(X_{1},X_{2};Y|S,\tilde{S}_{1}=\tilde{s}_{1})\nonumber\\
&&=I(X_{1},X_{2};Y|S,\tilde{S}_{1})=H(Y|S,\tilde{S}_{1})-H(Y|X_{1},X_{2},S,\tilde{S}_{1})\nonumber\\
&&\stackrel{(3)}=H(Y|S,\tilde{S}_{1},\tilde{S}_{2})-H(Y|X_{1},X_{2},S,\tilde{S}_{1},\tilde{S}_{2})=I(X_{1},X_{2};Y|S,\tilde{S}_{1},\tilde{S}_{2}),
\end{eqnarray}
where (1) is from the Markov chains $\tilde{S}_{2}\rightarrow (X_{2},S,\tilde{S}_{1})\rightarrow Y$ and
$\tilde{S}_{2}\rightarrow (X_{1},X_{2},S,\tilde{S}_{1})\rightarrow Y$, (2) is from the Markov chains
$\tilde{S}_{2}\rightarrow (X_{1},S,\tilde{S}_{1})\rightarrow Y$ and $\tilde{S}_{2}\rightarrow (X_{1},X_{2},S,\tilde{S}_{1})\rightarrow Y$, and (3)
is from $\tilde{S}_{2}\rightarrow (S,\tilde{S}_{1})\rightarrow Y$ and $\tilde{S}_{2}\rightarrow (X_{1},X_{2},S,\tilde{S}_{1})\rightarrow Y$.

\subsection{Equivocation analysis}\label{fm-3}

First, we give a lower bound
on $H(X_{1}^{N_{\tilde{s}_{1}}},X_{2}^{N_{\tilde{s}_{1}}}|W_{1,\tilde{s}_{1}},W_{2,\tilde{s}_{1}},
Z^{N_{\tilde{s}_{1}}},S^{N_{\tilde{s}_{1}}},\tilde{S}_{1}=\tilde{s}_{1})$, which will be used in the analysis of
the eavesdropper's equivocation about the transmitted messages $W_{1}$ and $W_{2}$. This conditional entropy can be lower bounded by Fano's inequality, which
needs the guarantee that given $z^{N}$, $s^{N}$, $w_{1}$ and $w_{2}$, the eavesdropper's decoding error probability of the dummy messages $w^{*}_{1}$ and $w^{*}_{2}$ tends to $0$. The eavesdropper's decoding scheme is described as follows.

Since the eavesdropper also knows the delayed feedback states $\tilde{S}_{1}$, he can use it to
demultiplex his received channel output $z^{N}$ and the state sequence $s^{N}$ into outputs with respect to the
sub-codebooks of the transmitters. Then, for each $\tilde{s}_{1}$ ($1\leq \tilde{s}_{1}\leq k$),
the eavesdropper has $z^{N_{\tilde{s}_{1}}}$, $s^{N_{\tilde{s}_{1}}}$, $w_{1,\tilde{s}_{1}}$ and
$w_{2,\tilde{s}_{1}}=(w_{2,\tilde{s}_{1},1},...,w_{2,\tilde{s}_{1},k})$, and he tries to find a unique
quadruple $(w_{1,\tilde{s}_{1}},\check{w}^{*}_{1,\tilde{s}_{1}},w_{2,\tilde{s}_{1}},\check{w}^{*}_{2,\tilde{s}_{1}}
=(\check{w}^{*}_{2,\tilde{s}_{1},1},...,\check{w}^{*}_{2,\tilde{s}_{1},k}))$
such that $(x_{1}^{N_{\tilde{s}_{1}}}(w_{1,\tilde{s}_{1}},\check{w}^{*}_{1,\tilde{s}_{1}}),
x_{2}^{N_{\tilde{s}_{1}}}(w_{2,\tilde{s}_{1}},\check{w}^{*}_{2,\tilde{s}_{1}}), z^{N_{\tilde{s}_{1}}}, s^{N_{\tilde{s}_{1}}})$
are strongly jointly typical sequences, i.e.,
\begin{eqnarray}\label{miyue4.1.sod}
&&(x_{1}^{N_{\tilde{s}_{1}}}(w_{1,\tilde{s}_{1}},\check{w}^{*}_{1,\tilde{s}_{1}}),
x_{2}^{N_{\tilde{s}_{1}}}(w_{2,\tilde{s}_{1}},\check{w}^{*}_{2,\tilde{s}_{1}}),
z^{N_{\tilde{s}_{1}}}, s^{N_{\tilde{s}_{1}}})\in T^{N_{\tilde{s}_{1}}}_{X_{1},X_{2},S,Z|\tilde{S}_{1}=\tilde{s}_{1}}(\epsilon).
\end{eqnarray}
If there exists such a unique quadruple, the eavesdropper declares that
$(w_{1,\tilde{s}_{1}},\check{w}^{*}_{1,\tilde{s}_{1}},w_{2,\tilde{s}_{1}},\check{w}^{*}_{2,\tilde{s}_{1}})$
is sent, otherwise he declares an error.
Based on the AEP, the construction of the codebooks, (\ref{hengda-1}), (\ref{miyue0}) and
(\ref{miyue0.xcc}), the eavesdropper's decoding error probability
$Pr\{(\check{w}^{*}_{1,\tilde{s}_{1}},\check{w}^{*}_{2,\tilde{s}_{1}})\neq
(w^{*}_{1,\tilde{s}_{1}},w^{*}_{2,\tilde{s}_{1}})\}$ tends to $0$ if
$N_{\tilde{s}_{1}}\rightarrow \infty$, and
\begin{eqnarray}\label{hengda-2.1.bangzi}
&&R^{*}_{1}(\tilde{s}_{1})\leq I(X_{1};Z|X_{2},S,\tilde{S}_{1}=\tilde{s}_{1}),
\end{eqnarray}
\begin{eqnarray}\label{hengda-3.1.bangzi}
&&\tilde{R}^{*}_{2}(\tilde{s}_{1})\leq I(X_{2};Z|X_{1},S,\tilde{S}_{1}=\tilde{s}_{1}),
\end{eqnarray}
\begin{eqnarray}\label{hengda-4.1.bangzi}
&&R^{*}_{1}(\tilde{s}_{1})+\tilde{R}^{*}_{2}(\tilde{s}_{1})
\leq I(X_{1},X_{2};Z|S,\tilde{S}_{1}=\tilde{s}_{1}),
\end{eqnarray}
where $\tilde{R}^{*}_{2}(\tilde{s}_{1})$ is given by (\ref{hengda-6}).
Combining (\ref{hengda-2.1}), (\ref{hengda-3.1}), (\ref{hengda-4.1}) with (\ref{c1.q1.rmb1}), (\ref{c1.q1}), (\ref{c1.q1.rmb11}) and (\ref{c1.q11}),
and letting $N_{\tilde{s}_{1}}\rightarrow \infty$, we have
\begin{eqnarray}\label{hengda-2.1.qqq}
&&R^{*}_{1}=\sum_{\tilde{s}_{1}=1}^{k}P_{\tilde{S}_{1}}(\tilde{s}_{1})R^{*}_{1}(\tilde{s}_{1})\nonumber\\
&&\leq \sum_{\tilde{s}_{1}=1}^{k}P_{\tilde{S}_{1}}(\tilde{s}_{1})I(X_{1};Z|X_{2},S,\tilde{S}_{1}=\tilde{s}_{1})\nonumber\\
&&=I(X_{1};Z|X_{2},S,\tilde{S}_{1})=H(Z|X_{2},S,\tilde{S}_{1})-H(Z|X_{1},X_{2},S,\tilde{S}_{1})\nonumber\\
&&\stackrel{(a)}=H(Z|X_{2},S,\tilde{S}_{1},\tilde{S}_{2})-H(Z|X_{1},X_{2},S,\tilde{S}_{1},\tilde{S}_{2})=I(X_{1};Z|X_{2},S,\tilde{S}_{1},\tilde{S}_{2}),
\end{eqnarray}
\begin{eqnarray}\label{hengda-3.1.qqq}
&&R^{*}_{2}=\sum_{\tilde{s}_{1}=1}^{k}P_{\tilde{S}_{1}}(\tilde{s}_{1})(\sum_{\tilde{s}_{2}=1}^{k}P_{\tilde{S}_{2}|\tilde{S}_{1}}(\tilde{s}_{2}|\tilde{s}_{1})
R^{*}_{2}(\tilde{s}_{1},\tilde{s}_{2}))\nonumber\\
&&\leq \sum_{\tilde{s}_{1}=1}^{k}P_{\tilde{S}_{1}}(\tilde{s}_{1})I(X_{2};Z|X_{1},S,\tilde{S}_{1}=\tilde{s}_{1})\nonumber\\
&&=I(X_{2};Z|X_{1},S,\tilde{S}_{1})=H(Z|X_{1},S,\tilde{S}_{1})-H(Z|X_{1},X_{2},S,\tilde{S}_{1})\nonumber\\
&&\stackrel{(b)}=H(Z|X_{1},S,\tilde{S}_{1},\tilde{S}_{2})-H(Z|X_{1},X_{2},S,\tilde{S}_{1},\tilde{S}_{2})=I(X_{2};Z|X_{1},S,\tilde{S}_{1},\tilde{S}_{2}),
\end{eqnarray}
\begin{eqnarray}\label{hengda-4.1.qqq}
&&R^{*}_{1}+R^{*}_{2}=\sum_{\tilde{s}_{1}=1}^{k}P_{\tilde{S}_{1}}(\tilde{s}_{1})(R^{*}_{1}(\tilde{s}_{1})+
\sum_{\tilde{s}_{2}=1}^{k}P_{\tilde{S}_{2}|\tilde{S}_{1}}(\tilde{s}_{2}|\tilde{s}_{1})
R^{*}_{2}(\tilde{s}_{1},\tilde{s}_{2}))\nonumber\\
&&\leq \sum_{\tilde{s}_{1}=1}^{k}P_{\tilde{S}_{1}}(\tilde{s}_{1})I(X_{1},X_{2};Z|S,\tilde{S}_{1}=\tilde{s}_{1})\nonumber\\
&&=I(X_{1},X_{2};Z|S,\tilde{S}_{1})=H(Z|S,\tilde{S}_{1})-H(Z|X_{1},X_{2},S,\tilde{S}_{1})\nonumber\\
&&\stackrel{(c)}=H(Z|S,\tilde{S}_{1},\tilde{S}_{2})-H(Z|X_{1},X_{2},S,\tilde{S}_{1},\tilde{S}_{2})=I(X_{1},X_{2};Z|S,\tilde{S}_{1},\tilde{S}_{2}),
\end{eqnarray}
where (a) is from the Markov chains $\tilde{S}_{2}\rightarrow (X_{2},S,\tilde{S}_{1})\rightarrow Z$ and
$\tilde{S}_{2}\rightarrow (X_{1},X_{2},S,\tilde{S}_{1})\rightarrow Z$, (b) is from the Markov chains
$\tilde{S}_{2}\rightarrow (X_{1},S,\tilde{S}_{1})\rightarrow Z$ and $\tilde{S}_{2}\rightarrow (X_{1},X_{2},S,\tilde{S}_{1})\rightarrow Z$, and (c)
is from $\tilde{S}_{2}\rightarrow (S,\tilde{S}_{1})\rightarrow Z$ and $\tilde{S}_{2}\rightarrow (X_{1},X_{2},S,\tilde{S}_{1})\rightarrow Z$.
Now it is easy to see that if (\ref{hengda-2.1.qqq}), (\ref{hengda-3.1.qqq}), (\ref{hengda-4.1.qqq}) are satisfied, the eavesdropper's decoding error probability of the dummy messages $w^{*}_{1}$ and $w^{*}_{2}$ tends to $0$. Applying Fano's inequality, hence we have
\begin{eqnarray}\label{dbian2}
&&\frac{1}{N_{\tilde{s}_{1}}}H(X_{1}^{N_{\tilde{s}_{1}}},X_{2}^{N_{\tilde{s}_{1}}}|W_{1,\tilde{s}_{1}},W_{2,\tilde{s}_{1}},
Z^{N_{\tilde{s}_{1}}},S^{N_{\tilde{s}_{1}}},\tilde{S}_{1}=\tilde{s}_{1})\leq \delta(\epsilon_{2}),
\end{eqnarray}
where $\delta(\epsilon_{2})\rightarrow 0$ as $N_{\tilde{s}_{1}}\rightarrow \infty$.

Now we bound the eavesdropper's equivocation $\Delta$,
\begin{eqnarray}\label{eapp5}
\Delta&=&\frac{1}{N}H(W_{1},W_{2}|Z^{N},S^{N})\nonumber\\
&=&\frac{1}{N}H(W_{1,1},W_{1,2},...,W_{1,k},W_{2,1,1},,W_{2,1,2},...,,W_{2,k,k}|Z^{N},S^{N})\nonumber\\
&\stackrel{(a)}=&\frac{1}{N}\sum_{\tilde{s}_{1}=1}^{k}H(W_{1,\tilde{s}_{1}},W_{2,\tilde{s}_{1}}|Z^{N},S^{N},
W_{1,1},...,W_{1,\tilde{s}_{1}-1},W_{2,1},...,W_{2,\tilde{s}_{1}-1})\nonumber\\
&\geq&\frac{1}{N}\sum_{\tilde{s}_{1}=1}^{k}H(W_{1,\tilde{s}_{1}},W_{2,\tilde{s}_{1}}|Z^{N},S^{N},
W_{1,1},...,W_{1,\tilde{s}_{1}-1},W_{2,1},...,W_{2,\tilde{s}_{1}-1},\tilde{S}_{1}=\tilde{s}_{1})\nonumber\\
&\stackrel{(b)}=&\frac{1}{N}\sum_{\tilde{s}_{1}=1}^{k}H(W_{1,\tilde{s}_{1}},W_{2,\tilde{s}_{1}}|Z^{N_{\tilde{s}_{1}}},S^{N_{\tilde{s}_{1}}},
\tilde{S}_{1}=\tilde{s}_{1})\nonumber\\
&=&\frac{1}{N}\sum_{\tilde{s}_{1}=1}^{k}(H(W_{1,\tilde{s}_{1}},W_{2,\tilde{s}_{1}},Z^{N_{\tilde{s}_{1}}},S^{N_{\tilde{s}_{1}}},
\tilde{S}_{1}=\tilde{s}_{1})-H(Z^{N_{\tilde{s}_{1}}},S^{N_{\tilde{s}_{1}}},\tilde{S}_{1}=\tilde{s}_{1}))\nonumber\\
&=&\frac{1}{N}\sum_{\tilde{s}_{1}=1}^{k}(H(W_{1,\tilde{s}_{1}},W_{2,\tilde{s}_{1}},X_{1}^{N_{\tilde{s}_{1}}},X_{2}^{N_{\tilde{s}_{1}}},
Z^{N_{\tilde{s}_{1}}},S^{N_{\tilde{s}_{1}}},\tilde{S}_{1}=\tilde{s}_{1})-
H(X_{1}^{N_{\tilde{s}_{1}}},X_{2}^{N_{\tilde{s}_{1}}}|W_{1,\tilde{s}_{1}},W_{2,\tilde{s}_{1}},Z^{N_{\tilde{s}_{1}}},
S^{N_{\tilde{s}_{1}}},\tilde{S}_{1}=\tilde{s}_{1})\nonumber\\
&&-H(Z^{N_{\tilde{s}_{1}}},S^{N_{\tilde{s}_{1}}},\tilde{S}_{1}=\tilde{s}_{1}))\nonumber\\
&\stackrel{(c)}=&\frac{1}{N}\sum_{\tilde{s}_{1}=1}^{k}(H(Z^{N_{\tilde{s}_{1}}}|X_{1}^{N_{\tilde{s}_{1}}},X_{2}^{N_{\tilde{s}_{1}}},
S^{N_{\tilde{s}_{1}}},\tilde{S}_{1}=\tilde{s}_{1})+H(X_{1}^{N_{\tilde{s}_{1}}},X_{2}^{N_{\tilde{s}_{1}}}|S^{N_{\tilde{s}_{1}}},\tilde{S}_{1}=\tilde{s}_{1})\nonumber\\
&&-H(X_{1}^{N_{\tilde{s}_{1}}},X_{2}^{N_{\tilde{s}_{1}}}|W_{1,\tilde{s}_{1}},W_{2,\tilde{s}_{1}},Z^{N_{\tilde{s}_{1}}},
S^{N_{\tilde{s}_{1}}},\tilde{S}_{1}=\tilde{s}_{1})
-H(Z^{N_{\tilde{s}_{1}}}|S^{N_{\tilde{s}_{1}}},\tilde{S}_{1}=\tilde{s}_{1}))\nonumber\\
&=&\frac{1}{N}\sum_{\tilde{s}_{1}=1}^{k}(H(Z^{N_{\tilde{s}_{1}}}|X_{1}^{N_{\tilde{s}_{1}}},X_{2}^{N_{\tilde{s}_{1}}},
S^{N_{\tilde{s}_{1}}},\tilde{S}_{1}=\tilde{s}_{1})+H(X_{1}^{N_{\tilde{s}_{1}}}|S^{N_{\tilde{s}_{1}}},\tilde{S}_{1}=\tilde{s}_{1})
+H(X_{2}^{N_{\tilde{s}_{1}}}|X_{1}^{N_{\tilde{s}_{1}}},S^{N_{\tilde{s}_{1}}},\tilde{S}_{1}=\tilde{s}_{1})\nonumber\\
&&-H(X_{1}^{N_{\tilde{s}_{1}}},X_{2}^{N_{\tilde{s}_{1}}}|W_{1,\tilde{s}_{1}},W_{2,\tilde{s}_{1}},Z^{N_{\tilde{s}_{1}}},S^{N_{\tilde{s}_{1}}},\tilde{S}_{1}=\tilde{s}_{1})
-H(Z^{N_{\tilde{s}_{1}}}|S^{N_{\tilde{s}_{1}}},\tilde{S}_{1}=\tilde{s}_{1}))\nonumber\\
&\stackrel{(d)}=&\frac{1}{N}\sum_{\tilde{s}_{1}=1}^{k}(H(Z^{N_{\tilde{s}_{1}}}|X_{1}^{N_{\tilde{s}_{1}}},X_{2}^{N_{\tilde{s}_{1}}},
S^{N_{\tilde{s}_{1}}},\tilde{S}_{1}=\tilde{s}_{1})+H(X_{1}^{N_{\tilde{s}_{1}}}|\tilde{S}_{1}=\tilde{s}_{1})
+H(X_{2}^{N_{\tilde{s}_{1}}}|\tilde{S}_{1}=\tilde{s}_{1})\nonumber\\
&&-H(X_{1}^{N_{\tilde{s}_{1}}},X_{2}^{N_{\tilde{s}_{1}}}|W_{1,\tilde{s}_{1}},W_{2,\tilde{s}_{1}},Z^{N_{\tilde{s}_{1}}},S^{N_{\tilde{s}_{1}}},\tilde{S}_{1}=\tilde{s}_{1})
-H(Z^{N_{\tilde{s}_{1}}}|S^{N_{\tilde{s}_{1}}},\tilde{S}_{1}=\tilde{s}_{1}))\nonumber\\
&\stackrel{(e)}=&\frac{1}{N}\sum_{\tilde{s}_{1}=1}^{k}(N_{\tilde{s}_{1}}H(Z|X_{1},X_{2},S,\tilde{S}_{1}=\tilde{s}_{1})
+H(X_{1}^{N_{\tilde{s}_{1}}}|\tilde{S}_{1}=\tilde{s}_{1})
+H(X_{2}^{N_{\tilde{s}_{1}}}|\tilde{S}_{1}=\tilde{s}_{1})\nonumber\\
&&-H(X_{1}^{N_{\tilde{s}_{1}}},X_{2}^{N_{\tilde{s}_{1}}}|W_{1,\tilde{s}_{1}},W_{2,\tilde{s}_{1}},Z^{N_{\tilde{s}_{1}}},S^{N_{\tilde{s}_{1}}},\tilde{S}_{1}=\tilde{s}_{1})
-H(Z^{N_{\tilde{s}_{1}}}|S^{N_{\tilde{s}_{1}}},\tilde{S}_{1}=\tilde{s}_{1}))\nonumber\\
&\stackrel{(f)}=&\frac{1}{N}\sum_{\tilde{s}_{1}=1}^{k}(N_{\tilde{s}_{1}}H(Z|X_{1},X_{2},S,\tilde{S}_{1}=\tilde{s}_{1})
+N_{\tilde{s}_{1}}(R_{1}(\tilde{s}_{1})+R^{*}_{1}(\tilde{s}_{1}))-1
+N_{\tilde{s}_{1}}(\tilde{R}_{2}(\tilde{s}_{1})+\tilde{R}^{*}_{2}(\tilde{s}_{1}))-1\nonumber\\
&&-H(X_{1}^{N_{\tilde{s}_{1}}},X_{2}^{N_{\tilde{s}_{1}}}|W_{1,\tilde{s}_{1}},W_{2,\tilde{s}_{1}},Z^{N_{\tilde{s}_{1}}},S^{N_{\tilde{s}_{1}}},\tilde{S}_{1}=\tilde{s}_{1})
-H(Z^{N_{\tilde{s}_{1}}}|S^{N_{\tilde{s}_{1}}},\tilde{S}_{1}=\tilde{s}_{1}))\nonumber\\
&\stackrel{(g)}\geq&\frac{1}{N}\sum_{\tilde{s}_{1}=1}^{k}(N_{\tilde{s}_{1}}H(Z|X_{1},X_{2},S,\tilde{S}_{1}=\tilde{s}_{1})
+N_{\tilde{s}_{1}}(R_{1}(\tilde{s}_{1})+R^{*}_{1}(\tilde{s}_{1}))-1+N_{\tilde{s}_{1}}(\tilde{R}_{2}(\tilde{s}_{1})+\tilde{R}^{*}_{2}(\tilde{s}_{1}))-1\nonumber\\
&&-N_{\tilde{s}_{1}}\delta(\epsilon_{2})
-N_{\tilde{s}_{1}}H(Z|S,\tilde{S}_{1}=\tilde{s}_{1}))\nonumber\\
&\stackrel{(h)}=&\frac{1}{N}\sum_{\tilde{s}_{1}=1}^{k}(N_{\tilde{s}_{1}}H(Z|X_{1},X_{2},S,\tilde{S}_{1}=\tilde{s}_{1},\tilde{S}_{2})
+N_{\tilde{s}_{1}}(R_{1}(\tilde{s}_{1})+R^{*}_{1}(\tilde{s}_{1}))-1+N_{\tilde{s}_{1}}(\tilde{R}_{2}(\tilde{s}_{1})+\tilde{R}^{*}_{2}(\tilde{s}_{1}))-1\nonumber\\
&&-N_{\tilde{s}_{1}}\delta(\epsilon_{2})
-N_{\tilde{s}_{1}}H(Z|S,\tilde{S}_{1}=\tilde{s}_{1},\tilde{S}_{2}))\nonumber\\
&\stackrel{(i)}=&\sum_{\tilde{s}_{1}=1}^{k}(P_{\tilde{S}_{1}}(\tilde{s}_{1})-\frac{\epsilon_{1}}{N})(R_{1}(\tilde{s}_{1})+R^{*}_{1}(\tilde{s}_{1})
+\tilde{R}_{2}(\tilde{s}_{1})+\tilde{R}^{*}_{2}(\tilde{s}_{1})-\frac{2}{N_{\tilde{s}_{1}}}+H(Z|X_{1},X_{2},S,\tilde{S}_{1}=\tilde{s}_{1},\tilde{S}_{2})\nonumber\\
&&-H(Z|S,\tilde{S}_{1}=\tilde{s}_{1},\tilde{S}_{2})-\delta(\epsilon_{2}))\nonumber\\
&\stackrel{(j)}=&\sum_{\tilde{s}_{1}=1}^{k}(P_{\tilde{S}_{1}}(\tilde{s}_{1})-\frac{\epsilon_{1}}{N})(R_{1}(\tilde{s}_{1})+R^{*}_{1}(\tilde{s}_{1})
+\sum_{\tilde{s}_{2}=1}^{k}P_{\tilde{S}_{2}|\tilde{S}_{1}}(\tilde{s}_{2}|\tilde{s}_{1})(R_{2}(\tilde{s}_{1},\tilde{s}_{2})+R^{*}_{2}(\tilde{s}_{1},\tilde{s}_{2}))\nonumber\\
&&-\sum_{\tilde{s}_{2}=1}^{k}\frac{\epsilon_{1}(1-kP_{\tilde{S}_{2}|\tilde{S}_{1}}(\tilde{s}_{2}|\tilde{s}_{1}))}{k(NP_{\tilde{S}_{1}}(\tilde{s}_{1})-\epsilon_{1})}
(R_{2}(\tilde{s}_{1},\tilde{s}_{2})+R^{*}_{2}(\tilde{s}_{1},\tilde{s}_{2}))-\frac{2}{N_{\tilde{s}_{1}}}-I(X_{1},X_{2};Z|S,\tilde{S}_{1}=\tilde{s}_{1},\tilde{S}_{2})-\delta(\epsilon_{2}))\nonumber\\
&\stackrel{(k)}=&R_{1}+R^{*}_{1}+R_{2}+R^{*}_{2}-I(X_{1},X_{2};Z|S,\tilde{S}_{1},\tilde{S}_{2})
-\sum_{\tilde{s}_{1}=1}^{k}P_{\tilde{S}_{1}}(\tilde{s}_{1})
\sum_{\tilde{s}_{2}=1}^{k}\frac{\epsilon_{1}(1-kP_{\tilde{S}_{2}|\tilde{S}_{1}}(\tilde{s}_{2}|\tilde{s}_{1}))}{k(NP_{\tilde{S}_{1}}(\tilde{s}_{1})-\epsilon_{1})}
(R_{2}(\tilde{s}_{1},\tilde{s}_{2})+R^{*}_{2}(\tilde{s}_{1},\tilde{s}_{2}))\nonumber\\
&&-\delta(\epsilon_{2})-\frac{\epsilon_{1}}{N}\sum_{\tilde{s}_{1}=1}^{k}(R_{1}(\tilde{s}_{1})+R^{*}_{1}(\tilde{s}_{1})
-I(X_{1},X_{2};Z|S,\tilde{S}_{1}=\tilde{s}_{1},\tilde{S}_{2}))\nonumber\\
&&-\frac{\epsilon_{1}}{N}\sum_{\tilde{s}_{1}=1}^{k}
\sum_{\tilde{s}_{2}=1}^{k}P_{\tilde{S}_{2}|\tilde{S}_{1}}(\tilde{s}_{2}|\tilde{s}_{1})
(R_{2}(\tilde{s}_{1},\tilde{s}_{2})+R^{*}_{2}(\tilde{s}_{1},\tilde{s}_{2}))\nonumber\\
&&-\frac{\epsilon_{1}}{N}\sum_{\tilde{s}_{1}=1}^{k}\sum_{\tilde{s}_{2}=1}^{k}\frac{\epsilon_{1}(kP_{\tilde{S}_{2}|\tilde{S}_{1}}(\tilde{s}_{2}|\tilde{s}_{1})-1)}{k(NP_{\tilde{S}_{1}}(\tilde{s}_{1})-\epsilon_{1})}
(R_{2}(\tilde{s}_{1},\tilde{s}_{2})+R^{*}_{2}(\tilde{s}_{1},\tilde{s}_{2}))-\frac{2k-k\epsilon_{1}\delta(\epsilon_{2})}{N},
\end{eqnarray}
where (a) is from the definition $W_{2,\tilde{s}_{1}}=(W_{2,\tilde{s}_{1},1},...,W_{2,\tilde{s}_{1},k})$ and the chain rule, (b) is from
the fact that the messages $W_{1,\tilde{s}_{1}}$ and $W_{2,\tilde{s}_{1}}$ depend only on the $\tilde{s}_{1}$-th sub-codebooks and the corresponding
channel inputs and outputs, i.e., the Markov chain
$(W_{1,1},...,W_{1,\tilde{s}_{1}-1},W_{2,1},...,W_{2,\tilde{s}_{1}-1},Z^{N_{1}},...,Z^{N_{\tilde{s}-1}},Z^{N_{\tilde{s}+1}},...,Z^{N_{k}},
S^{N_{1}},...,S^{N_{\tilde{s}-1}},\\S^{N_{\tilde{s}+1}},...,S^{N_{k}})
\rightarrow (Z^{N_{\tilde{s}}},S^{N_{\tilde{s}}},\tilde{S}_{1}=\tilde{s}_{1})\rightarrow (W_{1,\tilde{s}_{1}},W_{2,\tilde{s}_{1}})$ holds,
(c) is from $H(W_{1,\tilde{s}_{1}}|X_{1}^{N_{\tilde{s}_{1}}})=0$ and \\ $H(W_{2,\tilde{s}_{1}}|X_{2}^{N_{\tilde{s}_{1}}})=0$, (d) is from the fact that
given $\tilde{S}_{1}=\tilde{s}_{1}$, $X_{2}^{N_{\tilde{s}_{1}}}$ is independent of $X_{1}^{N_{\tilde{s}_{1}}}$ and $S^{N_{\tilde{s}_{1}}}$,
and given $\tilde{S}_{1}=\tilde{s}_{1}$, $X_{1}^{N_{\tilde{s}_{1}}}$ is independent of $S^{N_{\tilde{s}_{1}}}$,
(e) is from the fact that the channel is discrete memoryless, and the codewords $X_{1}^{N_{\tilde{s}_{1}}}$ and
$X_{2}^{N_{\tilde{s}_{1}}}$ are i.i.d. generated, (f) is from the fact that for each $\tilde{s}_{1}$,
there are $2^{N_{\tilde{s}_{1}}(R_{1}(\tilde{s}_{1})+R^{*}_{1}(\tilde{s}_{1}))}$ of $X_{1}^{N_{\tilde{s}_{1}}}$, and
$2^{N_{\tilde{s}_{1}}(\tilde{R}_{2}(\tilde{s}_{1})+\tilde{R}^{*}_{2}(\tilde{s}_{1}))}$ of $X_{2}^{N_{\tilde{s}_{1}}}$,
and applying a similar lemma in \cite{CK},
we have
\begin{eqnarray}\label{dbian1}
&&\frac{1}{N_{\tilde{s}}}H(X_{1}^{N_{\tilde{s}_{1}}}|\tilde{S}_{1}=\tilde{s}_{1})\geq
\frac{1}{N_{\tilde{s}}}\log 2^{N_{\tilde{s}_{1}}(R_{1}(\tilde{s}_{1})+R^{*}_{1}(\tilde{s}_{1}))}
-\frac{1}{N_{\tilde{s}}},
\end{eqnarray}
\begin{eqnarray}\label{dbian1.x}
&&\frac{1}{N_{\tilde{s}}}H(X_{2}^{N_{\tilde{s}_{1}}}|\tilde{S}_{1}=\tilde{s}_{1})\geq
\frac{1}{N_{\tilde{s}}}\log 2^{N_{\tilde{s}_{1}}(\tilde{R}_{2}(\tilde{s}_{1})+\tilde{R}^{*}_{2}(\tilde{s}_{1}))}
-\frac{1}{N_{\tilde{s}}},
\end{eqnarray}
where $\tilde{R}_{2}(\tilde{s}_{1})$ and $\tilde{R}^{*}_{2}(\tilde{s}_{1})$ are defined in (\ref{hengda-5}) and (\ref{hengda-6}), respectively,
(g) follows from (\ref{dbian2}), (h) is from the Markov chains
$\tilde{S}_{2}\rightarrow (X_{1},X_{2},S,\tilde{S}_{1}=\tilde{s}_{1})\rightarrow Z$ and
$\tilde{S}_{2}\rightarrow (S,\tilde{S}_{1}=\tilde{s}_{1})\rightarrow Z$,
(i) is from the definition (\ref{c3.q2}), (j) follows from (\ref{hengda-5}) and (\ref{hengda-6}), and (k) follows from the definitions of $R_{1}$, $R^{*}_{1}$,
$R_{2}$ and $R^{*}_{2}$, see
(\ref{c1.q1.rmb1}), (\ref{c1.q1}), (\ref{c1.q1.rmb11}) and (\ref{c1.q11}).

From (\ref{eapp5}), we can conclude that
\begin{eqnarray}\label{dbianbian3}
&&\Delta\geq R_{1}+R^{*}_{1}+R_{2}+R^{*}_{2}-I(X_{1},X_{2};Z|S,\tilde{S}_{1},\tilde{S}_{2})-\epsilon
\end{eqnarray}
for sufficiently large $N$. Letting
\begin{eqnarray}\label{dbianbian3.xx}
&&R^{*}_{1}+R^{*}_{2}\geq I(X_{1},X_{2};Z|S,\tilde{S}_{1},\tilde{S}_{2}),
\end{eqnarray}
$\Delta\geq R_{1}+R_{2}-\epsilon$ is proved.

Combining (\ref{hengda-2.1}), (\ref{hengda-3.1}), (\ref{hengda-4.1}), (\ref{hengda-2.1.qqq}), (\ref{hengda-3.1.qqq}) and (\ref{hengda-4.1.qqq})
with (\ref{dbianbian3.xx}), and applying Fourier-Motzkin elimination \cite{lall} to eliminate $R^{*}_{1}$ and $R^{*}_{2}$,
the region $\mathcal{R}^{*}$ is obtained.
Finally, using a standard time sharing technique presented in \cite[p.3438]{bash}, Theorem 3 is proved.

The proof of Theorem 3 is completed.

\section{Proof of Theorem 4}

Theorem 4 is proved by showing that for any achievable secrecy rate pair $(R_{1}, R_{2})$, the inequalities
$R_{1}\leq I(V_{1};Y|U,S,\tilde{S}_{1},\tilde{S}_{2})-I(V_{1};Z|U,S,\tilde{S}_{1},\tilde{S}_{2})$,
$R_{2}\leq I(V_{2};Y|U,S,\tilde{S}_{1},\tilde{S}_{2})-I(V_{2};Z|U,S,\tilde{S}_{1},\tilde{S}_{2})$ and
$R_{1}+R_{2}\leq I(V_{1},V_{2};Y|S,\tilde{S}_{1},\tilde{S}_{2},U)-I(V_{1},V_{2};Z|S,\tilde{S}_{1},\tilde{S}_{2},U)$ holds. Here
the random variables $U$, $V_{1}$, $V_{2}$, $S$, $\tilde{S}_{1}$, $\tilde{S}_{2}$, $Y$ and $Z$ are denoted by
\begin{eqnarray}\label{jmds1}
&&U\triangleq (Y^{J-1}, Z_{J+1}^{N}, S^{N}, J),\,\, V_{1}\triangleq (U, W_{1}), \,\, V_{2}\triangleq (U, W_{2}),\,\,Y\triangleq Y_{J},\,\,Z\triangleq Z_{J}\nonumber\\
&&S\triangleq S_{J},\,\, \tilde{S}_{1}\triangleq S_{J-d_{1}}, \,\, \tilde{S}_{2}\triangleq S_{J-d_{2}},
\end{eqnarray}
where the uniformly distributed random variable $J$ takes values in the set $\{1, 2, ,...,N\}$, and it is independent of
$Y^{N}$, $Z^{N}$, $W_{1}$, $W_{2}$ and $S^{N}$.

\textbf{Proof of $R_{1}\leq I(V_{1};Y|U,S,\tilde{S}_{1},\tilde{S}_{2})-I(V_{1};Z|U,S,\tilde{S}_{1},\tilde{S}_{2})$:}

First, note that the joint secrecy ensures the individual secrecy, and thus we have
\begin{eqnarray}\label{jmds2}
R_{1}-\epsilon&\leq&\frac{1}{N}H(W_{1}|Z^{N},S^{N})\nonumber\\
&=&\frac{1}{N}(H(W_{1})-I(W_{1};Z^{N},S^{N}))\stackrel{(a)}=\frac{1}{N}(H(W_{1})-I(W_{1};Z^{N}|S^{N}))\nonumber\\
&\stackrel{(b)}=&\frac{1}{N}(H(W_{1}|S^{N})-H(W_{1}|Y^{N},S^{N})+H(W_{1}|Y^{N},S^{N})-I(W_{1};Z^{N}|S^{N}))\nonumber\\
&\stackrel{(c)}\leq&\frac{1}{N}(I(W_{1};Y^{N}|S^{N})+\delta(P_{e})-I(W_{1};Z^{N}|S^{N}))\nonumber\\
&=&\frac{1}{N}\sum_{i=1}^{N}(I(W_{1};Y_{i}|Y^{i-1},S^{N})-I(W_{1};Z_{i}|Z_{i+1}^{N},S^{N}))+\frac{\delta(P_{e})}{N}\nonumber\\
&=&\frac{1}{N}\sum_{i=1}^{N}(I(W_{1};Y_{i}|Y^{i-1},S^{N},Z_{i+1}^{N})-I(W_{1};Z_{i}|Y^{i-1},S^{N},Z_{i+1}^{N})
+I(Y_{i};Z_{i+1}^{N}|Y^{i-1},S^{N})-I(Z_{i};Y^{i-1}|Z_{i+1}^{N},S^{N})\nonumber\\
&&-I(Y_{i};Z_{i+1}^{N}|W_{1},Y^{i-1},S^{N})+I(Z_{i};Y^{i-1}|W_{1},Z_{i+1}^{N},S^{N}))+\frac{\delta(P_{e})}{N}\nonumber\\
&\stackrel{(d)}=&\frac{1}{N}\sum_{i=1}^{N}(I(W_{1};Y_{i}|Y^{i-1},S^{N},Z_{i+1}^{N})-I(W_{1};Z_{i}|Y^{i-1},S^{N},Z_{i+1}^{N}))+\frac{\delta(P_{e})}{N}\nonumber\\
&\stackrel{(e)}=&\frac{1}{N}\sum_{i=1}^{N}(I(W_{1};Y_{i}|Y^{i-1},S^{N},Z_{i+1}^{N},S_{i},S_{i-d_{1}},S_{i-d_{2}})
-I(W_{1};Z_{i}|Y^{i-1},S^{N},Z_{i+1}^{N},S_{i},S_{i-d_{1}},S_{i-d_{2}}))+\frac{\delta(P_{e})}{N}\nonumber\\
&\stackrel{(f)}=&\frac{1}{N}\sum_{i=1}^{N}(I(V_{1,i};Y_{i}|U_{i},S_{i},S_{i-d_{1}},S_{i-d_{2}})
-I(V_{1,i};Z_{i}|U_{i},S_{i},S_{i-d_{1}},S_{i-d_{2}}))+\frac{\delta(P_{e})}{N}\nonumber\\
&\stackrel{(g)}=&\frac{1}{N}\sum_{i=1}^{N}(I(V_{1,i};Y_{i}|U_{i},S_{i},S_{i-d_{1}},S_{i-d_{2}},J=i)
-I(V_{1,i};Z_{i}|U_{i},S_{i},S_{i-d_{1}},S_{i-d_{2}},J=i))+\frac{\delta(P_{e})}{N}\nonumber\\
&=&I(V_{1,J};Y_{J}|U_{J},S_{J},S_{J-d_{1}},S_{J-d_{2}},J)
-I(V_{1,J};Z_{J}|U_{J},S_{J},S_{J-d_{1}},S_{J-d_{2}},J)+\frac{\delta(P_{e})}{N}\nonumber\\
&\stackrel{(h)}=&I(V_{1};Y|U,S,\tilde{S}_{1},\tilde{S}_{2})
-I(V_{1};Z|U,S,\tilde{S}_{1},\tilde{S}_{2})+\frac{\delta(P_{e})}{N}\nonumber\\
&\stackrel{(i)}\leq&I(V_{1};Y|U,S,\tilde{S}_{1},\tilde{S}_{2})
-I(V_{1};Z|U,S,\tilde{S}_{1},\tilde{S}_{2})+\frac{\delta(\epsilon)}{N},
\end{eqnarray}
where (a) and (b) follow from the fact that $W_{1}$ is independent of $S^{N}$, (c) follows from Fano's inequality,
(d) is from Csisz$\acute{a}$r's equality \cite{CK}, i.e.,
\begin{eqnarray}\label{jmds3.xxs}
&&I(Y_{i};Z_{i+1}^{N}|Y^{i-1},S^{N})=I(Z_{i};Y^{i-1}|Z_{i+1}^{N},S^{N}),
\end{eqnarray}
\begin{eqnarray}\label{jmds3.xxs1}
&&I(Y_{i};Z_{i+1}^{N}|W_{1},Y^{i-1},S^{N})=I(Z_{i};Y^{i-1}|W_{1},Z_{i+1}^{N},S^{N}),
\end{eqnarray}
(e) follows from the fact that $S_{i}$, $S_{i-d_{1}}$ and $S_{i-d_{2}}$ are included in $S^{N}$,
hence we have $H(S_{i},S_{i-d_{1}},S_{i-d_{2}}|S^{N})=0$, and here
note that $S_{i-d_{1}}=const$ (or $S_{i-d_{2}}=const$) when $i\leq d_{1}$ (or $i\leq d_{2}$),
(f) follows from the definitions $U_{i}=(Y^{i-1},S^{N},Z_{i+1}^{N})$ and $V_{1,i}=(W_{1},Y^{i-1},S^{N},Z_{i+1}^{N})$,
(g) follows from $J$ is a uniformly distributed random variable  which takes values in the set $\{1,2,...,N\}$, and it is
independent of $Y^{N}$, $Z^{N}$, $W_{1}$, $W_{2}$ and $S^{N}$, (h) is from the definitions in (\ref{jmds1}), and (i) follows from the fact that
$\delta(P_{e})$ is a monotonic increasing function of $P_{e}$ and $P_{e}\leq \epsilon$.
Then, letting $\epsilon\rightarrow 0$, the bound $R_{1}\leq I(V_{1};Y|U,S,\tilde{S}_{1},\tilde{S}_{2})-I(V_{1};Z|U,S,\tilde{S}_{1},\tilde{S}_{2})$ is obtained.

\textbf{Proof of $R_{1}\leq I(V_{1};Y|U,S,\tilde{S}_{1},\tilde{S}_{2})-I(V_{1};Z|U,S,\tilde{S}_{1},\tilde{S}_{2})$:}

The proof of $R_{2}\leq I(V_{2};Y|U,S,\tilde{S}_{1},\tilde{S}_{2})-I(V_{2};Z|U,S,\tilde{S}_{1},\tilde{S}_{2})$ is analogous to the proof of
$R_{1}\leq I(V_{1};Y|U,S,\tilde{S}_{1},\tilde{S}_{2})-I(V_{1};Z|U,S,\tilde{S}_{1},\tilde{S}_{2})$, and thus we omit the proof here.

\textbf{Proof of $R_{1}+R_{2}\leq I(V_{1},V_{2};Y|S,\tilde{S}_{1},\tilde{S}_{2},U)-I(V_{1},V_{2};Z|S,\tilde{S}_{1},\tilde{S}_{2},U)$:}

Note that
\begin{eqnarray}\label{jmds-xss1}
R_{1}+R_{2}-\epsilon&\leq&\frac{1}{N}H(W_{1},W_{2}|Z^{N},S^{N})\nonumber\\
&=&\frac{1}{N}(H(W_{1},W_{2})-I(W_{1},W_{2};Z^{N},S^{N}))\stackrel{(a)}=\frac{1}{N}(H(W_{1},W_{2})-I(W_{1},W_{2};Z^{N}|S^{N}))\nonumber\\
&\stackrel{(b)}=&\frac{1}{N}(H(W_{1},W_{2}|S^{N})-H(W_{1},W_{2}|Y^{N},S^{N})+H(W_{1},W_{2}|Y^{N},S^{N})-I(W_{1},W_{2};Z^{N}|S^{N}))\nonumber\\
&\stackrel{(c)}\leq&\frac{1}{N}(I(W_{1},W_{2};Y^{N}|S^{N})+\delta(P_{e})-I(W_{1},W_{2};Z^{N}|S^{N}))\nonumber\\
&=&\frac{1}{N}\sum_{i=1}^{N}(I(W_{1},W_{2};Y_{i}|Y^{i-1},S^{N})-I(W_{1},W_{2};Z_{i}|Z_{i+1}^{N},S^{N}))+\frac{\delta(P_{e})}{N}\nonumber\\
&=&\frac{1}{N}\sum_{i=1}^{N}(I(W_{1},W_{2};Y_{i}|Y^{i-1},S^{N},Z_{i+1}^{N})-I(W_{1},W_{2};Z_{i}|Y^{i-1},S^{N},Z_{i+1}^{N})
+I(Y_{i};Z_{i+1}^{N}|Y^{i-1},S^{N})\nonumber\\
&&-I(Z_{i};Y^{i-1}|Z_{i+1}^{N},S^{N})
-I(Y_{i};Z_{i+1}^{N}|W_{1},W_{2},Y^{i-1},S^{N})+I(Z_{i};Y^{i-1}|W_{1},W_{2},Z_{i+1}^{N},S^{N}))+\frac{\delta(P_{e})}{N}\nonumber\\
&\stackrel{(d)}=&\frac{1}{N}\sum_{i=1}^{N}(I(W_{1},W_{2};Y_{i}|Y^{i-1},S^{N},Z_{i+1}^{N})-I(W_{1},W_{2};Z_{i}|Y^{i-1},S^{N},Z_{i+1}^{N}))+\frac{\delta(P_{e})}{N}\nonumber\\
&\stackrel{(e)}=&\frac{1}{N}\sum_{i=1}^{N}(I(W_{1},W_{2};Y_{i}|Y^{i-1},S^{N},Z_{i+1}^{N},S_{i},S_{i-d_{1}},S_{i-d_{2}})\nonumber\\
&&-I(W_{1},W_{2};Z_{i}|Y^{i-1},S^{N},Z_{i+1}^{N},S_{i},S_{i-d_{1}},S_{i-d_{2}}))+\frac{\delta(P_{e})}{N}\nonumber\\
&\stackrel{(f)}=&\frac{1}{N}\sum_{i=1}^{N}(I(V_{1,i},V_{2,i};Y_{i}|U_{i},S_{i},S_{i-d_{1}},S_{i-d_{2}})
-I(V_{1,i},V_{2,i};Z_{i}|U_{i},S_{i},S_{i-d_{1}},S_{i-d_{2}}))+\frac{\delta(P_{e})}{N}\nonumber\\
&\stackrel{(g)}=&I(V_{1,J},V_{2,J};Y_{J}|U_{J},S_{J},S_{J-d_{1}},S_{J-d_{2}},J)
-I(V_{1,J},V_{2,J};Z_{J}|U_{J},S_{J},S_{J-d_{1}},S_{J-d_{2}},J)+\frac{\delta(P_{e})}{N}\nonumber\\
&\stackrel{(h)}=&I(V_{1},V_{2};Y|U,S,\tilde{S}_{1},\tilde{S}_{2})
-I(V_{1},V_{2};Z|U,S,\tilde{S}_{1},\tilde{S}_{2})+\frac{\delta(P_{e})}{N}\nonumber\\
&\stackrel{(i)}\leq&I(V_{1},V_{2};Y|U,S,\tilde{S}_{1},\tilde{S}_{2})
-I(V_{1},V_{2};Z|U,S,\tilde{S}_{1},\tilde{S}_{2})+\frac{\delta(\epsilon)}{N},
\end{eqnarray}
where (a) and (b) follow from the fact that $S^{N}$ is independent of $W_{1}$ and $W_{2}$, (c) follows from Fano's inequality,
(d) is from Csisz$\acute{a}$r's equality \cite{CK}, i.e.,
\begin{eqnarray}\label{jmds3.xxs}
&&I(Y_{i};Z_{i+1}^{N}|Y^{i-1},S^{N})=I(Z_{i};Y^{i-1}|Z_{i+1}^{N},S^{N}),
\end{eqnarray}
\begin{eqnarray}\label{jmds3.xxs1}
&&I(Y_{i};Z_{i+1}^{N}|W_{1},W_{2},Y^{i-1},S^{N})=I(Z_{i};Y^{i-1}|W_{1},W_{2},Z_{i+1}^{N},S^{N}),
\end{eqnarray}
(e) follows from the fact that $S_{i}$, $S_{i-d_{1}}$ and $S_{i-d_{2}}$ are included in $S^{N}$,
(f) follows from the definitions $U_{i}=(Y^{i-1},S^{N},Z_{i+1}^{N})$, $V_{1,i}=(W_{1},Y^{i-1},S^{N},Z_{i+1}^{N})$ and $V_{2,i}=(W_{2},Y^{i-1},S^{N},Z_{i+1}^{N})$,
(g) follows from $J$ is a uniformly distributed random variable which takes values in $\{1,2,...,N\}$, and it is
independent of $Y^{N}$, $Z^{N}$, $W_{1}$, $W_{2}$ and $S^{N}$, (h) is from the definitions in (\ref{jmds1}), and (i) follows from the fact that
$\delta(P_{e})$ is a monotonic increasing function of $P_{e}$ and $P_{e}\leq \epsilon$.
Then, letting $\epsilon\rightarrow 0$, the bound
$R_{1}+R_{2}\leq I(V_{1},V_{2};Y|S,\tilde{S}_{1},\tilde{S}_{2},U)-I(V_{1},V_{2};Z|S,\tilde{S}_{1},\tilde{S}_{2},U)$ is obtained.

The proof of Theorem 4 is completed.

\section{Proof of the Outer Bound on the Secrecy Capacity Region of the Degraded Case of the FS-MAC-WT with only Delayed State Feedback\label{appen3}}

In this section, we will show that for the degraded case $(X_{1}^{N},X_{2}^{N},S^{N})\rightarrow Y^{N}\rightarrow Z^{N}$,
all the achievable secrecy rate pairs $(R_{1},R_{2})$ of the FS-MAC-WT with only delayed state feedback are contained in the following region
\begin{eqnarray*}
&&\mathcal{C}^{d-out}_{s}=\{(R_{1}, R_{2}): 0\leq R_{1}\leq I(X_{1};Y|X_{2},S,\tilde{S}_{1},\tilde{S}_{2},Q)-I(X_{1};Z|S,\tilde{S}_{1},\tilde{S}_{2},Q),\\
&&0\leq R_{2}\leq I(X_{2};Y|X_{1},S,\tilde{S}_{1},\tilde{S}_{2},Q)-I(X_{2};Z|S,\tilde{S}_{1},\tilde{S}_{2},Q),\\
&&0\leq R_{1}+R_{2}\leq I(X_{1},X_{2};Y|S,\tilde{S}_{1},\tilde{S}_{2},Q)-I(X_{1},X_{2};Z|S,\tilde{S}_{1},\tilde{S}_{2},Q)\},
\end{eqnarray*}
where the joint probability satisfies
\begin{eqnarray}\label{dota1.rmx}
&&P_{QS\tilde{S}_{1}\tilde{S}_{2}X_{1}X_{2}YZ}(q,s,\tilde{s}_{1},\tilde{s}_{2},x_{1},x_{2},y,z)\nonumber\\
&&=P_{Z|Y}(z|y)P_{Y|X_{1},X_{2},S}(y|x_{1},x_{2},s)P_{X_{1}X_{2}S\tilde{S}_{1}\tilde{S}_{2}Q}(x_{1},x_{2},s,\tilde{s}_{1},\tilde{s}_{2},q).
\end{eqnarray}

\textbf{Proof of $R_{1}\leq I(X_{1};Y|X_{2},S,\tilde{S}_{1},\tilde{S}_{2},Q)-I(X_{1};Z|S,\tilde{S}_{1},\tilde{S}_{2},Q)$:}

Note that
\begin{eqnarray}\label{jmds2.daqindiguo1}
R_{1}-\epsilon&\leq&\frac{1}{N}H(W_{1}|Z^{N},S^{N})\nonumber\\
&=&\frac{1}{N}(H(W_{1}|Z^{N},S^{N})-H(W_{1}|Z^{N},S^{N},Y^{N},W_{2})+H(W_{1}|Z^{N},S^{N},Y^{N},W_{2}))\nonumber\\
&\stackrel{(a)}\leq&\frac{1}{N}(H(W_{1}|Z^{N},S^{N})-H(W_{1}|Z^{N},S^{N},Y^{N},W_{2})+\delta(P_{e}))\nonumber\\
&=&\frac{1}{N}(I(W_{1};Y^{N},W_{2}|Z^{N},S^{N})+\delta(P_{e}))\nonumber\\
&\stackrel{(b)}\leq&\frac{1}{N}(I(X_{1}^{N};Y^{N},W_{2}|Z^{N},S^{N})+\delta(P_{e}))\nonumber\\
&\stackrel{(c)}\leq&\frac{1}{N}(I(X_{1}^{N};Y^{N},X_{2}^{N}|Z^{N},S^{N})+\delta(P_{e}))\nonumber\\
&\stackrel{(d)}=&\frac{1}{N}(H(X_{1}^{N}|Z^{N},S^{N})-H(X_{1}^{N}|S^{N},Y^{N},X_{2}^{N})+\delta(P_{e}))\nonumber\\
&\stackrel{(e)}=&\frac{1}{N}(H(X_{1}^{N}|Z^{N},S^{N})-H(X_{1}^{N}|S^{N},Y^{N},X_{2}^{N})+H(X_{1}^{N}|X_{2}^{N},S^{N})
-H(X_{1}^{N}|S^{N})+\delta(P_{e}))\nonumber\\
&=&\frac{1}{N}(I(X_{1}^{N};Y^{N}|X_{2}^{N},S^{N})-I(X_{1}^{N};Z^{N}|S^{N})+\delta(P_{e}))\nonumber\\
&=&\frac{1}{N}\sum_{i=1}^{N}(H(Y_{i}|Y^{i-1},X_{2}^{N},S^{N})-H(Y_{i}|X_{1,i},X_{2,i},S_{i})\nonumber\\
&&-H(Z_{i}|Z^{i-1},S^{N})+H(Z_{i}|Z^{i-1},S^{N},X_{1}^{N}))+\frac{\delta(P_{e})}{N}\nonumber\\
&\stackrel{(f)}=&\frac{1}{N}\sum_{i=1}^{N}(H(Y_{i}|Y^{i-1},X_{2}^{N},S^{N},Z^{i-1})-H(Y_{i}|X_{1,i},X_{2,i},S_{i},Z^{i-1},S^{N})\nonumber\\
&&-H(Z_{i}|Z^{i-1},S^{N})+H(Z_{i}|Z^{i-1},S^{N},X_{1}^{N}))+\frac{\delta(P_{e})}{N}\nonumber\\
&\leq&\frac{1}{N}\sum_{i=1}^{N}(H(Y_{i}|X_{2,i},S^{N},Z^{i-1})-H(Y_{i}|X_{1,i},X_{2,i},S_{i},Z^{i-1},S^{N})\nonumber\\
&&-H(Z_{i}|Z^{i-1},S^{N})+H(Z_{i}|Z^{i-1},S^{N},X_{1,i}))+\frac{\delta(P_{e})}{N}\nonumber\\
&\stackrel{(g)}=&\frac{1}{N}\sum_{i=1}^{N}(H(Y_{i}|X_{2,i},S^{N},Z^{i-1},S_{i},S_{i-d_{1}},S_{i-d_{2}})-H(Y_{i}|X_{1,i},X_{2,i},S_{i},Z^{i-1},S^{N},S_{i-d_{1}},S_{i-d_{2}})\nonumber\\
&&-H(Z_{i}|Z^{i-1},S^{N},S_{i},S_{i-d_{1}},S_{i-d_{2}})+H(Z_{i}|Z^{i-1},S^{N},X_{1,i},S_{i},S_{i-d_{1}},S_{i-d_{2}}))+\frac{\delta(P_{e})}{N}\nonumber\\
&\stackrel{(h)}=&\frac{1}{N}\sum_{i=1}^{N}(H(Y_{i}|X_{2,i},Q_{i},S_{i},S_{i-d_{1}},S_{i-d_{2}})-H(Y_{i}|X_{1,i},X_{2,i},S_{i},Q_{i},S_{i-d_{1}},S_{i-d_{2}})\nonumber\\
&&-H(Z_{i}|Q_{i},S_{i},S_{i-d_{1}},S_{i-d_{2}})+H(Z_{i}|Q_{i},X_{1,i},S_{i},S_{i-d_{1}},S_{i-d_{2}}))+\frac{\delta(P_{e})}{N}\nonumber\\
&\stackrel{(i)}=&\frac{1}{N}\sum_{i=1}^{N}(H(Y_{i}|X_{2,i},Q_{i},S_{i},S_{i-d_{1}},S_{i-d_{2}},J=i)-H(Y_{i}|X_{1,i},X_{2,i},S_{i},Q_{i},S_{i-d_{1}},S_{i-d_{2}},J=i)\nonumber\\
&&-H(Z_{i}|Q_{i},S_{i},S_{i-d_{1}},S_{i-d_{2}},J=i)+H(Z_{i}|Q_{i},X_{1,i},S_{i},S_{i-d_{1}},S_{i-d_{2}},J=i))+\frac{\delta(P_{e})}{N}\nonumber\\
&\stackrel{(j)}=&H(Y_{J}|X_{2,J},Q_{J},S_{J},S_{J-d_{1}},S_{J-d_{2}},J)-H(Y_{J}|X_{1,J},X_{2,J},S_{J},Q_{J},S_{J-d_{1}},S_{J-d_{2}},J)\nonumber\\
&&-H(Z_{i}|Q_{J},S_{J},S_{J-d_{1}},S_{J-d_{2}},J)+H(Z_{J}|Q_{J},X_{1,J},S_{J},S_{J-d_{1}},S_{J-d_{2}},J)+\frac{\delta(P_{e})}{N}\nonumber\\
&\stackrel{(k)}=&I(X_{1};Y|X_{2},Q,S,\tilde{S}_{1},\tilde{S}_{2})-I(X_{1};Z|Q,S,\tilde{S}_{1},\tilde{S}_{2})+\frac{\delta(P_{e})}{N}\nonumber\\
&\stackrel{(l)}\leq&I(X_{1};Y|X_{2},Q,S,\tilde{S}_{1},\tilde{S}_{2})-I(X_{1};Z|Q,S,\tilde{S}_{1},\tilde{S}_{2})+\frac{\delta(\epsilon)}{N},
\end{eqnarray}
where (a) follows from Fano's inequality, (b) is from the fact that $H(W_{1}|X_{1}^{N})=0$, (c) is from $H(W_{2}|X_{2}^{N})=0$, (d) is from the Markov chain
$X_{1}^{N}\rightarrow (S^{N},Y^{N},X_{2}^{N})\rightarrow Z^{N}$, (e) follows from the fact that given $S^{N}$, $X_{1}^{N}$ is independent of
$X_{2}^{N}$, (f) follows from the Markov chains $Y_{i}\rightarrow (Y^{i-1},X_{2}^{N},S^{N})\rightarrow Z^{i-1}$,
$(Z^{i-1},S^{N})\rightarrow (X_{1,i},X_{2,i},S_{i})\rightarrow Y_{i}$, (g) follows from the fact that
$H(S_{i},S_{i-d_{1}},S_{i-d_{2}}|S^{N})=0$, and here
note that $S_{i-d_{1}}=const$ (or $S_{i-d_{2}}=const$) when $i\leq d_{1}$ (or $i\leq d_{2}$), (h) is from the definition
$Q_{i}=(Z^{i-1},S^{N})$, (i) and (j) follow from $J$ is a uniformly distributed random variable which takes values in the set $\{1,2,...,N\}$,
and it is independent of $X_{1}^{N}$, $X_{2}^{N}$, $Y^{N}$, $Z^{N}$, $W_{1}$, $W_{2}$ and $S^{N}$, (k) is from the definitions
$Q\triangleq (Q_{J},J)=(Z^{J-1},S^{N},J)$, $X_{1}\triangleq X_{1,J}$, $X_{2}\triangleq X_{2,J}$, $Y\triangleq Y_{J}$, $Z\triangleq Z_{J}$,
$S\triangleq S_{J}$, $\tilde{S}_{1}\triangleq S_{J-d_{1}}$ and $\tilde{S}_{2}\triangleq S_{J-d_{2}}$, and (l) follows from $\delta(P_{e})$
is a monotonic increasing function of $P_{e}$ and $P_{e}\leq \epsilon$.
Letting $\epsilon\rightarrow 0$, $R_{1}\leq I(X_{1};Y|X_{2},S,\tilde{S}_{1},\tilde{S}_{2},Q)-I(X_{1};Z|S,\tilde{S}_{1},\tilde{S}_{2},Q)$ is proved.

\textbf{Proof of $R_{2}\leq I(X_{2};Y|X_{1},S,\tilde{S}_{1},\tilde{S}_{2},Q)-I(X_{2};Z|S,\tilde{S}_{1},\tilde{S}_{2},Q)$:}

The proof of $R_{2}\leq I(X_{2};Y|X_{1},S,\tilde{S}_{1},\tilde{S}_{2},Q)-I(X_{2};Z|S,\tilde{S}_{1},\tilde{S}_{2},Q)$ is analogous to that of \\
$R_{1}\leq I(X_{1};Y|X_{2},S,\tilde{S}_{1},\tilde{S}_{2},Q)-I(X_{1};Z|S,\tilde{S}_{1},\tilde{S}_{2},Q)$, and thus we omit it here.

\textbf{Proof of $R_{1}+R_{2}\leq I(X_{1},X_{2};Y|S,\tilde{S}_{1},\tilde{S}_{2},Q)-I(X_{1},X_{2};Z|S,\tilde{S}_{1},\tilde{S}_{2},Q)$:}

Note that
\begin{eqnarray}\label{jmds2.daqindiguo1}
R_{1}+R_{2}-\epsilon&\leq&\frac{1}{N}H(W_{1},W_{2}|Z^{N},S^{N})\nonumber\\
&\stackrel{(1)}\leq&\frac{1}{N}(I(W_{1},W_{2};Y^{N}|Z^{N},S^{N})+\delta(P_{e}))\nonumber\\
&\stackrel{(2)}\leq&\frac{1}{N}(I(X_{1}^{N},X_{2}^{N};Y^{N}|Z^{N},S^{N})+\delta(P_{e}))\nonumber\\
&\stackrel{(3)}=&\frac{1}{N}(H(X_{1}^{N},X_{2}^{N}|Z^{N},S^{N})-H(X_{1}^{N},X_{2}^{N}|Y^{N},S^{N})-H(X_{1}^{N},X_{2}^{N}|S^{N})\nonumber\\
&&+H(X_{1}^{N},X_{2}^{N}|S^{N})+\delta(P_{e}))\nonumber\\
&=&\frac{1}{N}(I(X_{1}^{N},X_{2}^{N};Y^{N}|S^{N})-I(X_{1}^{N},X_{2}^{N};Z^{N}|S^{N})+\delta(P_{e}))\nonumber\\
&=&\frac{1}{N}\sum_{i=1}^{N}(H(Y_{i}|Y^{i-1},S^{N})-H(Y_{i}|X_{1,i},X_{2,i},S_{i})\nonumber\\
&&-H(Z_{i}|Z^{i-1},S^{N})+H(Z_{i}|X_{1,i},X_{2,i},S_{i}))+\frac{\delta(P_{e})}{N}\nonumber\\
&\stackrel{(4)}=&\frac{1}{N}\sum_{i=1}^{N}(H(Y_{i}|Y^{i-1},S^{N},Z^{i-1})-H(Y_{i}|X_{1,i},X_{2,i},S_{i},S^{N},Z^{i-1})\nonumber\\
&&-H(Z_{i}|Z^{i-1},S^{N})+H(Z_{i}|X_{1,i},X_{2,i},S_{i},S^{N},Z^{i-1}))+\frac{\delta(P_{e})}{N}\nonumber\\
&\leq&\frac{1}{N}\sum_{i=1}^{N}(H(Y_{i}|S^{N},Z^{i-1})-H(Y_{i}|X_{1,i},X_{2,i},S_{i},S^{N},Z^{i-1})\nonumber\\
&&-H(Z_{i}|Z^{i-1},S^{N})+H(Z_{i}|X_{1,i},X_{2,i},S_{i},S^{N},Z^{i-1}))+\frac{\delta(P_{e})}{N}\nonumber\\
&\stackrel{(5)}=&\frac{1}{N}\sum_{i=1}^{N}(H(Y_{i}|S^{N},Z^{i-1},S_{i},S_{i-d_{1}},S_{i-d_{2}})-H(Y_{i}|X_{1,i},X_{2,i},S_{i},S^{N},Z^{i-1},S_{i-d_{1}},S_{i-d_{2}})\nonumber\\
&&-H(Z_{i}|Z^{i-1},S^{N},S_{i},S_{i-d_{1}},S_{i-d_{2}})+H(Z_{i}|X_{1,i},X_{2,i},S_{i},S^{N},Z^{i-1},S_{i-d_{1}},S_{i-d_{2}}))+\frac{\delta(P_{e})}{N}\nonumber\\
&\stackrel{(6)}=&\frac{1}{N}\sum_{i=1}^{N}(H(Y_{i}|Q_{i},S_{i},S_{i-d_{1}},S_{i-d_{2}},J=i)-H(Y_{i}|X_{1,i},X_{2,i},S_{i},Q_{i},S_{i-d_{1}},S_{i-d_{2}},J=i)\nonumber\\
&&-H(Z_{i}|Q_{i},S_{i},S_{i-d_{1}},S_{i-d_{2}},J=i)+H(Z_{i}|X_{1,i},X_{2,i},S_{i},Q_{i},S_{i-d_{1}},S_{i-d_{2}},J=i))+\frac{\delta(P_{e})}{N}\nonumber\\
&\stackrel{(7)}=&I(X_{1},X_{2};Y|Q,S,\tilde{S}_{1},\tilde{S}_{2})-I(X_{1},X_{2};Z|Q,S,\tilde{S}_{1},\tilde{S}_{2})+\frac{\delta(P_{e})}{N}\nonumber\\
&\stackrel{(8)}\leq&I(X_{1},X_{2};Y|Q,S,\tilde{S}_{1},\tilde{S}_{2})-I(X_{1},X_{2};Z|Q,S,\tilde{S}_{1},\tilde{S}_{2})+\frac{\delta(\epsilon)}{N},
\end{eqnarray}
where (1) follows from Fano's inequality, (2) follows from the fact that $H(W_{1}|X_{1}^{N})=0$ and $H(W_{2}|X_{2}^{N})=0$, (3) is from the Markov chain
$(X_{1}^{N},X_{2}^{N})\rightarrow (S^{N},Y^{N})\rightarrow Z^{N}$, (4) follows from the Markov chains
$Y_{i}\rightarrow (Y^{i-1},S^{N})\rightarrow Z^{i-1}$,
$(Z^{i-1},S^{N})\rightarrow (X_{1,i},X_{2,i},S_{i})\rightarrow Y_{i}$ and $(Z^{i-1},S^{N})\rightarrow (X_{1,i},X_{2,i},S_{i})\rightarrow Z_{i}$,
(5) follows from the fact that
$H(S_{i},S_{i-d_{1}},S_{i-d_{2}}|S^{N})=0$, (6) is from the definition
$Q_{i}=(Z^{i-1},S^{N})$, and $J$ is a uniformly distributed random variable which takes values in the set $\{1,2,...,N\}$,
and it is independent of $X_{1}^{N}$, $X_{2}^{N}$, $Y^{N}$, $Z^{N}$, $W_{1}$, $W_{2}$ and $S^{N}$, (7) follows from
the definitions
$Q\triangleq (Q_{J},J)=(Z^{J-1},S^{N},J)$, $X_{1}\triangleq X_{1,J}$, $X_{2}\triangleq X_{2,J}$, $Y\triangleq Y_{J}$, $Z\triangleq Z_{J}$,
$S\triangleq S_{J}$, $\tilde{S}_{1}\triangleq S_{J-d_{1}}$ and $\tilde{S}_{2}\triangleq S_{J-d_{2}}$, and (8) follows from $\delta(P_{e})$
is a monotonic increasing function of $P_{e}$ and $P_{e}\leq \epsilon$.
Letting $\epsilon\rightarrow 0$, $R_{1}+R_{2}\leq I(X_{1},X_{2};Y|S,\tilde{S}_{1},\tilde{S}_{2},Q)-I(X_{1},X_{2};Z|S,\tilde{S}_{1},\tilde{S}_{2},Q)$ is proved.

The proof of the outer bound for the degraded case of the FS-MAC-WT with only delayed state feedback is completed.

\section{The Derivation of Corollary 2}

First, we explicitly compute the upper bound on the secrecy sum rate of $\mathcal{C}_{s}^{(dg-out)}$.
For the discrete memoryless degraded FS-MAC-WT with only delayed state feedback, we have shown that the outer bound $\mathcal{C}^{d-out}_{s}$ is given by
\begin{eqnarray*}
&&\mathcal{C}^{d-out}_{s}=\{(R_{1}, R_{2}): 0\leq R_{1}\leq I(X_{1};Y|X_{2},S,\tilde{S}_{1},\tilde{S}_{2},Q)-I(X_{1};Z|S,\tilde{S}_{1},\tilde{S}_{2},Q),\\
&&0\leq R_{2}\leq I(X_{2};Y|X_{1},S,\tilde{S}_{1},\tilde{S}_{2},Q)-I(X_{2};Z|S,\tilde{S}_{1},\tilde{S}_{2},Q),\\
&&0\leq R_{1}+R_{2}\leq I(X_{1},X_{2};Y|S,\tilde{S}_{1},\tilde{S}_{2},Q)-I(X_{1},X_{2};Z|S,\tilde{S}_{1},\tilde{S}_{2},Q)\}.
\end{eqnarray*}
Then for the sum rate, we have
\begin{eqnarray}\label{ex-1}
R_{1}+R_{2}&\leq&I(X_{1},X_{2};Y|S,\tilde{S}_{1},\tilde{S}_{2},Q)-I(X_{1},X_{2};Z|S,\tilde{S}_{1},\tilde{S}_{2},Q)\nonumber\\
&\stackrel{(1)}=&h(Y|S,\tilde{S}_{1},\tilde{S}_{2},Q)-h(Y|X_{1},X_{2},S,\tilde{S}_{1},\tilde{S}_{2})-h(Z|S,\tilde{S}_{1},\tilde{S}_{2},Q)
+h(Z|S,\tilde{S}_{1},\tilde{S}_{2},X_{1},X_{2})\nonumber\\
&\stackrel{(2)}\leq&h(Y|S,\tilde{S}_{1},\tilde{S}_{2})-h(Y|X_{1},X_{2},S,\tilde{S}_{1},\tilde{S}_{2})-h(Z|S,\tilde{S}_{1},\tilde{S}_{2})
+h(Z|S,\tilde{S}_{1},\tilde{S}_{2},X_{1},X_{2})\nonumber\\
&=&I(X_{1},X_{2};Y|S,\tilde{S}_{1},\tilde{S}_{2})-I(X_{1},X_{2};Z|S,\tilde{S}_{1},\tilde{S}_{2}),
\end{eqnarray}
where (1) is from the Markov chains $Q\rightarrow (S,\tilde{S}_{1},\tilde{S}_{2},X_{1},X_{2})\rightarrow Y$ and
$Q\rightarrow (S,\tilde{S}_{1},\tilde{S}_{2},X_{1},X_{2})\rightarrow Z$, and (2) is from
\begin{equation}\label{jiangyou1}
h(Y|S,\tilde{S}_{1},\tilde{S}_{2},Q)-h(Z|S,\tilde{S}_{1},\tilde{S}_{2},Q)\leq h(Y|S,\tilde{S}_{1},\tilde{S}_{2})-h(Z|S,\tilde{S}_{1},\tilde{S}_{2}).
\end{equation}
Here note that (\ref{jiangyou1}) can be re-written as
\begin{equation}\label{jiangyou2}
I(Q;Z|S,\tilde{S}_{1},\tilde{S}_{2})\leq I(Q;Y|S,\tilde{S}_{1},\tilde{S}_{2}),
\end{equation}
and from $Q\rightarrow (S,\tilde{S}_{1},\tilde{S}_{2},Y)\rightarrow Z$, it is easy to see that (\ref{jiangyou2}) holds.
Hence the secrecy sum rate of $\mathcal{C}_{s}^{(dg-out)}$ is bounded by
\begin{eqnarray}\label{ex-2}
&&R_{1}+R_{2}\leq I(X_{1},X_{2};Y|S,\tilde{S}_{1},\tilde{S}_{2})-I(X_{1},X_{2};Z|S,\tilde{S}_{1},\tilde{S}_{2}),
\end{eqnarray}
subject to the power constraints
\begin{eqnarray}\label{ex-3}
\sum_{\tilde{s}_{1}}\pi(\tilde{s}_{1})E[X_{1}^{2}|\tilde{s}_{1}]\leq \mathcal{P}_{1},
\end{eqnarray}
\begin{eqnarray}\label{ex-4}
\sum_{\tilde{s}_{1}}\pi(\tilde{s}_{1})\sum_{\tilde{s}_{2}}P_{\tilde{S}_{2}|\tilde{S}_{1}}(\tilde{s}_{2}|\tilde{s}_{1})
E[X_{2}^{2}|\tilde{s}_{1},\tilde{s}_{2}]\leq \mathcal{P}_{2}.
\end{eqnarray}
Similar to the power definition in \cite{vis},
let $\mathcal{P}_{1}(\tilde{s}_{1})$ and $\mathcal{P}_{2}(\tilde{s}_{1},\tilde{s}_{2})$ be the power respectively allocated to
the states $\tilde{s}_{1}$ and $\tilde{s}_{2}$, i.e., $\mathcal{P}_{1}(\tilde{s}_{1})=E[X_{1}^{2}|\tilde{s}_{1}]$ and
$\mathcal{P}_{2}(\tilde{s}_{1},\tilde{s}_{2})=E[X_{2}^{2}|\tilde{s}_{1},\tilde{s}_{2}]$.
Moreover, let $h(Z )$ be the differential entropy of the continuous random variable $Z$.
Then, we can bound $I(X_{1},X_{2};Y|S,\tilde{S}_{1},\tilde{S}_{2})-I(X_{1},X_{2};Z|S,\tilde{S}_{1},\tilde{S}_{2})$ in (\ref{ex-2}) as follows.
\begin{eqnarray}\label{ex-5}
&&I(X_{1},X_{2};Y|S,\tilde{S}_{1},\tilde{S}_{2})-I(X_{1},X_{2};Z|S,\tilde{S}_{1},\tilde{S}_{2})\nonumber\\
&&=\sum_{\tilde{s}_{1}}\pi(\tilde{s}_{1})\sum_{\tilde{s}_{2}}P_{\tilde{S}_{2}|\tilde{S}_{1}}(\tilde{s}_{2}|\tilde{s}_{1})
\sum_{s}P_{S|\tilde{S}_{2}}(s|\tilde{s}_{2})(I(X_{1},X_{2};Y|S=s,\tilde{S}_{1}=\tilde{s}_{1},\tilde{S}_{2}=\tilde{s}_{2})-\nonumber\\
&&I(X_{1},X_{2};Z|S=s,\tilde{S}_{1}=\tilde{s}_{1},\tilde{S}_{2}=\tilde{s}_{2}))\nonumber\\
&&=\sum_{\tilde{s}_{1}}\pi(\tilde{s}_{1})\sum_{\tilde{s}_{2}}P_{\tilde{S}_{2}|\tilde{S}_{1}}(\tilde{s}_{2}|\tilde{s}_{1})
\sum_{s}P_{S|\tilde{S}_{2}}(s|\tilde{s}_{2})(h(Y|s,\tilde{s}_{1},\tilde{s}_{2})-h(N_{s}|s)-h(Z|s,\tilde{s}_{1},\tilde{s}_{2})+h(h_{3}(s)N_{s}+N_{w}|s))\nonumber\\
&&\stackrel{(b)}\leq \sum_{\tilde{s}_{1}}\pi(\tilde{s}_{1})\sum_{\tilde{s}_{2}}P_{\tilde{S}_{2}|\tilde{S}_{1}}(\tilde{s}_{2}|\tilde{s}_{1})
\sum_{s}P_{S|\tilde{S}_{2}}(s|\tilde{s}_{2})(h(Y|s,\tilde{s}_{1},\tilde{s}_{2})-h(N_{s}|s)\nonumber\\
&&-\frac{1}{2}\log(2^{2h(h_{3}(s)Y|s,\tilde{s}_{1},\tilde{s}_{2})}+2^{2h(N_{w})})+h(h_{3}(s)N_{s}+N_{w}|s))\nonumber\\
&&\stackrel{(c)}=\sum_{\tilde{s}_{1}}\pi(\tilde{s}_{1})\sum_{\tilde{s}_{2}}P_{\tilde{S}_{2}|\tilde{S}_{1}}(\tilde{s}_{2}|\tilde{s}_{1})
\sum_{s}P_{S|\tilde{S}_{2}}(s|\tilde{s}_{2})(h(Y|s,\tilde{s}_{1},\tilde{s}_{2})-h(N_{s}|s)\nonumber\\
&&-\frac{1}{2}\log(2^{2h(Y|s,\tilde{s}_{1},\tilde{s}_{2})}h^{2}_{3}(s)+2^{2h(N_{w})})+h(h_{3}(s)N_{s}+N_{w}|s))\nonumber\\
&&\stackrel{(d)}=\sum_{\tilde{s}_{1}}\pi(\tilde{s}_{1})\sum_{\tilde{s}_{2}}P_{\tilde{S}_{2}|\tilde{S}_{1}}(\tilde{s}_{2}|\tilde{s}_{1})
\sum_{s}P_{S|\tilde{S}_{2}}(s|\tilde{s}_{2})(h(Y|s,\tilde{s}_{1},\tilde{s}_{2})-\frac{1}{2}\log(2\pi e\sigma^{2}_{s})\nonumber\\
&&-\frac{1}{2}\log(2^{2h(Y|s,\tilde{s}_{1},\tilde{s}_{2})}h^{2}_{3}(s)+2\pi e\sigma^{2}_{w})+\frac{1}{2}\log(2\pi e(h^{2}_{3}(s)\sigma^{2}_{s}+\sigma^{2}_{w}))\nonumber\\
&&\stackrel{(e)}\leq\sum_{\tilde{s}_{1}}\pi(\tilde{s}_{1})\sum_{\tilde{s}_{2}}P_{\tilde{S}_{2}|\tilde{S}_{1}}(\tilde{s}_{2}|\tilde{s}_{1})
\sum_{s}P_{S|\tilde{S}_{2}}(s|\tilde{s}_{2})(\frac{1}{2}\log(2\pi eE[(h_{1}(s)X_{1}+h_{2}(s)X_{2}+N_{s})^{2}|s,\tilde{s}_{1},\tilde{s}_{2}])
-\frac{1}{2}\log(2\pi e\sigma^{2}_{s})\nonumber\\
&&-\frac{1}{2}\log(2\pi eE[(h_{1}(s)X_{1}+h_{2}(s)X_{2}+N_{s})^{2}|s,\tilde{s}_{1},\tilde{s}_{2}]h^{2}_{3}(s)
+2\pi e\sigma^{2}_{w})+\frac{1}{2}\log(2\pi e(h^{2}_{3}(s)\sigma^{2}_{s}+\sigma^{2}_{w}))\nonumber\\
&&\stackrel{(f)}=\sum_{\tilde{s}_{1}}\pi(\tilde{s}_{1})\sum_{\tilde{s}_{2}}P_{\tilde{S}_{2}|\tilde{S}_{1}}(\tilde{s}_{2}|\tilde{s}_{1})
\sum_{s}P_{S|\tilde{S}_{2}}(s|\tilde{s}_{2})(\frac{1}{2}\log(2\pi e(h^{2}_{1}(s)\mathcal{P}_{1}(\tilde{s}_{1})
+h^{2}_{2}(s)\mathcal{P}_{2}(\tilde{s}_{1},\tilde{s}_{2})+\sigma^{2}_{s}))-\frac{1}{2}\log(2\pi e\sigma^{2}_{s})\nonumber\\
&&-\frac{1}{2}\log(2\pi e(h^{2}_{1}(s)\mathcal{P}_{1}(\tilde{s}_{1})
+h^{2}_{2}(s)\mathcal{P}_{2}(\tilde{s}_{1},\tilde{s}_{2})+\sigma^{2}_{s})h^{2}_{3}(s)
+2\pi e\sigma^{2}_{w})+\frac{1}{2}\log(2\pi e(h^{2}_{3}(s)\sigma^{2}_{s}+\sigma^{2}_{w}))\nonumber\\
&&=\sum_{\tilde{s}_{1}}\pi(\tilde{s}_{1})\sum_{\tilde{s}_{2}}P_{\tilde{S}_{2}|\tilde{S}_{1}}(\tilde{s}_{2}|\tilde{s}_{1})
\sum_{s}P_{S|\tilde{S}_{2}}(s|\tilde{s}_{2})(\frac{1}{2}\log(1+\frac{h^{2}_{1}(s)\mathcal{P}_{1}(\tilde{s}_{1})
+h^{2}_{2}(s)\mathcal{P}_{2}(\tilde{s}_{1},\tilde{s}_{2})}{\sigma^{2}_{s}})\nonumber\\
&&-\frac{1}{2}\log(1+\frac{h^{2}_{3}(s)h^{2}_{1}(s)\mathcal{P}_{1}(\tilde{s}_{1})
+h^{2}_{3}(s)h^{2}_{2}(s)\mathcal{P}_{2}(\tilde{s}_{1},\tilde{s}_{2})}{h^{2}_{3}(s)\sigma^{2}_{s}+\sigma^{2}_{w}})),
\end{eqnarray}
where (b) follows from the entropy power inequality $2^{2h(h_{3}(s)Y+N_{w}|s,\tilde{s}_{1},\tilde{s}_{2})}\geq
2^{2h(h_{3}(s)Y|s,\tilde{s}_{1},\tilde{s}_{2})}+2^{2h(N_{w}|s,\tilde{s}_{1},\tilde{s}_{2})}$ and the fact that $N_{w}$ is independent of
$S$, $\tilde{S}_{1}$ and $\tilde{S}_{2}$, (c) follows from the property $h(aX)=h(X)+\log a$ for a constant $a$, (d) follows from
$N_{s}\sim \mathcal{N}(0, \sigma^{2}_{s})$ and $N_{w}\sim \mathcal{N}(0, \sigma^{2}_{w})$, (e) follows from
$h(Y|s,\tilde{s}_{1},\tilde{s}_{2})\leq \frac{1}{2}\log(2\pi eE[(h_{1}(s)X_{1}+h_{2}(s)X_{2}+N_{s})^{2}|s,\tilde{s}_{1},\tilde{s}_{2}])$
and the fact that $h(Y|s,\tilde{s}_{1},\tilde{s}_{2})-\frac{1}{2}\log(2^{2h(Y|s,\tilde{s}_{1},\tilde{s}_{2})}h^{2}_{3}(s)+2\pi e\sigma^{2}_{w})$ is increasing while
$h(Y|s,\tilde{s}_{1},\tilde{s}_{2})$ is increasing,
and (f) follows from the definitions $\mathcal{P}_{1}(\tilde{s}_{1})=E[X_{1}^{2}|\tilde{s}_{1}]$ and
$\mathcal{P}_{2}(\tilde{s}_{1},\tilde{s}_{2})=E[X_{2}^{2}|\tilde{s}_{1},\tilde{s}_{2}]$.

Now for the degraded Gaussian fading FS-MAC-WT with only delayed state feedback, we have the following result on
the secrecy sum rate of $\mathcal{C}_{s}^{(dg-out)}$:
\begin{eqnarray}\label{ex-6}
&&R_{1}+R_{2}\nonumber\\
&&\leq
\sum_{\tilde{s}_{1}}\pi(\tilde{s}_{1})\sum_{\tilde{s}_{2}}P_{\tilde{S}_{2}|\tilde{S}_{1}}(\tilde{s}_{2}|\tilde{s}_{1})
\sum_{s}P_{S|\tilde{S}_{2}}(s|\tilde{s}_{2})(\frac{1}{2}\log(1+\frac{h^{2}_{1}(s)\mathcal{P}_{1}(\tilde{s}_{1})
+h^{2}_{2}(s)\mathcal{P}_{2}(\tilde{s}_{1},\tilde{s}_{2})}{\sigma^{2}_{s}})\nonumber\\
&&-\frac{1}{2}\log(1+\frac{h^{2}_{3}(s)h^{2}_{1}(s)\mathcal{P}_{1}(\tilde{s}_{1})
+h^{2}_{3}(s)h^{2}_{2}(s)\mathcal{P}_{2}(\tilde{s}_{1},\tilde{s}_{2})}{h^{2}_{3}(s)\sigma^{2}_{s}+\sigma^{2}_{w}}))\nonumber\\
&&=
\sum_{\tilde{s}_{1}}\pi(\tilde{s}_{1})\sum_{\tilde{s}_{2}}K^{d_{1}-d_{2}}(\tilde{s}_{2},\tilde{s}_{1})
\sum_{s}K^{d_{2}}(s,\tilde{s}_{2})(\frac{1}{2}\log(1+\frac{h^{2}_{1}(s)\mathcal{P}_{1}(\tilde{s}_{1})
+h^{2}_{2}(s)\mathcal{P}_{2}(\tilde{s}_{1},\tilde{s}_{2})}{\sigma^{2}_{s}})\nonumber\\
&&-\frac{1}{2}\log(1+\frac{h^{2}_{3}(s)h^{2}_{1}(s)\mathcal{P}_{1}(\tilde{s}_{1})
+h^{2}_{3}(s)h^{2}_{2}(s)\mathcal{P}_{2}(\tilde{s}_{1},\tilde{s}_{2})}{h^{2}_{3}(s)\sigma^{2}_{s}+\sigma^{2}_{w}}))
\end{eqnarray}
subject to the power constraints
\begin{eqnarray}\label{ex-7}
\sum_{\tilde{s}_{1}}\pi(\tilde{s}_{1})\mathcal{P}_{1}(\tilde{s}_{1})\leq \mathcal{P}_{1},
\end{eqnarray}
\begin{eqnarray}\label{ex-8}
\sum_{\tilde{s}_{1}}\pi(\tilde{s}_{1})\sum_{\tilde{s}_{2}}P_{\tilde{S}_{2}|\tilde{S}_{1}}(\tilde{s}_{2}|\tilde{s}_{1})
\mathcal{P}_{2}(\tilde{s}_{1},\tilde{s}_{2})\leq \mathcal{P}_{2}.
\end{eqnarray}
Then, analogously, the transmission rate $R_{1}$ in $\mathcal{C}_{s}^{(dg-out)}$ can be upper bounded by
\begin{eqnarray}\label{ex-9}
&&R_{1}\leq
\sum_{\tilde{s}_{1}}\pi(\tilde{s}_{1})\sum_{\tilde{s}_{2}}K^{d_{1}-d_{2}}(\tilde{s}_{2},\tilde{s}_{1})
\sum_{s}K^{d_{2}}(s,\tilde{s}_{2})(\frac{1}{2}\log(1+\frac{h^{2}_{1}(s)\mathcal{P}_{1}(\tilde{s}_{1})}{\sigma^{2}_{s}})\nonumber\\
&&-\frac{1}{2}\log(\frac{h^{2}_{1}(s)\mathcal{P}_{1}(\tilde{s}_{1})+\sigma^{2}_{s}+\sigma^{2}_{w}}
{h^{2}_{3}(s)h^{2}_{2}(s)\mathcal{P}_{2}(\tilde{s}_{1},\tilde{s}_{2})+h^{2}_{3}\sigma^{2}_{s}+\sigma^{2}_{w}}))
\end{eqnarray}
subject to the power constraint in (\ref{ex-7}) and (\ref{ex-8}), and $R_{2}$ in $\mathcal{C}_{s}^{(dg-out)}$ can be upper bounded by
\begin{eqnarray}\label{ex-10}
&&R_{2}\leq
\sum_{\tilde{s}_{1}}\pi(\tilde{s}_{1})\sum_{\tilde{s}_{2}}K^{d_{1}-d_{2}}(\tilde{s}_{2},\tilde{s}_{1})
\sum_{s}K^{d_{2}}(s,\tilde{s}_{2})(\frac{1}{2}\log(1+\frac{h^{2}_{2}(s)\mathcal{P}_{2}(\tilde{s}_{1},\tilde{s}_{2})}{\sigma^{2}_{s}})\nonumber\\
&&-\frac{1}{2}\log(\frac{h^{2}_{2}(s)\mathcal{P}_{2}(\tilde{s}_{1},\tilde{s}_{2})+\sigma^{2}_{s}+\sigma^{2}_{w}}
{h^{2}_{3}(s)h^{2}_{1}(s)\mathcal{P}_{1}(\tilde{s}_{1})+h^{2}_{3}\sigma^{2}_{s}+\sigma^{2}_{w}}))
\end{eqnarray}
subject to the power constraint in (\ref{ex-7}) and (\ref{ex-8}).

\section{The Derivation of Corollary 4}

First, we compute the upper bound on the secrecy sum rate of $\mathcal{C}_{sf}^{(dg-out)}$.
From Theorem 2, we know that
\begin{eqnarray}\label{ex-1.2}
R_{1}+R_{2}&\leq&\min\{I(V_{1},V_{2};Y|U,S,\tilde{S}_{1},\tilde{S}_{2}),H(Y|Z,U,S,\tilde{S}_{1},\tilde{S}_{2})\}\nonumber\\
&=&\min\{H(Y|U,S,\tilde{S}_{1},\tilde{S}_{2})-H(Y|V_{1},V_{2},U,S,\tilde{S}_{1},\tilde{S}_{2}),H(Y|Z,U,S,\tilde{S}_{1},\tilde{S}_{2})\}\nonumber\\
&\leq&\min\{H(Y|U,S,\tilde{S}_{1},\tilde{S}_{2})-H(Y|X_{1},X_{2},V_{1},V_{2},U,S,\tilde{S}_{1},\tilde{S}_{2}),H(Y|Z,U,S,\tilde{S}_{1},\tilde{S}_{2})\}\nonumber\\
&\stackrel{(1)}=&\min\{H(Y|U,S,\tilde{S}_{1},\tilde{S}_{2})-H(Y|X_{1},X_{2},S,\tilde{S}_{1},\tilde{S}_{2}),H(Y|Z,U,S,\tilde{S}_{1},\tilde{S}_{2})\}\nonumber\\
&\leq&\min\{H(Y|S,\tilde{S}_{1},\tilde{S}_{2})-H(Y|X_{1},X_{2},S,\tilde{S}_{1},\tilde{S}_{2}),H(Y|Z,S,\tilde{S}_{1},\tilde{S}_{2})\}\nonumber\\
&=&\min\{I(X_{1},X_{2};Y|S,\tilde{S}_{1},\tilde{S}_{2}),H(Y|Z,S,\tilde{S}_{1},\tilde{S}_{2})\},
\end{eqnarray}
where (1) is from the Markov chain $(V_{1},V_{2},U)\rightarrow (S,\tilde{S}_{1},\tilde{S}_{2},X_{1},X_{2})\rightarrow Y$.
Now it remains to compute $I(X_{1},X_{2};Y|S,\tilde{S}_{1},\tilde{S}_{2})$ and $H(Y|Z,S,\tilde{S}_{1},\tilde{S}_{2})$ in (\ref{ex-1.2}), respectively.
First, we bound the conditional mutual information $I(X_{1},X_{2};Y|S,\tilde{S}_{1},\tilde{S}_{2})$, and it is given by
\begin{eqnarray}\label{ex-5.2}
&&I(X_{1},X_{2};Y|S,\tilde{S}_{1},\tilde{S}_{2})\nonumber\\
&&=\sum_{\tilde{s}_{1}}\pi(\tilde{s}_{1})\sum_{\tilde{s}_{2}}P_{\tilde{S}_{2}|\tilde{S}_{1}}(\tilde{s}_{2}|\tilde{s}_{1})
\sum_{s}P_{S|\tilde{S}_{2}}(s|\tilde{s}_{2})I(X_{1},X_{2};Y|S=s,\tilde{S}_{1}=\tilde{s}_{1},\tilde{S}_{2}=\tilde{s}_{2})\nonumber\\
&&=\sum_{\tilde{s}_{1}}\pi(\tilde{s}_{1})\sum_{\tilde{s}_{2}}P_{\tilde{S}_{2}|\tilde{S}_{1}}(\tilde{s}_{2}|\tilde{s}_{1})
\sum_{s}P_{S|\tilde{S}_{2}}(s|\tilde{s}_{2})(h(h_{1}(s)X_{1}+h_{2}(s)X_{2}+N_{s}|s,\tilde{s}_{1},\tilde{s}_{2})-h(N_{s}|s))\nonumber\\
&&\stackrel{(2)}\leq \sum_{\tilde{s}_{1}}\pi(\tilde{s}_{1})\sum_{\tilde{s}_{2}}P_{\tilde{S}_{2}|\tilde{S}_{1}}(\tilde{s}_{2}|\tilde{s}_{1})
\sum_{s}P_{S|\tilde{S}_{2}}(s|\tilde{s}_{2})\frac{1}{2}\log(1+\frac{h^{2}_{1}(s)\mathcal{P}_{1}(\tilde{s}_{1})
+h^{2}_{2}(s)\mathcal{P}_{2}(\tilde{s}_{1},\tilde{s}_{2})}{\sigma^{2}_{s}}),
\end{eqnarray}
where (2) is from the definitions $\mathcal{P}_{1}(\tilde{s}_{1})=E[X_{1}^{2}|\tilde{s}_{1}]$ and
$\mathcal{P}_{2}(\tilde{s}_{1},\tilde{s}_{2})=E[X_{2}^{2}|\tilde{s}_{1},\tilde{s}_{2}]$.

Then, we bound the conditional entropy $H(Y|Z,S,\tilde{S}_{1},\tilde{S}_{2})$, and it is given by
\begin{eqnarray}\label{ex-6.2}
&&H(Y|Z,S,\tilde{S}_{1},\tilde{S}_{2})\nonumber\\
&&=\sum_{\tilde{s}_{1}}\pi(\tilde{s}_{1})\sum_{\tilde{s}_{2}}P_{\tilde{S}_{2}|\tilde{S}_{1}}(\tilde{s}_{2}|\tilde{s}_{1})
\sum_{s}P_{S|\tilde{S}_{2}}(s|\tilde{s}_{2})h(Y|Z,S=s,\tilde{S}_{1}=\tilde{s}_{1},\tilde{S}_{2}=\tilde{s}_{2})\nonumber\\
&&=\sum_{\tilde{s}_{1}}\pi(\tilde{s}_{1})\sum_{\tilde{s}_{2}}P_{\tilde{S}_{2}|\tilde{S}_{1}}(\tilde{s}_{2}|\tilde{s}_{1})
\sum_{s}P_{S|\tilde{S}_{2}}(s|\tilde{s}_{2})(h(Y,Z,S=s,\tilde{S}_{1}=\tilde{s}_{1},\tilde{S}_{2}=\tilde{s}_{2})
-h(Z,S=s,\tilde{S}_{1}=\tilde{s}_{1},\tilde{S}_{2}=\tilde{s}_{2}))\nonumber\\
&&\stackrel{(4)}=\sum_{\tilde{s}_{1}}\pi(\tilde{s}_{1})\sum_{\tilde{s}_{2}}P_{\tilde{S}_{2}|\tilde{S}_{1}}(\tilde{s}_{2}|\tilde{s}_{1})
\sum_{s}P_{S|\tilde{S}_{2}}(s|\tilde{s}_{2})(h(Z|Y)+h(Y,S=s,\tilde{S}_{1}=\tilde{s}_{1},\tilde{S}_{2}=\tilde{s}_{2})
-h(Z,S=s,\tilde{S}_{1}=\tilde{s}_{1},\tilde{S}_{2}=\tilde{s}_{2}))\nonumber\\
&&=\sum_{\tilde{s}_{1}}\pi(\tilde{s}_{1})\sum_{\tilde{s}_{2}}P_{\tilde{S}_{2}|\tilde{S}_{1}}(\tilde{s}_{2}|\tilde{s}_{1})
\sum_{s}P_{S|\tilde{S}_{2}}(s|\tilde{s}_{2})(h(Z|Y)+h(Y|S=s,\tilde{S}_{1}=\tilde{s}_{1},\tilde{S}_{2}=\tilde{s}_{2})
-h(Z|S=s,\tilde{S}_{1}=\tilde{s}_{1},\tilde{S}_{2}=\tilde{s}_{2}))\nonumber\\
&&\stackrel{(5)}\leq\sum_{\tilde{s}_{1}}\pi(\tilde{s}_{1})\sum_{\tilde{s}_{2}}P_{\tilde{S}_{2}|\tilde{S}_{1}}(\tilde{s}_{2}|\tilde{s}_{1})
\sum_{s}P_{S|\tilde{S}_{2}}(s|\tilde{s}_{2})(h(N_{w})+h(Y|S=s,\tilde{S}_{1}=\tilde{s}_{1},\tilde{S}_{2}=\tilde{s}_{2})\nonumber\\
&&-\frac{1}{2}\log(2^{2h(Y|s,\tilde{s}_{1},\tilde{s}_{2})}h^{2}_{3}(s)+2^{2h(N_{w})}))\nonumber\\
&&\stackrel{(6)}\leq \sum_{\tilde{s}_{1}}\pi(\tilde{s}_{1})\sum_{\tilde{s}_{2}}P_{\tilde{S}_{2}|\tilde{S}_{1}}(\tilde{s}_{2}|\tilde{s}_{1})
\sum_{s}P_{S|\tilde{S}_{2}}(s|\tilde{s}_{2})(\frac{1}{2}\log(2\pi e\sigma^{2}_{w})+\frac{1}{2}\log(2\pi e(h^{2}_{1}(s)\mathcal{P}_{1}(\tilde{s}_{1})
+h^{2}_{2}(s)\mathcal{P}_{2}(\tilde{s}_{1},\tilde{s}_{2})+\sigma^{2}_{s}))\nonumber\\
&&-\frac{1}{2}\log(2\pi e(h^{2}_{1}(s)\mathcal{P}_{1}(\tilde{s}_{1})
+h^{2}_{2}(s)\mathcal{P}_{2}(\tilde{s}_{1},\tilde{s}_{2})+\sigma^{2}_{s})h^{2}_{3}(s)+2\pi e\sigma^{2}_{w})\nonumber\\
&&=\sum_{\tilde{s}_{1}}\pi(\tilde{s}_{1})\sum_{\tilde{s}_{2}}P_{\tilde{S}_{2}|\tilde{S}_{1}}(\tilde{s}_{2}|\tilde{s}_{1})
\sum_{s}P_{S|\tilde{S}_{2}}(s|\tilde{s}_{2})(\frac{1}{2}\log(2\pi e\sigma^{2}_{w})\nonumber\\
&&+\frac{1}{2}\log(\frac{h^{2}_{1}(s)\mathcal{P}_{1}(\tilde{s}_{1})
+h^{2}_{2}(s)\mathcal{P}_{2}(\tilde{s}_{1},\tilde{s}_{2})+\sigma^{2}_{s}}{h^{2}_{3}(s)(h^{2}_{1}(s)\mathcal{P}_{1}(\tilde{s}_{1})
+h^{2}_{2}(s)\mathcal{P}_{2}(\tilde{s}_{1},\tilde{s}_{2})+\sigma^{2}_{s})+\sigma^{2}_{w}})),
\end{eqnarray}
where (4) is from the Markov chain $(S,\tilde{S}_{1},\tilde{S}_{2})\rightarrow Y\rightarrow Z$, (5) is from
the entropy power inequality $2^{2h(h_{3}(s)Y+N_{w}|s,\tilde{s}_{1},\tilde{s}_{2})}\geq
2^{2h(h_{3}(s)Y|s,\tilde{s}_{1},\tilde{s}_{2})}+2^{2h(N_{w}|s,\tilde{s}_{1},\tilde{s}_{2})}$, the property $h(aX)=h(X)+\log a$ for a constant $a$
and the fact that $N_{w}$ is independent of $S$, $\tilde{S}_{1}$ and $\tilde{S}_{2}$, and (6) is from $h(N_{w})=\frac{1}{2}\log(2\pi e\sigma^{2}_{w})$,
$h(Y|s,\tilde{s}_{1},\tilde{s}_{2})\leq \frac{1}{2}\log(2\pi eE[(h_{1}(s)X_{1}+h_{2}(s)X_{2}+N_{s})^{2}|s,\tilde{s}_{1},\tilde{s}_{2}])$,
$\mathcal{P}_{1}(\tilde{s}_{1})=E[X_{1}^{2}|\tilde{s}_{1}]$,
$\mathcal{P}_{2}(\tilde{s}_{1},\tilde{s}_{2})=E[X_{2}^{2}|\tilde{s}_{1},\tilde{s}_{2}]$
and the fact that $h(Y|s,\tilde{s}_{1},\tilde{s}_{2})-\frac{1}{2}\log(2^{2h(Y|s,\tilde{s}_{1},\tilde{s}_{2})}h^{2}_{3}(s)+2^{2h(N_{w})})$ is increasing while
$h(Y|s,\tilde{s}_{1},\tilde{s}_{2})$ is increasing.

From (\ref{ex-1.2}), (\ref{ex-5.2}) and (\ref{ex-6.2}), we have the following result:
\begin{eqnarray}\label{ex-comoduo1}
&&R_{1}+R_{2}\nonumber\\
&&\leq \min\{
\sum_{\tilde{s}_{1}}\pi(\tilde{s}_{1})\sum_{\tilde{s}_{2}}K^{d_{1}-d_{2}}(\tilde{s}_{2},\tilde{s}_{1})
\sum_{s}K^{d_{2}}(s,\tilde{s}_{2})\frac{1}{2}\log(1+\frac{h^{2}_{1}(s)\mathcal{P}_{1}(\tilde{s}_{1})
+h^{2}_{2}(s)\mathcal{P}_{2}(\tilde{s}_{1},\tilde{s}_{2})}{\sigma^{2}_{s}}),\nonumber\\
&&\sum_{\tilde{s}_{1}}\pi(\tilde{s}_{1})\sum_{\tilde{s}_{2}}K^{d_{1}-d_{2}}(\tilde{s}_{2},\tilde{s}_{1})
\sum_{s}K^{d_{2}}(s,\tilde{s}_{2})(\frac{1}{2}\log(2\pi e\sigma^{2}_{w})\nonumber\\
&&+\frac{1}{2}\log(\frac{h^{2}_{1}(s)\mathcal{P}_{1}(\tilde{s}_{1})
+h^{2}_{2}(s)\mathcal{P}_{2}(\tilde{s}_{1},\tilde{s}_{2})+\sigma^{2}_{s}}{h^{2}_{3}(s)(h^{2}_{1}(s)\mathcal{P}_{1}(\tilde{s}_{1})
+h^{2}_{2}(s)\mathcal{P}_{2}(\tilde{s}_{1},\tilde{s}_{2})+\sigma^{2}_{s})+\sigma^{2}_{w}}))\}
\end{eqnarray}
subject to the power constraints in (\ref{ex-7}) and (\ref{ex-8}).

Then, analogously, the transmission rate $R_{1}$ in $\mathcal{C}_{sf}^{(dg-out)}$ can be upper bounded by
\begin{eqnarray}\label{ex-9.2}
&&R_{1}\leq
\sum_{\tilde{s}_{1}}\pi(\tilde{s}_{1})\sum_{\tilde{s}_{2}}K^{d_{1}-d_{2}}(\tilde{s}_{2},\tilde{s}_{1})
\sum_{s}K^{d_{2}}(s,\tilde{s}_{2})\frac{1}{2}\log(1+\frac{h^{2}_{1}(s)\mathcal{P}_{1}(\tilde{s}_{1})
+h^{2}_{2}(s)\mathcal{P}_{2}(\tilde{s}_{1},\tilde{s}_{2})}{\sigma^{2}_{s}})\nonumber\\
\end{eqnarray}
subject to the power constraints in (\ref{ex-7}) and (\ref{ex-8}), and $R_{2}$ in $\mathcal{C}_{sf}^{(dg-out)}$ can be upper bounded by
\begin{eqnarray}\label{ex-10.2}
&&R_{2}\leq
\sum_{\tilde{s}_{1}}\pi(\tilde{s}_{1})\sum_{\tilde{s}_{2}}K^{d_{1}-d_{2}}(\tilde{s}_{2},\tilde{s}_{1})
\sum_{s}K^{d_{2}}(s,\tilde{s}_{2})\frac{1}{2}\log(1+\frac{h^{2}_{1}(s)\mathcal{P}_{1}(\tilde{s}_{1})
+h^{2}_{2}(s)\mathcal{P}_{2}(\tilde{s}_{1},\tilde{s}_{2})}{\sigma^{2}_{s}})\nonumber\\
\end{eqnarray}
subject to the power constraints in (\ref{ex-7}) and (\ref{ex-8}).

\end{document}